\documentclass[aps,prd,twocolumn,nofootinbib,superscriptaddress]{revtex4-2}

\usepackage{amsmath,amssymb}
\usepackage{graphicx}
\usepackage{hyperref}
\usepackage{xcolor}
\hypersetup{hidelinks}

\begin{document}

\title{Convexity criterion and radial-profile response for off-shell Kerr
geometries: a fuzzy-dark-matter profile as an analytic benchmark}

\author{Jingxu Wu}
\email{wuxj@my.msu.ru} 
\altaffiliation{These authors contributed equally to this work.}

\author{Jie Shi}
\email{shitcze@my.msu.ru} 
\altaffiliation{These authors contributed equally to this work.}

\affiliation{Faculty of Physics, Lomonosov Moscow State University, Moscow 119991, Russia}

\author{Liangyu Luo}
\email{niu985@sina.com} 
\affiliation{School of International Education, Sechenov First Moscow State Medical University, Moscow 119991, Russia}

\date{\today}

\begin{abstract}
We establish a sufficient one-minimum criterion for the off-shell Kerr family $\Delta(r) = r^2 - 2rm(r) + a^2$ with a positive, nondecreasing mass profile $m(r)$, showing that $1 - 2m'(r) - rm''(r) > 0$ ensures strict convexity and determines root counts for $\Delta$. Using a fuzzy-dark-matter-inspired benchmark satisfying this bound, we derive first-order responses for the outer horizon, extremal branch, photon sphere, and shadow functional under general deformations $m/M_{\text{ADM}} = 1 + \varepsilon h$. We demonstrate that static horizon and photon responses are profile-controlled, spin-odd shadow displacements are completion-dependent, and scale-consistent weak-field limits render local profile-gradient effects negligible ($\ll 10^{-20}$), confirming the strong-field box as a formal radial-profile benchmark rather than a self-consistent rotating scalar-field solution.
\end{abstract}

\maketitle

\section{Introduction}
\label{sec:introduction}

The nature of dark matter remains one of the central open problems in
gravitational physics and cosmology. Among the proposed alternatives to
particle dark matter, ultralight bosonic fields provide a particularly
interesting possibility because their macroscopic de Broglie wavelength can
modify the structure of self-gravitating systems on galactic and
subgalactic scales. In the fuzzy-dark-matter scenario, numerical
Schr\"odinger--Poisson simulations predict a central solitonic core surrounded
by an extended halo, with the inner density commonly represented by a
smooth, finite-density profile
\begin{equation}
\rho_{\rm FDM}(r)
=
\rho_c
\left[
1+\alpha
\left(
\frac{r}{r_c}
\right)^2
\right]^{-8},
\qquad
\alpha=2^{1/8}-1.
\label{eq:intro_fdm_profile}
\end{equation}
This profile is motivated by the wave-dark-matter literature
\cite{WidrowKaiser1993,MatosUrenaLopez2000,Hu2000,Peebles2000,
Schive2014Nature,Schive2014PRL,Marsh2016,Hui2017,Hui2021,
RoblesMatos2012,Schwabe2016,Levkov2018,Eggemeier2019}.
If a massive black hole is embedded inside such a core, the geometry in the
strong-field region is affected not only by the black-hole mass, but also by
the amount of environmental mass enclosed within the horizon and photon
region.

Most studies of black holes in dark-matter environments adopt one of two
approaches. The first treats the environment perturbatively on a fixed Kerr
background, which is appropriate when its gravitational influence is weak.
The second introduces a phenomenological modification of a static or
rotating metric and studies the resulting horizons, circular orbits,
lensing, or shadows
\cite{BarausseCardosoPani2014,Cardoso2019,XuHouWang2018,
HouXuWang2018,Haroon2019,PerlickTsupko2022,PantigOvgun2023}.
Both strategies are useful, but they can obscure an important distinction:
strong-field observables are controlled by the local radial mass function
and its derivatives, whereas the asymptotic spacetime is characterized by
the total ADM mass. Replacing an extended distribution by a constant mass
can therefore preserve the asymptotic charge while eliminating the radial
information responsible for the environmental response.

The rotating problem is especially subtle. Exact stationary and
axisymmetric black-hole solutions sourced by realistic scalar
configurations are generally not available in closed form. A frequently
used alternative is the Newman--Janis algorithm, originally introduced to
generate the Kerr geometry from the Schwarzschild solution
\cite{NewmanJanis1965}. The method has subsequently been applied to many
regular and matter-supported black-hole metrics
\cite{GursesGursey1975,DrakeSzekeres2000,BambiModesto2013,
AzregAinou2014,Toshmatov2014,NevesSaa2014,Toshmatov2017,
Dymnikova2006,Hayward2006,Frolov2016,BeltracchiGondolo2021,
BeltracchiGondolo2021II,SimpsonVisser2022}.
However, the
complexification step is not unique, and a rotating metric obtained in this
way need not solve the same matter equations as the original static seed.
A Newman--Janis geometry must therefore be interpreted carefully: its
algebraic consistency, asymptotic charges, effective source, horizon
structure, and limiting cases should be checked explicitly before its
optical properties are analyzed.

In this work, we construct an effective Kerr-like geometry with a compact
FDM-inspired radial profile. We first integrate the density
profile in Eq.~\eqref{eq:intro_fdm_profile} exactly and retain the complete
enclosed mass,
\begin{equation}
m(r)
=
M_\bullet+M_{\rm FDM}(r),
\label{eq:intro_mass_function}
\end{equation}
rather than replacing the environmental contribution by a constant. The
rotating extension is then defined by the radial function
\begin{equation}
\Delta(r)
=
r^2-2r\,m(r)+a^2.
\label{eq:intro_delta}
\end{equation}
The resulting spacetime is asymptotically Kerr, with
\begin{equation}
M_{\rm ADM}
=
M_\bullet+M_{\rm sol},
\qquad
J_{\rm ADM}
=
aM_{\rm ADM},
\label{eq:intro_adm_charges}
\end{equation}
but differs from Kerr near the black hole because
\(m(r)\neq M_{\rm ADM}\) at finite radius.

The purpose of this construction is not to claim a new self-consistent rotating
Einstein--Klein--Gordon solution. Instead, we use it as a controlled
effective geometry in which the consequences of a finite radial mass
profile can be isolated analytically. This distinction is essential. The
static density uniquely determines the enclosed mass through
\begin{equation}
m'(r)=4\pi r^2\rho(r),
\end{equation}
but it does not determine a unique static pressure closure or redshift
function. Similarly, the Newman--Janis prescription does not uniquely fix
the rotating matter source. We therefore calculate the effective Einstein
tensor of the rotating metric and treat the resulting density and pressures
as consistency diagnostics of the chosen ansatz.

A first objective is to determine when the black-hole classification is
complete for an entire class of radial mass functions, rather than for one
numerical scan.  Defining
\begin{equation}
H(r)=r-m(r)-rm'(r),
\qquad
\Delta'(r)=2H(r),
\end{equation}
one has
\begin{equation}
H'(r)=1-2m'(r)-rm''(r).
\end{equation}
We prove that standard central and asymptotic limits together with
\(H'>0\) make \(\Delta\) strictly convex on \(r>0\).  Thus it has a unique
positive-radius minimum.  For \(a\ne0\) it has no more than two positive
roots; for \(a=0\), the origin is an excluded factor root at the singular
inner boundary and exactly
one simple positive root remains.  The
FDM-inspired profile is then shown to satisfy this general theorem on a
domain substantially wider than the production box.  Extremality is
governed by
\begin{equation}
\Delta(r_e)=0,
\qquad
\Delta'(r_e)=0.
\label{eq:intro_extremality_conditions}
\end{equation}
These yield
\begin{equation}
r_e
=
m(r_e)+r_e m'(r_e),
\qquad
a_{\rm ext}^2
=
r_e^2
\left[
1-2m'(r_e)
\right].
\label{eq:intro_extremality_result}
\end{equation}
These equations are general double-root relations, while uniqueness of
their positive solution follows from the convexity theorem under its stated
hypotheses.  They do not modify a universal angular-momentum inequality. We also determine
the region in which the effective stress tensor
satisfies the weak energy condition throughout the exterior. Horizon
existence and exterior energy-condition satisfaction are shown to define
distinct restrictions on the parameter space.

The derivative identity for \(H'\) is elementary.  The substantive use of
it here is the global completeness result obtained after adding the
central and ADM derivative limits: the criterion certifies when no
secondary horizon extrema or additional extremal branches can have been
missed.  Its portability and its relation to profile-specific
Kerr--Schild, rotating regular-black-hole, and source-derived
multihorizon constructions are set out explicitly in
Sec.~\ref{sec:horizon_extremality}.

A second objective is to derive the photon region and shadow without
numerically integrating individual null geodesics. Because the environmental
modification is confined to the radial function \(\Delta(r)\), the null
Hamilton--Jacobi equation retains a separable Kerr-like form. The critical
impact parameters of unstable spherical photon orbits can therefore be
written as
\begin{equation}
\xi_c(r_p)
=
\frac{
(r_p^2+a^2)\Delta'(r_p)-4r_p\Delta(r_p)
}{
a\Delta'(r_p)
},
\label{eq:intro_xi}
\end{equation}
and
\begin{equation}
\eta_c(r_p)
=
\frac{
16r_p^2\Delta(r_p)
}{
[\Delta'(r_p)]^2
}
-
[\xi_c(r_p)-a]^2.
\label{eq:intro_eta}
\end{equation}
The corresponding critical curve is compared with that of a Kerr black
hole having the same ADM mass, ADM angular momentum, and observer
inclination. This fixed-charge comparison separates genuine radial-profile
effects from a trivial change of mass normalization.

Pantig and \"Ovg\"un studied a closely related soliton-shaped density in a
static, spherical black-hole geometry and calculated its shadow, disk, and
weak-lensing phenomenology \cite{PantigOvgun2023}. Relative to that work,
the present additions are explicitly geometric: a general convexity
theorem for monotone off-shell Kerr mass functions, its strict
profile-specific corollary, fixed-ADM normalization, one selected rotating
extension, and a factorization of the effective-source WEC. Carter
separability, the general critical-impact-parameter formulas, the
\(m(r)\)-Kerr stress tensor, and the algebraic double-root equations are
reused off-shell structures and are not claimed as new results.

Two recent source-based studies sharpen this comparison.  Datta and
Singha derive the frame-dragging function from the linearized
\(t\phi\) Einstein equation for a slowly rotating anisotropic environment
with an explicit fluid angular velocity, and propagate it to light rings,
the ISCO, and epicyclic frequencies \cite{DattaSingha2026}.  Their
\(\omega(r)\) is therefore fixed after a matter rotation law and pressure
closure are supplied.  Here the static one-function closure is different,
the Newman--Janis completion is not source derived, and the slow-rotation
family is used to expose and quantify that missing information.  The
results not contained in their construction are the horizon convexity
theorem and the arbitrary-\(h(r)\) response theory; conversely, the present
work does not provide their source-based orbital-frequency calculation.

Fonseca \emph{et al.} construct self-consistent Einstein-cluster
environments for Hernquist, NFW, and Jaffe profiles and find additional
light rings, secondary horizons, and trapped scalar-wave structures at
high compactness \cite{Fonseca2026}.  Their pressure closure and redshift
equation differ from the selected ansatz here.  Their results are
complementary to the theorem: the theorem supplies a sufficient domain
that excludes additional horizon extrema, whereas their ultracompact
branches demonstrate that such structures can occur after its hypotheses
fail.  Section~\ref{sec:horizon_extremality} makes this relation explicit
for representative Hernquist and Jaffe mass functions and gives a smooth
four-root counterexample.

The distinction becomes sharper beyond the one-function family.
Source-derived static environments generally require an independent
lapse/redshift equation and a pressure closure
\cite{CardosoDestounisDuqueMacedoMaselli2022,Datta2024}.  At finite spin,
Fernandes and Cardoso instead solve the Einstein equations for stationary
axisymmetric black holes supported by an anisotropic fluid and find
spin-enhanced environmental deviations in geodesics, shadows, and energy
conditions \cite{FernandesCardoso2025}.  Bound geodesics in those
source-derived rotating environments need not retain a Carter-like
constant; Destounis and Fernandes find nonintegrability and chaotic layers
\cite{DestounisFernandes2026}.  Separability here is therefore a
structural property of the selected radial-\(\Delta\) completion, not a
generic prediction for rotating matter environments.  Correspondingly,
the finite-spin critical curves below are conditional illustrations
within that completion.

A third objective is to establish the domain of validity of simplified
descriptions. We compare the full radial-profile calculation within the
selected ansatz with a
first-order expansion in the total soliton mass fraction,
\begin{equation}
f_{\rm sol}
=
\frac{M_{\rm sol}}{M_{\rm ADM}},
\label{eq:intro_fsol}
\end{equation}
and with the constant-ADM-mass replacement
\begin{equation}
m(r)\longrightarrow M_{\rm ADM}.
\label{eq:intro_constant_mass}
\end{equation}
The perturbative expansion retains the leading response of the local mass
profile and is accurate when \(f_{\rm sol}\) is sufficiently small and the
geometry is not too close to extremality. By contrast, the constant-mass
replacement returns the Kerr geometry identically and removes all dependence on
\(m'(r)\) and \(m''(r)\). It is therefore a useful reference baseline, but
not a faithful approximation to the distributed environment.

The bookkeeping chain used throughout the paper is
\begin{equation}
\begin{aligned}
\rho_{\rm FDM}(r)&\longrightarrow m(r)
\longrightarrow [m(r),m'(r),m''(r)]
\\
&\longrightarrow
[r_+,a_{\rm ext},r_{\rm ph},\xi_c,\eta_c,{\cal C}_{\rm sh}].
\end{aligned}
\label{eq:intro_main_chain}
\end{equation}
This chain organizes the calculation; it is not itself a new strong-field
mechanism. The principal general contributions are the convexity theorem
and the profile-functional first-order response within the separable
one-function radial-\(\Delta\) family.  The FDM-inspired
specialization supplies a strict analytic corollary, the closed mass
integral and WEC factor, the fixed-charge comparison, and the controlled
one-sided shadow derivative.

Table~\ref{tab:known_new} separates the off-shell Kerr structures used here
from the profile-specific results.  This distinction is important because
separability and the general \(m(r)\) stress tensor are established features
of Kerr--Schild and off-shell Carter geometries
\cite{GursesGursey1975,BenentiFrancaviglia1979,FrolovKubiznak2007,
Krtous2007,Houri2007,FrolovKrtousKubiznak2017,Johannsen2013,
KonoplyaRezzollaZhidenko2016,BeltracchiGondolo2021}.

\begin{table*}[t]
\caption{Relation between established general results and the additions of
this work.  ``General'' below is restricted to the separable one-function
radial-\(\Delta\) family; closed-form or nonperturbative applications are
within the selected effective ansatz.}
\label{tab:known_new}
\begin{ruledtabular}
\begin{tabular}{p{0.25\textwidth}p{0.31\textwidth}p{0.34\textwidth}}
Topic & Established general result & Use or addition in this work \\
\hline
Rotating geometry & Kerr-like metric with
\(\Delta=r^2-2rm(r)+a^2\), Carter separability, and a type-I effective
source & Fixed complexification, closed FDM-inspired mass integral, and
fixed-ADM normalization \\
Energy conditions & Frame eigenvalues expressed through
\(m'\) and \(m''\) & Factorized WEC polynomial, unique
\(r_{\rm WEC}\), and amplitude/sign distinction \\
Horizons & Positive zeros and algebraic double-root conditions for a
general \(\Delta\) & A sufficient strict-convexity theorem for a class of
monotone \(m(r)\), a globally complete root classification, multi-profile
tests, and a four-root counterexample \\
Photon region & General spherical-orbit impact parameters for radial
\(\Delta\) & Endpoint-refined critical curves and fixed-charge Kerr
comparison for the selected profile \\
Profile response & Local perturbation of a radial potential & First-order
\(h,h'\) response of horizons, the extremal branch, photon orbits, and the
fixed-angle shadow functional within the radial-\(\Delta\) family \\
Completion and stability & Slow-rotation freedom and test-field
diagnostics & Separation of profile-determined and completion-dependent
responses, an explicit displacement band, and a static massless-scalar
mode-stability criterion \\
\end{tabular}
\end{ruledtabular}
\end{table*}

The paper is organized as follows. Section~\ref{sec:static_geometry}
introduces the analytic radial-mass benchmark and its selected static seed.
Section~\ref{sec:rotating_geometry} introduces the rotating
Newman--Janis extension and determines its asymptotic charges.
Section~\ref{sec:rotating_effective_source} analyzes the effective source
and the energy conditions. Section~\ref{sec:horizon_extremality} proves
the general horizon theorem, specializes it to the FDM-inspired profile,
and presents the phase diagram. Section~\ref{sec:photon_dynamics} derives the spherical photon
orbits, critical impact parameters, and shadow geometry.
Section~\ref{sec:approximation_hierarchy} gives the general first-order
profile response and compares it with the full-profile and constant-mass
descriptions. The main results and
limitations are summarized in
Sec.~\ref{sec:discussion_conclusions}. Technical derivations and numerical
checks are collected in the appendices: the closure nonuniqueness is
derived in Appendix~\ref{app:static_closure}, Newman--Janis algebra and
curvature/causality checks are given in Appendix~\ref{app:nja_checks}, the
mass and energy-condition reductions are documented in
Appendix~\ref{app:mass_energy}, and the all-root and endpoint-refined
algorithms are specified in Appendix~\ref{app:numerics}.

\section{Static effective geometry generated from the FDM profile}
\label{sec:static_geometry}

We begin by constructing a static and spherically symmetric seed geometry
from a fuzzy-dark-matter-inspired radial density profile.  An important
distinction must be made at the outset.  A prescribed density determines
the Misner--Sharp mass function through the temporal Einstein equation,
but it does not uniquely determine the redshift function or the complete
spacetime metric.  The static geometry used below therefore contains two
logically separate ingredients: the empirical soliton-like density
profile and an additional one-function closure condition.  This
distinction will be essential when the geometry is extended to the
rotating case.

Throughout the article, geometrized units
\begin{equation}
G=c=1
\end{equation}
are used unless physical units are restored explicitly.  The central
black-hole mass parameter is denoted by \(M_{\bullet}\), while
\(M_{\rm FDM}(r)\) denotes the effective dark-matter mass enclosed within
the areal radius \(r\).  The total radial mass function is
\begin{equation}
m(r)=M_{\bullet}+M_{\rm FDM}(r).
\label{eq:total_radial_mass}
\end{equation}

\subsection{Closed-form radial mass within the selected ansatz}
\label{subsec:exact_mass}

We adopt the commonly used soliton-like density profile motivated by
Schr\"odinger--Poisson simulations of fuzzy dark matter
\cite{Schive2014Nature,Schive2014PRL},
\begin{equation}
\begin{aligned}
\rho_{\rm FDM}(r)
&=\rho_{c}
\left[1+\alpha\left(\frac{r}{r_{c}}\right)^{2}\right]^{-8},
\\
\alpha&=2^{1/8}-1\simeq 0.09051.
\end{aligned}
\label{eq:fdm_density}
\end{equation}
Here \(\rho_{c}\) is the central density and \(r_{c}\) is the
half-density radius, since
\begin{equation}
\rho_{\rm FDM}(r_{c})=\frac{\rho_{c}}{2}.
\end{equation}
In practical numerical calculations one may use the rounded value
\(\alpha=0.091\), but the closed expression above is preferable for
analytic manipulations.

The enclosed mass is defined by
\begin{equation}
M_{\rm FDM}(r)
=
4\pi
\int_{0}^{r}
\rho_{\rm FDM}(s)s^{2}\,ds.
\label{eq:fdm_mass_definition}
\end{equation}
Introducing the dimensionless variable
\begin{equation}
x=\frac{r}{r_{c}},
\qquad
y=\sqrt{\alpha}\,x,
\label{eq:dimensionless_xy}
\end{equation}
Eq.~\eqref{eq:fdm_mass_definition} becomes
\begin{equation}
M_{\rm FDM}(r)
=
\frac{4\pi\rho_{c}r_{c}^{3}}{\alpha^{3/2}}
\,{\cal I}(y),
\qquad
{\cal I}(y)
=
\int_{0}^{y}
\frac{z^{2}}{(1+z^{2})^{8}}\,dz.
\label{eq:mass_integral_dimensionless}
\end{equation}
The integral admits a closed elementary representation,
\begin{equation}
{\cal I}(y)
=
\frac{33}{2048}\arctan y
+
\frac{y\,{\cal P}(y^{2})}
{215040(1+y^{2})^{7}},
\label{eq:elementary_mass_integral}
\end{equation}
where
\begin{align}
{\cal P}(u)
={}&
-3465
+48580u
+92323u^{2}
+101376u^{3}
\nonumber\\
&\quad
+65373u^{4}
+23100u^{5}
+3465u^{6}.
\label{eq:mass_polynomial}
\end{align}
The elementary expression is useful for studying asymptotic limits and
for avoiding repeated evaluation of special functions.  Direct
differentiation gives
\begin{equation}
\frac{d{\cal I}}{dy}
=
\frac{y^{2}}{(1+y^{2})^{8}},
\qquad
{\cal I}(0)=0,
\label{eq:elementary_mass_check}
\end{equation}
which proves that Eq.~\eqref{eq:elementary_mass_integral} is the required
primitive.

For symbolic differentiation and stable numerical evaluation close to
the origin, the same mass can be written in hypergeometric form,
\begin{equation}
M_{\rm FDM}(r)
=
\frac{4\pi}{3}\rho_{c}r^{3}
\,{}_2F_{1}
\left(
\frac{3}{2},8;
\frac{5}{2};
-\alpha\frac{r^{2}}{r_{c}^{2}}
\right).
\label{eq:fdm_mass_hypergeometric}
\end{equation}
Equivalently, defining
\begin{equation}
z(r)
=
\frac{\alpha r^{2}}
{r_{c}^{2}+\alpha r^{2}},
\label{eq:beta_argument}
\end{equation}
one obtains the incomplete-beta representation
\begin{equation}
M_{\rm FDM}(r)
=
\frac{2\pi\rho_{c}r_{c}^{3}}{\alpha^{3/2}}
B_{z(r)}
\left(
\frac{3}{2},
\frac{13}{2}
\right).
\label{eq:fdm_mass_beta}
\end{equation}
Equations~\eqref{eq:elementary_mass_integral},
\eqref{eq:fdm_mass_hypergeometric}, and
\eqref{eq:fdm_mass_beta} are analytically equivalent.  In particular,
each representation obeys the defining consistency relation
\begin{equation}
M_{\rm FDM}'(r)
=
4\pi r^{2}\rho_{\rm FDM}(r).
\label{eq:mass_density_identity}
\end{equation}

Near the center, the closed profile mass has the regular expansion
\begin{align}
M_{\rm FDM}(r)
={}&
\frac{4\pi}{3}\rho_{c}r^{3}
\left[
1
-\frac{24\alpha}{5}
\frac{r^{2}}{r_{c}^{2}}
+\frac{108\alpha^{2}}{7}
\frac{r^{4}}{r_{c}^{4}}
\right.
\nonumber\\
&\left.
\hspace{2.8cm}
-40\alpha^{3}
\frac{r^{6}}{r_{c}^{6}}
+
{\cal O}
\left(
\frac{r^{8}}{r_{c}^{8}}
\right)
\right].
\label{eq:mass_central_expansion}
\end{align}
Thus the environmental contribution behaves as \(r^{3}\) at small
radius and does not introduce an additional matter singularity at the
center.

The total mass of the profile is finite.  Taking \(r\rightarrow\infty\)
in any of the closed representations gives
\begin{align}
M_{\rm sol}
\equiv
M_{\rm FDM}(\infty)
&=
\frac{2\pi\rho_{c}r_{c}^{3}}{\alpha^{3/2}}
B
\left(
\frac{3}{2},
\frac{13}{2}
\right)
\nonumber\\
&=
\frac{33\pi^{2}}
{1024\alpha^{3/2}}
\rho_{c}r_{c}^{3}.
\label{eq:total_soliton_mass}
\end{align}
The approach to the asymptotic value is rapid:
\begin{equation}
M_{\rm FDM}(r)
=
M_{\rm sol}
-
\frac{4\pi\rho_{c}r_{c}^{16}}
{13\alpha^{8}r^{13}}
+
{\cal O}(r^{-15}).
\label{eq:mass_asymptotic_expansion}
\end{equation}
Consequently, the exterior geometry is asymptotically Schwarzschild at
leading order, with total asymptotic mass
\begin{equation}
M_{\rm ADM}
=
M_{\bullet}+M_{\rm sol}.
\label{eq:static_adm_mass}
\end{equation}
The interpretation of \(M_{\rm ADM}\) and its relation to the angular
momentum of the rotating extension will be discussed in
Sec.~\ref{sec:rotating_geometry}.

\subsection{Physical scaling and benchmark status}
\label{subsec:physical_scaling}

The four quantities \(\rho_c,r_c,M_{\rm sol}\), and the boson mass
\(m_\phi\) are not independent for a genuine non-self-interacting FDM
ground state.  In a commonly used normalization of the
Schr\"odinger--Poisson scaling symmetry,
\begin{equation}
\begin{aligned}
 \rho_c &\simeq 1.93\times10^{7}
 m_{22}^{-2}\left(\frac{r_c}{\mathrm{kpc}}\right)^{-4}
 \frac{M_\odot}{\mathrm{kpc}^{3}},\\
 m_{22}&\equiv\frac{m_\phi}{10^{-22}\,\mathrm{eV}},
\end{aligned}
\label{eq:fdm_physical_scaling}
\end{equation}
up to convention-dependent order-unity factors in the definition of the
core radius \cite{Schive2014PRL,MarshPop2015,Bar2018,Hui2021}.
Combining Eq.~\eqref{eq:fdm_physical_scaling} with
Eq.~\eqref{eq:total_soliton_mass} gives
\begin{equation}
 M_{\rm sol}\simeq 2.25\times10^{8}
 m_{22}^{-2}\left(\frac{r_c}{\mathrm{kpc}}\right)^{-1}M_\odot.
\label{eq:fdm_mass_radius_scaling}
\end{equation}
Cosmological core--halo relations impose a further, scattered relation to
the host halo \cite{Schive2014PRL,Mocz2017,Veltmaat2018,Mocz2019,
MaySpringel2021,Chan2022}.  A central black hole changes the problem again:
it squeezes the core and, in the black-hole-dominated limit, drives the
solution toward a hydrogenic profile rather than preserving
Eq.~\eqref{eq:fdm_density} \cite{DaviesMocz2020}.

These relations show why the strong-field parameter box used below is not
presented as an astrophysical FDM prediction.  For example, taking
\(M_\bullet=4.0\times10^6M_\odot\), \(r_c=3M_{\rm ADM}\), and
\(f_{\rm sol}=0.1\) gives
\(M_{\rm ADM}=M_\bullet/(1-f_{\rm sol})
\simeq4.444\times10^6M_\odot\) and would require approximately
\begin{equation}
\begin{aligned}
 r_c&=\frac{3GM_{\rm ADM}}{c^2}
 \simeq6.38\times10^{-7}\,\mathrm{pc},\\
 m_\phi&\simeq8.9\times10^{-17}\,\mathrm{eV},
\end{aligned}
\label{eq:formal_sgra_mapping}
\end{equation}
if the isolated-soliton scaling were imposed.  The corresponding
gravitational fine-structure parameter is
\begin{equation}
\alpha_g\equiv\frac{GM_\bullet m_\phi}{\hbar c}\simeq2.7,
\label{eq:formal_sgra_gravitational_coupling}
\end{equation}
which is larger than unity and lies outside
the controlled nonrelativistic regime in which the empirical profile was
calibrated.  Thus no row of the scans with \(r_c/M_{\rm ADM}=2,3,5,10\)
and \(f_{\rm sol}\leq0.3\) is claimed to represent an undisturbed FDM
soliton.  They are deliberately formal benchmarks of a compact radial
profile.  Establishing an astrophysical realization would require solving
the black-hole--soliton system self-consistently and then matching it to a
host halo; that task is outside the present ansatz.

A controlled weak-field scale mapping does exist, but it lies far outside
the strong-field box.  As one explicit example, choose
\begin{equation}
M_\bullet=4.0\times10^6M_\odot,\qquad
m_\phi=10^{-19}\ {\rm eV},\qquad
f_{\rm sol}=10^{-2}.
\label{eq:controlled_mapping_input}
\end{equation}
The isolated-soliton relation then gives
\begin{equation}
\begin{aligned}
M_{\rm ADM}&=4.0404\times10^6M_\odot,&
M_{\rm sol}&=4.0404\times10^4M_\odot,\\
r_c&\simeq5.57\ {\rm pc},&
\frac{r_c}{M_{\rm ADM}}&\simeq2.88\times10^7,\\
\alpha_g&\simeq3.03\times10^{-3}.&&
\end{aligned}
\label{eq:controlled_mapping_output}
\end{equation}
Thus \(\alpha_g\ll1\), \(r_c\gg M_{\rm ADM}\), and
\(f_{\rm sol}\ll1\) hold simultaneously.  Near the photon sphere,
\({\cal F}(3M_{\rm ADM}/r_c)\simeq4.1\times10^{-22}\), so the local
profile gradients are negligible.  In the static selected ansatz this
gives
\begin{equation}
\begin{aligned}
\frac{r_+}{M_{\rm ADM}}&\simeq1.980,&
\frac{r_{\rm ph}}{M_{\rm ADM}}&\simeq2.970,\\
\frac{R_{\rm sh}}{M_{\rm ADM}}&\simeq5.14419.&&
\end{aligned}
\label{eq:controlled_mapping_observables}
\end{equation}
These are one-percent shifts relative to a Schwarzschild geometry
normalized by the \emph{total} ADM mass, but they reduce to below
\(10^{-20}\) at fixed central mass: the apparent change is almost entirely
charge normalization, not a strong-field density-gradient effect.  Black-hole
compression will in addition alter the inner scalar profile.  Fully
relativistic evolutions also show the time-dependent accretion of a
bosonic/FDM soliton by a central black hole
\cite{CardosoIkedaVicenteZilhao2022}.  This example
therefore supplies a controlled physical scale map and also demonstrates
why the compact production box cannot be inferred from ordinary FDM
parameters.

\paragraph{Scope and interpretation.}
\label{par:model_scope}
The results fall into three distinct layers.  The convexity criterion and
the arbitrary-\(h(r)\) response identities apply to every radial
\(\Delta=r^2-2rm(r)+a^2\) satisfying their stated assumptions.  Static
horizon, photon-sphere, and test-field statements are fixed by the
selected one-function closure and its radial mass profile.  Finite-spin
critical curves, the numerical \(j_{\rm ext}\), and the spin-odd shadow
displacement are instead conditional on the selected Newman--Janis
completion.  These layers should not be assigned the same physical
status.

Four limitations then apply throughout and are not repeated in every
later section: the metric is not an Einstein--Klein--Gordon solution; the
compact profile is not claimed to be a realizable, unperturbed FDM core;
the constant-mass replacement is only a Kerr reference; and energy
conditions diagnose the effective Einstein source rather than a
fundamental scalar theory.  Because no parameter region satisfying
relativistic scalar boundary conditions and black-hole compression is
constructed here, all production points are formal compact-profile
benchmarks.  The weak-field scale map is used separately to distinguish
the fixed-ADM normalization effect from the genuinely local
profile-gradient effect.
Where the adjective ``exact'' is retained below for a mathematical
relation, it means exact only within the selected effective ansatz.

\subsection{Static closure, effective source, and geometric properties}
\label{subsec:static_closure}

The density profile in Eq.~\eqref{eq:fdm_density} does not by itself
specify a unique static spacetime.  To exhibit this nonuniqueness
explicitly, consider the general static and spherically symmetric metric
in areal-radius coordinates,
\begin{equation}
ds^{2}
=
-e^{2\Phi(r)}dt^{2}
+
\frac{dr^{2}}
{1-2m(r)/r}
+
r^{2}
\left(
d\theta^{2}
+
\sin^{2}\theta\,d\phi^{2}
\right),
\label{eq:general_static_metric}
\end{equation}
together with an anisotropic effective stress tensor
\begin{equation}
T^{\mu}{}_{\nu}
=
{\rm diag}
\left(
-\rho,
p_{r},
p_{\perp},
p_{\perp}
\right).
\label{eq:anisotropic_stress_tensor}
\end{equation}
The independent \(tt\) and \(rr\) components of the Einstein equations
are
\begin{equation}
m'(r)
=
4\pi r^{2}\rho(r),
\label{eq:einstein_mass_equation}
\end{equation}
and
\begin{equation}
\Phi'(r)
=
\frac{
m(r)+4\pi r^{3}p_{r}(r)
}{
r\,[r-2m(r)]
}.
\label{eq:einstein_redshift_equation}
\end{equation}
Equation~\eqref{eq:einstein_mass_equation} shows that specifying
\(\rho(r)\) fixes \(m(r)\), whereas
Eq.~\eqref{eq:einstein_redshift_equation} shows that the redshift
function remains undetermined until one supplies the radial pressure,
an equation of state, or an equivalent closure relation.

In the present effective construction we impose the one-function
Schwarzschild-gauge closure
\begin{equation}
e^{2\Phi(r)}
=
1-\frac{2m(r)}{r}
\equiv f(r).
\label{eq:schwarzschild_gauge_closure}
\end{equation}
The static seed metric therefore becomes
\begin{equation}
ds^{2}
=
-f(r)dt^{2}
+
\frac{dr^{2}}{f(r)}
+
r^{2}
\left(
d\theta^{2}
+
\sin^{2}\theta\,d\phi^{2}
\right),
\label{eq:static_seed_metric}
\end{equation}
with
\begin{equation}
f(r)
=
1
-
\frac{2M_{\bullet}}{r}
-
\frac{2M_{\rm FDM}(r)}{r}.
\label{eq:static_lapse}
\end{equation}
Differentiating Eq.~\eqref{eq:schwarzschild_gauge_closure} gives
\begin{equation}
\Phi'(r)
=
\frac{
m(r)-rm'(r)
}{
r\,[r-2m(r)]
}.
\label{eq:closure_phi_derivative}
\end{equation}
Comparison with Eq.~\eqref{eq:einstein_redshift_equation}, together
with Eq.~\eqref{eq:einstein_mass_equation}, yields
\begin{equation}
p_{r}(r)=-\rho(r).
\label{eq:radial_equation_of_state}
\end{equation}
Thus the condition \(g_{tt}g_{rr}=-1\) is not merely a coordinate
choice after the areal radius has been fixed.  It is a physical closure
condition selecting a particular anisotropic effective source.

For \(r>0\), the constant term \(M_{\bullet}\) drops out of all radial
derivatives, and the effective density is exactly
\begin{equation}
\rho(r)
=
\frac{m'(r)}{4\pi r^{2}}
=
\rho_{\rm FDM}(r).
\label{eq:effective_density_static}
\end{equation}
The transverse pressure follows either from the angular Einstein
equation or from stress-energy conservation,
\begin{equation}
\nabla_{\mu}T^{\mu}{}_{r}=0.
\end{equation}
Using \(p_{r}=-\rho\), one obtains
\begin{equation}
p_{\perp}(r)
=
-\rho(r)
-
\frac{r}{2}\rho'(r)
=
-\frac{m''(r)}{8\pi r}.
\label{eq:transverse_pressure_static}
\end{equation}
For the profile in Eq.~\eqref{eq:fdm_density},
\begin{equation}
\frac{\rho'(r)}{\rho(r)}
=
-
\frac{
16\alpha r
}{
r_{c}^{2}+\alpha r^{2}
}.
\label{eq:density_log_derivative}
\end{equation}
Defining
\begin{equation}
u
=
\alpha\frac{r^{2}}{r_{c}^{2}},
\label{eq:static_u_variable}
\end{equation}
the pressure ratios take the compact form
\begin{equation}
\frac{p_{r}}{\rho}
=
-1,
\qquad
\frac{p_{\perp}}{\rho}
=
\frac{7u-1}{1+u}.
\label{eq:static_pressure_ratios}
\end{equation}

These expressions allow the static energy conditions to be evaluated
analytically.  Since \(\rho\geq0\),
\begin{equation}
\rho+p_{r}=0,
\label{eq:static_radial_nec}
\end{equation}
so the radial null-energy condition is saturated.  In the transverse
direction,
\begin{equation}
\rho+p_{\perp}
=
\frac{8u}{1+u}\rho
\geq0.
\label{eq:static_transverse_nec}
\end{equation}
The static effective source therefore satisfies the null and weak
energy conditions for all \(r>0\).

For reference, the static strong-energy condition changes sign at
\(r_c/\sqrt{7\alpha}\), while the dominant-energy condition holds only
for \(r\leq r_c/\sqrt{3\alpha}\). Neither is used as a parameter prior.
The rotating WEC is analyzed separately in an orthonormal frame.

The regularity properties of the environmental sector can also be read
off directly.  Near the center,
\begin{equation}
f(r)
=
1
-
\frac{2M_{\bullet}}{r}
-
\frac{8\pi}{3}\rho_{c}r^{2}
+
\frac{64\pi\alpha}{5}
\frac{\rho_{c}r^{4}}{r_{c}^{2}}
+
{\cal O}
\left(
\frac{r^{6}}{r_{c}^{4}}
\right).
\label{eq:static_lapse_center}
\end{equation}
The matter contribution is regular and has the leading form of a
positive-density de Sitter-like core.  Nevertheless, when
\(M_{\bullet}\neq0\), the term \(2M_{\bullet}/r\) retains the usual
central black-hole singularity.  The soliton-like profile regularizes
the environmental density but does not regularize the black-hole
curvature singularity.

For the one-function metric, the Ricci scalar is
\begin{equation}
R
=
\frac{4m'(r)}{r^{2}}
+
\frac{2m''(r)}{r}
=
8\pi
\left[
4\rho(r)+r\rho'(r)
\right].
\label{eq:static_ricci_scalar}
\end{equation}
At the origin, the regular matter contribution approaches
\begin{equation}
R_{\rm FDM}(0)=32\pi\rho_{c},
\label{eq:ricci_center}
\end{equation}
whereas at large radius it decays as \(r^{-16}\).  Using
Eq.~\eqref{eq:mass_asymptotic_expansion}, the lapse behaves as
\begin{equation}
f(r)
=
1
-
\frac{2M_{\rm ADM}}{r}
+
{\cal O}(r^{-14}),
\label{eq:static_lapse_asymptotic}
\end{equation}
which establishes asymptotic flatness and identifies
\(M_{\rm ADM}\) as the total static mass.

\paragraph{Static test-scalar stability.}
A limited but rigorous stability statement can be made without assuming a
fundamental matter model.  For a massless test scalar,
\(\Box\Psi=0\), write
\(\Psi=e^{-i\omega t}Y_{\ell m}(\theta,\phi)u(r)/r\) and introduce
\(dr_*/dr=f^{-1}\).  The radial equation is
\begin{equation}
\frac{d^2u}{dr_*^2}
+\left[\omega^2-V_\ell(r)\right]u=0,
\label{eq:static_scalar_radial}
\end{equation}
with
\begin{equation}
V_\ell(r)
=f(r)\left[
\frac{\ell(\ell+1)}{r^2}
+\frac{2[m(r)-rm'(r)]}{r^3}
\right].
\label{eq:static_scalar_potential}
\end{equation}
If \(f>0\) outside the outer horizon and
\begin{equation}
m(r)-rm'(r)\geq0\qquad (r\geq r_+),
\label{eq:static_scalar_stability_condition}
\end{equation}
then \(V_\ell\geq0\).  Multiplication of
Eq.~\eqref{eq:static_scalar_radial} by \(u^*\), integration over
\(r_*\), and one integration by parts give
\begin{equation}
\begin{aligned}
\int_{-\infty}^{+\infty}
\left(
\left|\frac{du}{dr_*}\right|^2
+V_\ell|u|^2
\right)dr_*
=\omega^2
\int_{-\infty}^{+\infty}|u|^2dr_* ,
\end{aligned}
\label{eq:static_scalar_energy_identity}
\end{equation}
provided the boundary term
\([u^*du/dr_*]_{-\infty}^{+\infty}\) vanishes.  For a putative growing
mode \({\rm Im}\,\omega>0\), the ingoing horizon solution
\(u\sim e^{-i\omega r_*}\) decays as \(r_*\to-\infty\), while the outgoing
solution \(u\sim e^{+i\omega r_*}\) decays as
\(r_*\to+\infty\).  The boundary term therefore vanishes and both norms
are finite.

The left-hand side of
Eq.~\eqref{eq:static_scalar_energy_identity} is real and nonnegative.
Hence \(\omega^2\) must be real and nonnegative.  If
\({\rm Re}\,\omega\ne0\), a growing mode would instead give
\({\rm Im}(\omega^2)=2{\rm Re}\,\omega\,{\rm Im}\,\omega\ne0\); if
\({\rm Re}\,\omega=0\), it would give
\(\omega^2=-({\rm Im}\,\omega)^2<0\).  Both are contradictions.
The argument includes \(\ell=0\), because the second term in
Eq.~\eqref{eq:static_scalar_potential} remains nonnegative under
Eq.~\eqref{eq:static_scalar_stability_condition}.  At \(\omega=0\), the
identity forces \(du/dr_*=0\) and \(V_\ell u=0\); the decaying boundary
conditions then leave only the trivial solution, so there is no
normalizable threshold mode.

This establishes mode stability only: it excludes exponentially growing
massless test-scalar modes, but does not prove boundedness, decay, or
stability of gravitational, matter, or rotating perturbations.  It also
uses the static selected closure \(g_{tt}g_{rr}=-1\)
\cite{Chandrasekhar1983}.

For the present profile,
\begin{equation}
\mu-x\mu'
=1-f_{\rm sol}
+f_{\rm sol}\left[{\cal F}(y)-y{\cal F}'(y)\right],
\qquad y=\frac{x}{\widehat r_c}.
\label{eq:static_scalar_profile_bound}
\end{equation}
The bracket has one minimum because its derivative is
\(-y{\cal F}''(y)\), and
\({\cal F}''(y)=0\) only at
\(y=(7\alpha)^{-1/2}=1.25635\).  At that point,
\begin{equation}
\min_{y\geq0}\left[{\cal F}(y)-y{\cal F}'(y)\right]
=-0.347979.
\end{equation}
Consequently,
\begin{equation}
\mu-x\mu'\geq1-1.347979f_{\rm sol}
\geq0.595606
\label{eq:static_scalar_uniform_margin}
\end{equation}
throughout \(0\leq f_{\rm sol}\leq0.3\), independently of
\(\widehat r_c\).  Every static black-hole point in the production box
therefore satisfies the sufficient massless-scalar stability criterion
with a uniform positive margin.

Two limiting cases provide immediate checks:
\begin{align}
\rho_{c}\rightarrow0
&:
&
f(r)
&\rightarrow
1-\frac{2M_{\bullet}}{r},
\label{eq:schwarzschild_limit}
\\
M_{\bullet}\rightarrow0
&:
&
f(r)
&\rightarrow
1-\frac{2M_{\rm FDM}(r)}{r}.
\label{eq:pure_effective_core_limit}
\end{align}
The first limit reproduces the Schwarzschild geometry.  The second
describes a horizon-bearing or horizonless anisotropic-fluid geometry,
depending on whether \(2M_{\rm FDM}(r)/r\) reaches unity.  It should not
be identified automatically with a relativistic boson-star or
Einstein--Klein--Gordon soliton solution, since its pressure sector is
fixed by the closure
Eq.~\eqref{eq:schwarzschild_gauge_closure}, rather than by an underlying
complex scalar field.

We therefore use the term \emph{FDM-inspired} in a precise and limited
sense: the radial density entering the mass function has the functional
form of the empirical FDM soliton profile.  The metric
Eq.~\eqref{eq:static_seed_metric} is an effective anisotropic-fluid
realization of that profile.  It is neither claimed to be the unique
geometry compatible with the density nor a self-consistent solution of
the Einstein--Klein--Gordon equations in the presence of a central black
hole.  The rotating construction introduced in the following section
inherits this effective character.

\section{Rotating extension and asymptotic charges}
\label{sec:rotating_geometry}

We now construct a stationary and axisymmetric extension of the static
effective geometry derived in Sec.~\ref{sec:static_geometry}.  The
construction is based on a specified Newman--Janis prescription applied to
the one-function seed metric
\begin{equation}
ds^{2}
=
-f(r)\,dt^{2}
+
\frac{dr^{2}}{f(r)}
+
r^{2}d\Omega^{2},
\qquad
f(r)
=
1-\frac{2m(r)}{r},
\label{eq:sec3_static_seed}
\end{equation}
where
\begin{equation}
m(r)
=
M_{\bullet}+M_{\rm FDM}(r).
\label{eq:sec3_radial_mass}
\end{equation}
Here \(M_{\bullet}\) denotes the central mass parameter, while
\(M_{\rm FDM}(r)\) is the closed-form enclosed mass generated by the adopted
FDM-inspired density profile.
The constant-mass limit is the Kerr solution
\cite{Kerr1963,Wald1984,Poisson2004}.

The Newman--Janis algorithm is not unique for a nonvacuum seed geometry.
In particular, different complexification prescriptions may lead to
inequivalent rotating metrics and inequivalent effective sources.  We
therefore regard the prescription specified below as part of the
definition of the model.  The resulting spacetime is an effective
Kerr-like geometry and is not claimed to be a self-consistent rotating
solution of the Einstein--Klein--Gordon equations.

\subsection{Metric construction and charges}
\label{subsec:nja_prescription}

We first introduce an outgoing Eddington--Finkelstein coordinate \(u\)
through
\begin{equation}
du
=
dt-\frac{dr}{f(r)}.
\label{eq:sec3_ef_coordinate}
\end{equation}
The static seed metric becomes
\begin{equation}
ds^{2}
=
-f(r)\,du^{2}
-
2\,du\,dr
+
r^{2}
\left(
d\theta^{2}
+
\sin^{2}\theta\,d\varphi^{2}
\right).
\label{eq:sec3_ef_metric}
\end{equation}
A convenient null tetrad is
\begin{align}
\ell^{\mu}
&=
\delta^{\mu}_{r},
\label{eq:sec3_null_l}
\\
n^{\mu}
&=
\delta^{\mu}_{u}
-
\frac{f(r)}{2}\delta^{\mu}_{r},
\label{eq:sec3_null_n}
\\
m^{\mu}
&=
\frac{1}{\sqrt{2}\,r}
\left(
\delta^{\mu}_{\theta}
+
\frac{i}{\sin\theta}\delta^{\mu}_{\varphi}
\right),
\label{eq:sec3_null_m}
\end{align}
with \(\bar m^{\mu}\) given by complex conjugation.  The inverse metric is
reconstructed as
\begin{equation}
g^{\mu\nu}
=
-\ell^{\mu}n^{\nu}
-\ell^{\nu}n^{\mu}
+
m^{\mu}\bar m^{\nu}
+
m^{\nu}\bar m^{\mu}.
\label{eq:sec3_inverse_tetrad}
\end{equation}

We adopt the complex coordinate transformation
\begin{equation}
u
\longrightarrow
u-ia\cos\theta,
\qquad
r
\longrightarrow
r+ia\cos\theta,
\label{eq:sec3_complex_transformation}
\end{equation}
where \(a\) is a real rotation parameter with dimensions of length.  The
real section is defined by the replacements
\begin{equation}
r^{2}
\longrightarrow
\Sigma,
\qquad
\frac{1}{r}
\longrightarrow
\frac{r}{\Sigma},
\qquad
m(r)
\longrightarrow
m(\operatorname{Re}r),
\label{eq:sec3_real_section_rules}
\end{equation}
with
\begin{equation}
\Sigma
=
r^{2}+a^{2}\cos^{2}\theta.
\label{eq:sec3_sigma}
\end{equation}
After the real section is imposed, \(r\) is again taken to be real and
the mass function is evaluated at this real radial coordinate.  Thus the
complexified lapse is represented by
\begin{equation}
\widetilde f(r,\theta)
=
1-\frac{2r\,m(r)}{\Sigma}.
\label{eq:sec3_complexified_lapse}
\end{equation}

To obtain Boyer--Lindquist-type coordinates, we perform the differential
transformation
\begin{align}
du
&=
dt
-
\frac{r^{2}+a^{2}}{\Delta(r)}\,dr,
\label{eq:sec3_bl_time}
\\
d\varphi
&=
d\phi
-
\frac{a}{\Delta(r)}\,dr,
\label{eq:sec3_bl_phi}
\end{align}
where the generalized radial function is
\begin{equation}
\Delta(r)
=
r^{2}-2r\,m(r)+a^{2}.
\label{eq:sec3_delta}
\end{equation}
The essential feature of this prescription is that
\(\Delta\) remains a function of \(r\) alone.  This property will later
permit Hamilton--Jacobi separation for null geodesics.

The resulting rotating line element is
\begin{align}
ds^{2}
={}&
-
\left(
1-\frac{2r\,m(r)}{\Sigma}
\right)dt^{2}
-
\frac{4ar\,m(r)\sin^{2}\theta}{\Sigma}
\,dt\,d\phi
\nonumber\\
&+
\frac{\Sigma}{\Delta(r)}\,dr^{2}
+
\Sigma\,d\theta^{2}
\nonumber\\
&+
\left[
r^{2}+a^{2}
+
\frac{2a^{2}r\,m(r)\sin^{2}\theta}{\Sigma}
\right]
\sin^{2}\theta\,d\phi^{2}.
\label{eq:sec3_rotating_metric}
\end{align}
For later use, its nonvanishing covariant components are
\begin{align}
g_{tt}
&=
-
\left(
1-\frac{2r\,m(r)}{\Sigma}
\right),
\label{eq:sec3_gtt}
\\
g_{t\phi}
&=
-
\frac{2ar\,m(r)\sin^{2}\theta}{\Sigma},
\label{eq:sec3_gtphi}
\\
g_{rr}
&=
\frac{\Sigma}{\Delta(r)},
\label{eq:sec3_grr}
\\
g_{\theta\theta}
&=
\Sigma,
\label{eq:sec3_gthetatheta}
\\
g_{\phi\phi}
&=
\left[
r^{2}+a^{2}
+
\frac{2a^{2}r\,m(r)\sin^{2}\theta}{\Sigma}
\right]\sin^{2}\theta.
\label{eq:sec3_gphiphi}
\end{align}

The determinant, inverse metric, Boyer--Lindquist transformation, static
limit, and constant-\(m\) Kerr limit are recorded once in
Appendix~\ref{app:nja_checks}, where they are also checked symbolically.

The closed-form enclosed mass derived in Sec.~\ref{sec:static_geometry}
approaches a finite asymptotic value,
\begin{equation}
M_{\rm FDM}(r)
=
M_{\rm sol}
-
\frac{
4\pi\rho_{c}r_{c}^{16}
}{
13\alpha^{8}r^{13}
}
+
{\cal O}(r^{-15}).
\label{eq:sec3_mass_asymptotic}
\end{equation}
Consequently,
\begin{equation}
m(r)
=
M_{\rm ADM}
-
\frac{
4\pi\rho_{c}r_{c}^{16}
}{
13\alpha^{8}r^{13}
}
+
{\cal O}(r^{-15}),
\label{eq:sec3_radial_mass_asymptotic}
\end{equation}
where
\begin{equation}
M_{\rm ADM}
=
M_{\bullet}+M_{\rm sol}.
\label{eq:sec3_adm_mass}
\end{equation}

The leading asymptotic metric components are therefore
\begin{align}
g_{tt}
&=
-1
+
\frac{2M_{\rm ADM}}{r}
+
{\cal O}(r^{-2}),
\label{eq:sec3_asymptotic_gtt}
\\
g_{rr}
&=
1
+
\frac{2M_{\rm ADM}}{r}
+
{\cal O}(r^{-2}),
\label{eq:sec3_asymptotic_grr}
\\
g_{t\phi}
&=
-
\frac{
2aM_{\rm ADM}\sin^{2}\theta
}{r}
+
{\cal O}(r^{-2}).
\label{eq:sec3_asymptotic_gtphi}
\end{align}
Comparison with the standard asymptotically flat stationary expansion,
\begin{equation}
g_{t\phi}
=
-
\frac{
2J_{\rm ADM}\sin^{2}\theta
}{r}
+
{\cal O}(r^{-2}),
\label{eq:sec3_standard_spin_expansion}
\end{equation}
gives
\begin{equation}
{
J_{\rm ADM}
=
aM_{\rm ADM}
}.
\label{eq:sec3_adm_angular_momentum}
\end{equation}
Thus \(a\) is the specific angular momentum parameter of the complete
effective spacetime, rather than of the central mass parameter alone.
The natural dimensionless spin is
\begin{equation}
{
j_{\rm ADM}
\equiv
\frac{
J_{\rm ADM}
}{
M_{\rm ADM}^{2}
}
=
\frac{a}{M_{\rm ADM}}
}.
\label{eq:sec3_dimensionless_adm_spin}
\end{equation}

This distinction becomes important whenever the total environmental mass
is not negligible relative to \(M_{\bullet}\).  Defining
\begin{equation}
f_{\rm sol}
\equiv
\frac{M_{\rm sol}}{M_{\rm ADM}},
\label{eq:sec3_soliton_fraction}
\end{equation}
one has
\begin{equation}
\frac{M_{\bullet}}{M_{\rm ADM}}
=
1-f_{\rm sol}.
\label{eq:sec3_bh_mass_fraction}
\end{equation}
The rotation parameter normalized by the central mass is related to the
physical ADM spin through
\begin{equation}
\frac{a}{M_{\bullet}}
=
\frac{
j_{\rm ADM}
}{
1-f_{\rm sol}
}.
\label{eq:sec3_spin_conversion}
\end{equation}
Therefore \(a/M_{\bullet}\) can exceed unity even when
\(j_{\rm ADM}<1\).  Such a value does not by itself imply a
superextremal spacetime, because the extremality condition must be
formulated using the complete radial function \(\Delta(r)\) and the ADM
normalization.  The generalized extremality problem will be analyzed in
Sec.~\ref{sec:horizon_extremality}.

\subsection{Radial response and model scope}
\label{subsec:frame_dragging_response}

It is useful to separate the total asymptotic mass from its radial
distribution.  Let
\begin{equation}
{\cal F}(x)
\equiv
\frac{
M_{\rm FDM}(r)
}{
M_{\rm sol}
},
\qquad
x=\frac{r}{r_{c}}.
\label{eq:sec3_mass_fraction_function}
\end{equation}
For the adopted profile,
\begin{equation}
{\cal F}(x)
=
I_{z(x)}
\left(
\frac{3}{2},
\frac{13}{2}
\right),
\label{eq:sec3_regularized_beta_mass}
\end{equation}
where \(I_z(a,b)\) is the regularized incomplete beta function and
\begin{equation}
z(x)
=
\frac{
\alpha x^{2}
}{
1+\alpha x^{2}
}.
\label{eq:sec3_z_function}
\end{equation}
The complete radial mass function can then be written in the
dimensionless form
\begin{equation}
{
\frac{m(r)}{M_{\rm ADM}}
=
1-f_{\rm sol}
+
f_{\rm sol}
{\cal F}
\left(
\frac{r}{r_{c}}
\right)
}.
\label{eq:sec3_dimensionless_mass_profile}
\end{equation}
At small radius,
\begin{equation}
\frac{m(r)}{M_{\rm ADM}}
\longrightarrow
1-f_{\rm sol},
\label{eq:sec3_mass_small_radius}
\end{equation}
whereas asymptotically
\begin{equation}
\frac{m(r)}{M_{\rm ADM}}
\longrightarrow
1.
\label{eq:sec3_mass_large_radius}
\end{equation}

To quantify the radial correction to the rotational sector, we compare
the effective metric with a Kerr spacetime having the same
\(M_{\rm ADM}\) and the same \(a\).  The Kerr frame-dragging component is
\begin{equation}
g_{t\phi}^{\rm Kerr}
=
-
\frac{
2aM_{\rm ADM}r\sin^{2}\theta
}{
\Sigma
}.
\label{eq:sec3_kerr_gtphi}
\end{equation}
We define
\begin{equation}
\delta_{\rm FD}(r,\theta)
\equiv
\frac{
g_{t\phi}(r,\theta)
-
g_{t\phi}^{\rm Kerr}(r,\theta)
}{
g_{t\phi}^{\rm Kerr}(r,\theta)
}.
\label{eq:sec3_frame_dragging_definition}
\end{equation}
Using Eqs.~\eqref{eq:sec3_gtphi} and
\eqref{eq:sec3_kerr_gtphi}, this quantity reduces exactly to
\begin{align}
\delta_{\rm FD}(r,\theta)
&=
\frac{m(r)}{M_{\rm ADM}}-1
\nonumber\\
&=
-f_{\rm sol}
\left[
1-
{\cal F}
\left(
\frac{r}{r_{c}}
\right)
\right].
\label{eq:sec3_frame_dragging_exact}
\end{align}
For fixed \(M_{\rm ADM}\) and \(a\), the relative correction is
independent of \(\theta\), although the metric component itself retains
its usual angular dependence.  In the equatorial plane,
\begin{equation}
g_{t\phi}(r,\pi/2)
=
-\frac{2a\,m(r)}{r},
\label{eq:sec3_equatorial_gtphi}
\end{equation}
and Eq.~\eqref{eq:sec3_frame_dragging_exact} directly measures the
relative change of this component.

Because
\begin{equation}
m(r)\leq M_{\rm ADM},
\end{equation}
the correction satisfies
\begin{equation}
-f_{\rm sol}
\leq
\delta_{\rm FD}(r,\theta)
\leq
0.
\label{eq:sec3_frame_dragging_bounds}
\end{equation}
The negative sign means that, at fixed ADM mass and angular momentum, the
magnitude of \(g_{t\phi}\) is reduced at radii that do not enclose the
full environmental mass.  The Kerr behavior is recovered at infinity,
\begin{equation}
\lim_{r\rightarrow\infty}
\delta_{\rm FD}(r,\theta)
=
0.
\label{eq:sec3_frame_dragging_asymptotic}
\end{equation}

\paragraph{Slow-rotation completion uncertainty.}
The freedom in the rotating completion can be exposed without choosing a
second complexification.  The most general circular, equatorially
symmetric first-order slow-rotation extension of the same static seed may
be written, after fixing the areal-radius gauge, as
\begin{equation}
ds^2=ds_0^2
-2a r^2\omega(r)\sin^2\theta\,dt\,d\phi
+{\cal O}(a^2),
\label{eq:sec3_slow_rotation_general}
\end{equation}
where \(ds_0^2\) is Eq.~\eqref{eq:sec3_static_seed}.  This is the
Hartle slow-rotation sector \cite{Hartle1967,HartleThorne1968}.  The
selected Newman--Janis metric gives
\begin{equation}
\omega_{\rm NJ}(r)=\frac{2m(r)}{r^3}.
\label{eq:sec3_slow_rotation_nj}
\end{equation}
By contrast, the static density and closure determine neither the
azimuthal matter current nor a unique \(\omega(r)\).  An equally
asymptotically normalized completion can be parameterized as
\begin{equation}
\omega(r)=\frac{2m(r)}{r^3}
+\lambda\,\delta\omega(r),
\qquad
\lim_{r\to\infty}r^3\delta\omega(r)=0,
\label{eq:sec3_slow_rotation_family}
\end{equation}
where \(\lambda\) measures completion uncertainty.  Determining
\(\delta\omega\) requires an independent equation for the rotating source,
not merely \(m'(r)=4\pi r^2\rho\).
Datta and Singha supply precisely such extra information: after choosing
an anisotropic pressure closure and a fluid angular velocity, they solve
the linearized \(t\phi\) Einstein equation for \(\omega(r)\)
\cite{DattaSingha2026}.  Their construction is source based and therefore
physically stronger in the slow-rotation regime.  The purpose here is
different: to display the completion freedom left by the selected static
closure and propagate it into the shadow observable.

This comparison separates robust and conditional responses.  At
\({\cal O}(a)\), \(g^{rr}=f+{\cal O}(a^2)\), so changing
\(\delta\omega\) does not shift the static outer horizon:
\begin{equation}
\left.\delta_\lambda r_+\right|_{{\cal O}(a)}=0.
\label{eq:sec3_completion_horizon}
\end{equation}
The static photon-sphere and shadow-radius responses are likewise fixed by
\(m(r)\).  The spin-odd shadow displacement, however, depends directly on
\(\omega(r)\) and is completion dependent already at
\({\cal O}(a)\).  Reflection symmetry makes the area-equivalent shadow
radius even in \(a\), so its first completion-dependent term is
\({\cal O}(a^2)\).  Finally, a Hartle expansion cannot test the
near-extremal branch \(a\sim M_{\rm ADM}\); the reported
\(j_{\rm ext}\) remains specific to the selected off-shell completion.
Schematically, for fixed static \(m(r)\),
\begin{equation}
\begin{aligned}
\delta_\lambda r_+&={\cal O}(\lambda a^2),&
\delta_\lambda R_A&={\cal O}(\lambda a^2),\\
\delta_\lambda D_{\rm sh}&={\cal O}(\lambda a).&&
\end{aligned}
\label{eq:sec3_completion_error_budget}
\end{equation}
where \(D_{\rm sh}\) denotes a spin-odd horizontal displacement.

The displacement dependence can be made quantitative.  Define
\begin{equation}
\bar\omega(x)=M_{\rm ADM}^2\omega(r),\qquad
F_0(x)=1-\frac{2\mu(x)}{x},
\end{equation}
and let \(x_0\) be the static photon radius,
\(x_0F_0'(x_0)-2F_0(x_0)=0\), with
\(\widehat b_0^2=x_0^2/F_0(x_0)\).  Separating the null Hamilton--Jacobi
equation to first order in \(j_{\rm ADM}\) gives
\begin{equation}
\frac{{\cal R}}{E^2}
=1-2j_{\rm ADM}\bar\omega(x)\widehat\xi
-\frac{F_0(x)\widehat b^2}{x^2}
+{\cal O}(j_{\rm ADM}^2).
\label{eq:sec3_slow_rotation_radial_potential}
\end{equation}
Here \(\widehat\xi=L_z/(EM_{\rm ADM})\) and
\(\widehat b^2={\cal K}/(E^2M_{\rm ADM}^2)\), with \({\cal K}\) the
total angular separation constant of the spherical background.
The circular-orbit conditions then yield a translated circle,
\begin{equation}
\frac{X_c}{M_{\rm ADM}}
=j_{\rm ADM}\widehat b_0^2\bar\omega(x_0)\sin\iota
+{\cal O}(j_{\rm ADM}^2),
\label{eq:sec3_slow_rotation_shadow_center}
\end{equation}
so only the value of the completion at the static photon radius enters at
this order.

For an explicit same-charge, same-Kerr-limit benchmark, choose
\begin{equation}
\begin{aligned}
\bar\omega_\kappa(x)
&=\frac{2\mu(x)}{x^3}\\
&\quad+\kappa\frac{2[1-\mu(x)]}{x^3}
\frac{\widehat L}{x+\widehat L},\\[-0.2ex]
&\qquad |\kappa|\leq1,\quad\widehat L=3.
\end{aligned}
\label{eq:sec3_explicit_alternative_completion}
\end{equation}
The additional term vanishes when \(f_{\rm sol}=0\), decays faster than
\(x^{-3}\), and leaves \(J_{\rm ADM}\) unchanged.  It is not claimed to
solve a specified matter equation; it is a controlled completion-uncertainty
probe whose amplitude is tied to the unenclosed mass fraction.  Its
half-width for the displacement is
\begin{equation}
\frac{\delta_\kappa D_{\rm sh}}{M_{\rm ADM}}
=j_{\rm ADM}\widehat b_0^2
\frac{2[1-\mu(x_0)]}{x_0^3}
\frac{\widehat L}{x_0+\widehat L}\sin\iota.
\label{eq:sec3_completion_band}
\end{equation}
For \(\widehat r_c=3\), \(j_{\rm ADM}=0.2\), and
\(\iota=60^\circ\), this gives \(0.01687\) at
\(f_{\rm sol}=0.1\) and \(0.04343\) at \(f_{\rm sol}=0.2\).
The band is displayed in Fig.~\ref{fig:shadow_relative_kerr}.  Thus the
slow-rotation comparison does not remove the nonuniqueness: it quantifies
both the order and a representative numerical size.

\begin{figure*}[t]
\centering
\includegraphics[
width=0.98\textwidth
]{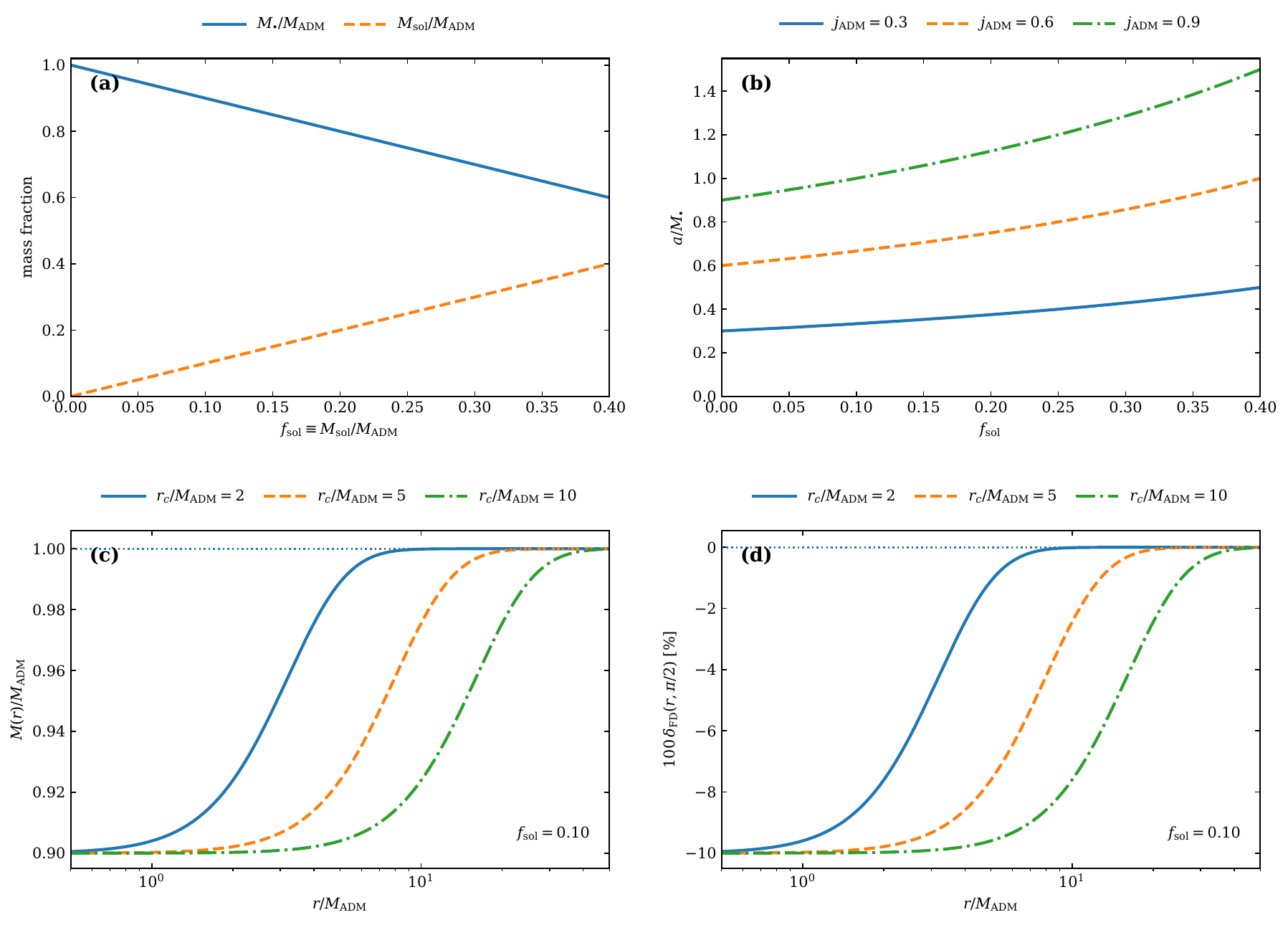}
\caption{
Asymptotic normalization and radial rotational response of the
FDM-inspired Kerr-like geometry.
Panel (a) shows the decomposition of the ADM mass into the central
black-hole contribution and the effective soliton contribution.
Panel (b) compares the central-mass-normalized parameter
$a/M_{\bullet}$ with the physical ADM spin
$j_{\rm ADM}=a/M_{\rm ADM}$ for three fixed values of
$j_{\rm ADM}$.
Panel (c) displays the dimensionless radial mass function for
$f_{\rm sol}=0.10$ and several core radii.
Panel (d) shows the relative correction
$\delta_{\rm FD}$ to the equatorial frame-dragging metric component,
with respect to a Kerr spacetime having the same ADM mass and angular
momentum.
Panels (a,b) extend to \(f_{\rm sol}=0.4\) only to illustrate charge
normalization; every production scan and quantitative bound in the paper
uses \(f_{\rm sol}\leq0.3\).
The correction is largest in the inner region and vanishes
asymptotically as the complete soliton mass becomes enclosed.
}
\label{fig:asymptotic_charges}
\end{figure*}

Figure~\ref{fig:asymptotic_charges} summarizes the mass normalization and
the radial rotational response.

Panel~(a) of Fig.~\ref{fig:asymptotic_charges} illustrates the mass
identity
\begin{equation}
M_{\rm ADM}
=
M_{\bullet}+M_{\rm sol}.
\end{equation}
Increasing \(f_{\rm sol}\) at fixed \(M_{\rm ADM}\) decreases the central
mass fraction linearly.  Panel~(b) demonstrates the corresponding change
in spin normalization.  For example, a configuration with
\(j_{\rm ADM}=0.9\) and \(f_{\rm sol}=0.4\) has
\begin{equation}
\frac{a}{M_{\bullet}}
=
\frac{0.9}{0.6}
=
1.5,
\label{eq:sec3_spin_example}
\end{equation}
although the physical dimensionless ADM spin remains \(0.9\).  This
example shows why \(a/M_{\bullet}\) should not be used as the primary
spin variable for the full spacetime.

Panel~(c) shows the radial mass function for a representative total
environmental fraction \(f_{\rm sol}=0.10\).  A compact core reaches the
asymptotic mass at smaller \(r/M_{\rm ADM}\), whereas a more extended
core retains a substantial fraction of its mass outside the strong-field
region.  Consequently, two geometries with the same ADM mass and the
same asymptotic spin can differ significantly in their local values of
\(m(r)\).

Panel~(d) displays the corresponding frame-dragging correction.  For the
chosen illustration, the inner-region correction approaches
approximately \(-10\%\), reflecting the fact that the central mass
fraction is \(1-f_{\rm sol}=0.90\).  The correction then approaches zero
as \(r\) becomes larger than the core scale.  The figure is intended as a
dimensionless theoretical diagnostic and does not represent an
astrophysical constraint on the soliton fraction.

The asymptotic relation
\begin{equation}
J_{\rm ADM}=aM_{\rm ADM}
\end{equation}
has an important physical implication.  The Newman--Janis construction
does not leave the environmental component strictly static while
rotating only the central mass parameter.  Instead, the full radial mass
function \(m(r)\) enters the off-diagonal metric component,
\begin{equation}
g_{t\phi}
=
-\frac{
2ar\,m(r)\sin^{2}\theta
}{
\Sigma
},
\end{equation}
so that the effective source supporting the geometry must carry a
stationary axisymmetric momentum distribution.

Accordingly, the spacetime should be interpreted as a rotating effective
geometry generated from an FDM-inspired radial mass profile, rather than
as the unique geometry of a Kerr black hole embedded in an otherwise
static scalar soliton.  Determining whether a fundamental rotating scalar
configuration can reproduce the same metric would require solving the
stationary axisymmetric Einstein--Klein--Gordon system with appropriate
boundary conditions.

The present construction nevertheless has several useful properties.
First, it is asymptotically flat with well-defined ADM mass and angular
momentum.  Second, it reduces exactly to the static effective geometry
when \(a=0\).  Third, it reduces to Kerr when the environmental density
vanishes.  Fourth, the generalized radial function
\(\Delta(r)\) depends only on \(r\), which preserves the canonical
separable structure of the null Hamilton--Jacobi equation.

The rotating geometry also introduces new consistency questions that are
absent in the static case.  In particular, the Newman--Janis procedure
changes the effective pressure and momentum sectors, and the static
relations
\begin{equation}
p_r=-\rho,
\qquad
p_{\perp}
=
-\rho-\frac{r}{2}\rho'
\end{equation}
cannot be transferred directly to the rotating spacetime.  The complete
effective stress tensor must instead be reconstructed from the Einstein
tensor in an orthonormal frame.  This analysis is carried out in
Sec.~\ref{sec:rotating_effective_source}.

\section{Effective source and energy conditions}
\label{sec:rotating_effective_source}

The rotating metric constructed in
Sec.~\ref{sec:rotating_geometry} is not a vacuum geometry whenever
\(m'(r)\neq0\).  Its matter content must therefore be reconstructed from
the Einstein tensor rather than inferred directly from the static
density profile.  This distinction is essential because the
Newman--Janis transformation modifies not only the metric but also the
effective pressure and momentum sectors.

We work with the signature \((-+++)\) and use
\begin{equation}
G_{\mu\nu}=8\pi T_{\mu\nu}.
\label{eq:sec4_einstein_equation}
\end{equation}
All local energy conditions discussed below are evaluated in an
orthonormal frame outside coordinate singularities and away from the
surface \(\Sigma=0\).

\subsection{Stress tensor and weak energy condition}
\label{subsec:carter_effective_source}

For the rotating line element
Eq.~\eqref{eq:sec3_rotating_metric}, a convenient orthonormal coframe is
\begin{align}
\vartheta^{(0)}
&=
\sqrt{\frac{\Delta}{\Sigma}}
\left(
dt-a\sin^{2}\theta\,d\phi
\right),
\label{eq:sec4_coframe_0}
\\
\vartheta^{(1)}
&=
\sqrt{\frac{\Sigma}{\Delta}}\,dr,
\label{eq:sec4_coframe_1}
\\
\vartheta^{(2)}
&=
\sqrt{\Sigma}\,d\theta,
\label{eq:sec4_coframe_2}
\\
\vartheta^{(3)}
&=
\frac{\sin\theta}{\sqrt{\Sigma}}
\left[
\left(r^{2}+a^{2}\right)d\phi-a\,dt
\right],
\label{eq:sec4_coframe_3}
\end{align}
where
\begin{equation}
\Sigma=r^{2}+a^{2}\cos^{2}\theta,
\qquad
\Delta=r^{2}-2r\,m(r)+a^{2}.
\label{eq:sec4_sigma_delta}
\end{equation}
The metric is then
\begin{equation}
ds^{2}
=
-\left(\vartheta^{(0)}\right)^{2}
+\left(\vartheta^{(1)}\right)^{2}
+\left(\vartheta^{(2)}\right)^{2}
+\left(\vartheta^{(3)}\right)^{2}.
\label{eq:sec4_orthonormal_metric}
\end{equation}

In this frame, the Einstein tensor is diagonal and can be written as an
effective anisotropic stress tensor,
\begin{equation}
T^{(a)}{}_{(b)}
=
\operatorname{diag}
\left(
-\rho_{\rm rot},
p_{r},
p_{\perp},
p_{\perp}
\right).
\label{eq:sec4_stress_diagonal}
\end{equation}
Direct evaluation gives
\begin{equation}
8\pi\rho_{\rm rot}
=
\frac{2r^{2}m'(r)}{\Sigma^{2}},
\label{eq:sec4_rotating_density}
\end{equation}
\begin{equation}
8\pi p_{r}
=
-\frac{2r^{2}m'(r)}{\Sigma^{2}},
\label{eq:sec4_rotating_radial_pressure}
\end{equation}
and
\begin{equation}
8\pi p_{\perp}
=
-
\frac{
r\Sigma m''(r)
+
2a^{2}\cos^{2}\theta\,m'(r)
}{
\Sigma^{2}
}.
\label{eq:sec4_rotating_transverse_pressure}
\end{equation}
Consequently,
\begin{equation}
{
p_{r}=-\rho_{\rm rot}
}
\label{eq:sec4_radial_saturation}
\end{equation}
also holds in the rotating orthonormal frame.

The constant central contribution \(M_{\bullet}\) drops out of
\(m'(r)\) and \(m''(r)\).  Since
\begin{equation}
m'(r)
=
M_{\rm FDM}'(r)
=
4\pi r^{2}\rho_{\rm FDM}(r),
\label{eq:sec4_mass_derivative}
\end{equation}
the rotating-frame energy density can be expressed as
\begin{equation}
{
\rho_{\rm rot}(r,\theta)
=
\frac{r^{4}}{\Sigma^{2}}
\rho_{\rm FDM}(r)
}.
\label{eq:sec4_density_profile_relation}
\end{equation}
Thus \(\rho_{\rm rot}\geq0\) wherever
\(\rho_{\rm FDM}\geq0\).  On the equatorial plane,
\(\Sigma=r^{2}\), and therefore
\begin{equation}
\rho_{\rm rot}(r,\pi/2)
=
\rho_{\rm FDM}(r).
\label{eq:sec4_equatorial_density}
\end{equation}
Away from the equator, rotation suppresses the local effective energy
density by the factor \(r^{4}/\Sigma^{2}\).

For the adopted profile,
\begin{equation}
\rho_{\rm FDM}(r)
=
\rho_{c}
\left(
1+\alpha\frac{r^{2}}{r_{c}^{2}}
\right)^{-8},
\label{eq:sec4_density_repeated}
\end{equation}
the ratio of the transverse pressure to the energy density becomes
\begin{equation}
{
\frac{p_{\perp}}{\rho_{\rm rot}}
=
\frac{
7\alpha r^{4}
-r^{2}r_{c}^{2}
+
2a^{2}\cos^{2}\theta
\left(
3\alpha r^{2}-r_{c}^{2}
\right)
}{
r^{2}
\left(
r_{c}^{2}+\alpha r^{2}
\right)
}
}.
\label{eq:sec4_pressure_ratio}
\end{equation}
In the static limit \(a\rightarrow0\), this reduces to
\begin{equation}
\frac{p_{\perp}}{\rho}
=
\frac{
7\alpha r^{2}-r_{c}^{2}
}{
\alpha r^{2}+r_{c}^{2}
},
\label{eq:sec4_pressure_static_limit}
\end{equation}
in agreement with the static effective source derived in
Sec.~\ref{sec:static_geometry}.

The frame is real and orthonormal in each stationary block with
\(\Delta>0\) and \(\Sigma>0\); a horizon-penetrating frame is required on
\(\Delta=0\).  The mixed tensor has eigenvalues
\[
\{-\rho_{\rm rot},-\rho_{\rm rot},p_\perp,p_\perp\},
\]
and is Hawking--Ellis type I (Segre type \([(11)(11)]\)) at generic
points, with enhanced degeneracy where \(p_\perp=-\rho_{\rm rot}\).
Direct substitution of Eqs.~\eqref{eq:sec4_rotating_density}--
\eqref{eq:sec4_rotating_transverse_pressure} into the connection of
Eq.~\eqref{eq:sec3_rotating_metric} gives
\begin{equation}
\nabla_\mu T^{\mu\nu}_{\rm eff}=0
\qquad(\nu=t,r,\theta,\phi)
\label{eq:sec4_conservation}
\end{equation}
identically for differentiable \(m(r)\).  This is also required by the
contracted Bianchi identity, but the component substitution checks that no
frame or sign convention has been lost.

Two compact curvature diagnostics are
\begin{equation}
R=\frac{2\,[r m''(r)+2m'(r)]}{\Sigma}
=16\pi(\rho_{\rm rot}-p_\perp),
\label{eq:sec4_rotating_ricci}
\end{equation}
and
\begin{equation}
R_{\mu\nu}R^{\mu\nu}
=128\pi^2\left(\rho_{\rm rot}^2+p_\perp^2\right).
\label{eq:sec4_rotating_ricci_squared}
\end{equation}
A computer-algebra reduction of the Kretschmann scalar has the form
\begin{equation}
R_{\mu\nu\rho\sigma}R^{\mu\nu\rho\sigma}
=\frac{{\cal P}_{K}(r,\cos\theta;m,m',m'')}{\Sigma^{6}},
\label{eq:sec4_kretschmann_structure}
\end{equation}
where \({\cal P}_{K}\) is polynomial in its displayed smooth arguments.
The complete expression is too long to aid the printed argument, but it is
not left implicit: the accompanying open CAS script
\texttt{kretschmann\_cas.wls} writes the full scalar and the machine-readable
\(\Sigma^6\) numerator to \texttt{Kretschmann\_numerator\_Sigma6.txt}.
Appendix~\ref{app:nja_checks} gives the independent limiting checks. For the
present smooth profile there is no curvature
singularity at a zero of \(\Delta\), and no additional curvature
singularity for \(r>0\); the Kerr-like ring \(\Sigma=0\) remains.

Finally,
\begin{equation}
g_{\phi\phi}
=\frac{\sin^2\theta}{\Sigma}
\left[(r^2+a^2)\Sigma+2a^2r\,m(r)\sin^2\theta\right]>0
\label{eq:sec4_no_ctc}
\end{equation}
for \(r>0\), \(m(r)>0\), and \(0<\theta<\pi\).  Hence the domain outside
the outer positive Killing horizon contains no azimuthal
\(g_{\phi\phi}<0\) closed timelike curves.  These checks strengthen the
internal consistency of the ansatz but do not supply a fundamental scalar
matter model \cite{NevesSaa2014,Toshmatov2017,
BeltracchiGondolo2021,SimpsonVisser2022}.

For a diagonal anisotropic stress tensor, the weak energy condition
requires
\begin{equation}
\rho_{\rm rot}\geq0,
\qquad
\rho_{\rm rot}+p_{r}\geq0,
\qquad
\rho_{\rm rot}+p_{\perp}\geq0.
\label{eq:sec4_wec_general}
\end{equation}
We use the standard pointwise energy-condition definitions
\cite{HawkingEllis1973,Visser1996,BarceloVisser2000}.
The first condition follows from
Eq.~\eqref{eq:sec4_density_profile_relation}.  The radial condition is
identically saturated,
\begin{equation}
\rho_{\rm rot}+p_{r}=0.
\label{eq:sec4_radial_nec_saturated}
\end{equation}
The entire nontrivial WEC content is therefore contained in the
transverse combination.

Using Eqs.~\eqref{eq:sec4_rotating_density} and
\eqref{eq:sec4_rotating_transverse_pressure}, one obtains
\begin{equation}
8\pi
\left(
\rho_{\rm rot}+p_{\perp}
\right)
=
\frac{
2\left(r^{2}-a^{2}\cos^{2}\theta\right)m'
-r\Sigma m''
}{
\Sigma^{2}
}.
\label{eq:sec4_transverse_nec_general}
\end{equation}
After substituting the FDM-inspired density profile, this expression has
the closed factorization
\begin{equation}
\rho_{\rm rot}+p_{\perp}
=
\frac{
2\rho_{c}r^{2}r_{c}^{16}
}{
\Sigma^{2}
\left(
r_{c}^{2}+\alpha r^{2}
\right)^{9}
}
\,
{\cal W}(r,\theta),
\label{eq:sec4_transverse_nec_factorized}
\end{equation}
where
\begin{equation}
{
{\cal W}(r,\theta)
=
4\alpha r^{4}
+
3\alpha a^{2}r^{2}\cos^{2}\theta
-
a^{2}r_{c}^{2}\cos^{2}\theta
}.
\label{eq:sec4_wec_factor}
\end{equation}
For \(\rho_c>0\), equivalently \(f_{\rm sol}>0\), all factors multiplying
\({\cal W}\) in
Eq.~\eqref{eq:sec4_transverse_nec_factorized} are non-negative for
\(r>0\).  The sign of the transverse NEC and hence of the WEC is
therefore determined entirely by
\begin{equation}
{\cal W}(r,\theta)\geq0.
\label{eq:sec4_wec_polynomial_condition}
\end{equation}

On the equatorial plane,
\begin{equation}
{\cal W}(r,\pi/2)
=
4\alpha r^{4}>0,
\label{eq:sec4_equatorial_wec}
\end{equation}
so the WEC is always satisfied there for \(r>0\).  Possible violations
are confined to non-equatorial directions.

The angular dependence can be examined through
\begin{equation}
\frac{\partial{\cal W}}
{\partial(\cos^{2}\theta)}
=
a^{2}
\left(
3\alpha r^{2}-r_{c}^{2}
\right).
\label{eq:sec4_wec_angular_derivative}
\end{equation}
When
\begin{equation}
r^{2}
<
\frac{r_{c}^{2}}{3\alpha},
\label{eq:sec4_inner_angular_region}
\end{equation}
the minimum occurs on the rotation axis,
\(\cos^{2}\theta=1\).  At larger radii, the minimum moves to the
equatorial plane, where \({\cal W}>0\).  It follows that the rotation
axis provides the globally most restrictive WEC condition.

On the axis,
\begin{equation}
{\cal W}_{\rm ax}(r)
=
4\alpha r^{4}
+
3\alpha a^{2}r^{2}
-
a^{2}r_{c}^{2}.
\label{eq:sec4_axial_wec_factor}
\end{equation}
For every nonzero \(a\),
\begin{equation}
{\cal W}_{\rm ax}(0)
=
-a^{2}r_{c}^{2}<0,
\label{eq:sec4_axial_origin_violation}
\end{equation}
whereas
\({\cal W}_{\rm ax}(r)\rightarrow+\infty\) as \(r\rightarrow\infty\).
There is therefore a unique positive transition radius for a nonzero
effective-source amplitude.

For \(f_{\rm sol}>0\), solving
\begin{equation}
{\cal W}_{\rm ax}(r_{\rm WEC})=0
\label{eq:sec4_wec_radius_definition}
\end{equation}
gives
\begin{equation}
{
r_{\rm WEC}^{2}
=
\frac{
-3\alpha a^{2}
+
\sqrt{
9\alpha^{2}a^{4}
+
16\alpha a^{2}r_{c}^{2}
}
}{
8\alpha
}
}.
\label{eq:sec4_wec_radius}
\end{equation}
For \(a=0\), the natural continuous definition is
\begin{equation}
r_{\rm WEC}=0,
\label{eq:sec4_wec_static_value}
\end{equation}
consistent with global static effective-Einstein-source WEC satisfaction.
This radius is a root of
the normalized sign factor, not a matter boundary in vacuum.

Introducing
\begin{equation}
q\equiv\frac{a}{r_{c}},
\label{eq:sec4_q_definition}
\end{equation}
the result can be written as
\begin{equation}
\left(
\frac{r_{\rm WEC}}{r_{c}}
\right)^{2}
=
\frac{
-3\alpha q^{2}
+
\sqrt{
9\alpha^{2}q^{4}
+
16\alpha q^{2}
}
}{
8\alpha
}.
\label{eq:sec4_dimensionless_wec_radius}
\end{equation}
The critical radius increases monotonically with \(a/r_{c}\) and
approaches
\begin{equation}
\lim_{a/r_{c}\rightarrow\infty}
\frac{r_{\rm WEC}}{r_{c}}
=
\frac{1}{\sqrt{3\alpha}}.
\label{eq:sec4_wec_large_rotation_limit}
\end{equation}

A local effective-source WEC violation near the central region does not necessarily imply
a violation in the domain of outer communication.  If the violating
region lies entirely inside the outer Killing horizon of the stationary
effective geometry, the complete black-hole exterior can still satisfy
the WEC.

Let \(r_{+}\) denote the largest positive root of
\begin{equation}
\Delta(r)
=
r^{2}-2r\,m(r)+a^{2}=0.
\label{eq:sec4_horizon_equation}
\end{equation}
For \(f_{\rm sol}>0\), the WEC is satisfied for every
\begin{equation}
r\geq r_{\rm WEC},
\end{equation}
the necessary and sufficient condition for effective-Einstein-source WEC
satisfaction throughout
the complete exterior is
\begin{equation}
{
r_{+}\geq r_{\rm WEC}
}.
\label{eq:sec4_exterior_wec_condition}
\end{equation}
Equivalently, for \(f_{\rm sol}>0\) define the exterior-WEC margin
\begin{equation}
{\cal E}_{\rm WEC}
\equiv
\frac{
r_{+}-r_{\rm WEC}
}{
M_{\rm ADM}
}.
\label{eq:sec4_wec_margin}
\end{equation}
Thus, \({\cal E}_{\rm WEC}>0\) means that the complete black-hole
exterior satisfies the WEC, \({\cal E}_{\rm WEC}=0\) means that the WEC
boundary touches the outer horizon, and \({\cal E}_{\rm WEC}<0\) means
that a polar WEC-violating region extends outside the horizon.
If \(\Delta(r)\) has no positive outer root, the configuration is
classified as horizonless and
\({\cal E}_{\rm WEC}\) is not assigned a black-hole-exterior
interpretation.

For numerical analysis, introduce
\begin{equation}
x
=
\frac{r}{M_{\rm ADM}},
\qquad
\widehat r_{c}
=
\frac{r_{c}}{M_{\rm ADM}},
\qquad
j_{\rm ADM}
=
\frac{a}{M_{\rm ADM}}.
\label{eq:sec4_dimensionless_variables}
\end{equation}
Using the soliton fraction
\begin{equation}
f_{\rm sol}
=
\frac{M_{\rm sol}}{M_{\rm ADM}},
\label{eq:sec4_soliton_fraction}
\end{equation}
the radial mass function is
\begin{equation}
\frac{m(r)}{M_{\rm ADM}}
=
1-f_{\rm sol}
+
f_{\rm sol}
{\cal F}
\left(
\frac{x}{\widehat r_{c}}
\right),
\label{eq:sec4_dimensionless_mass_function}
\end{equation}
where \({\cal F}\) was defined in
Eq.~\eqref{eq:sec3_mass_fraction_function}.  The dimensionless horizon
function becomes
\begin{equation}
\widehat\Delta(x)
=
x^{2}
-
2x
\left[
1-f_{\rm sol}
+
f_{\rm sol}
{\cal F}
\left(
\frac{x}{\widehat r_{c}}
\right)
\right]
+
j_{\rm ADM}^{2}.
\label{eq:sec4_dimensionless_delta}
\end{equation}
The outer horizon \(x_{+}\) is the largest positive zero of
\(\widehat\Delta\).

In the same variables,
\begin{equation}
\left(
\frac{r_{\rm WEC}}{M_{\rm ADM}}
\right)^{2}
=
\frac{
-3\alpha j_{\rm ADM}^{2}
+
\sqrt{
9\alpha^{2}j_{\rm ADM}^{4}
+
16\alpha j_{\rm ADM}^{2}\widehat r_{c}^{2}
}
}{
8\alpha
}.
\label{eq:sec4_dimensionless_wec_global}
\end{equation}
Although \(r_{\rm WEC}\) itself depends only on \(a\) and \(r_{c}\), the
exterior condition depends additionally on \(f_{\rm sol}\), because the
horizon location is controlled by the complete radial mass function.
The apparent independence of the sign-change radius from \(f_{\rm sol}\)
must not be confused with an amplitude-independent violation:
\begin{equation}
T^{(a)}{}_{(b)}\propto \rho_c\propto M_{\rm sol}
\propto f_{\rm sol}.
\label{eq:sec4_wec_amplitude}
\end{equation}
Thus \(f_{\rm sol}\to0^+\) sends every effective-source eigenvalue and the
absolute WEC-violation amplitude to zero even though the normalized factor
\({\cal W}=0\) retains a formal location. At the boundary
\(f_{\rm sol}=0\) itself, \(T_{\mu\nu}^{\rm eff}=0\) identically and the
vacuum Kerr geometry satisfies the WEC trivially for every subextremal spin;
\(r_{\rm WEC}\) has no physical or classificatory meaning there. For
nonzero source amplitude we therefore call \(r_+\geq r_{\rm WEC}\) a
\emph{sign-based effective-source consistency domain}, not a
physical-viability condition.

\begin{figure*}[t]
\centering
\includegraphics[
width=0.98\textwidth
]{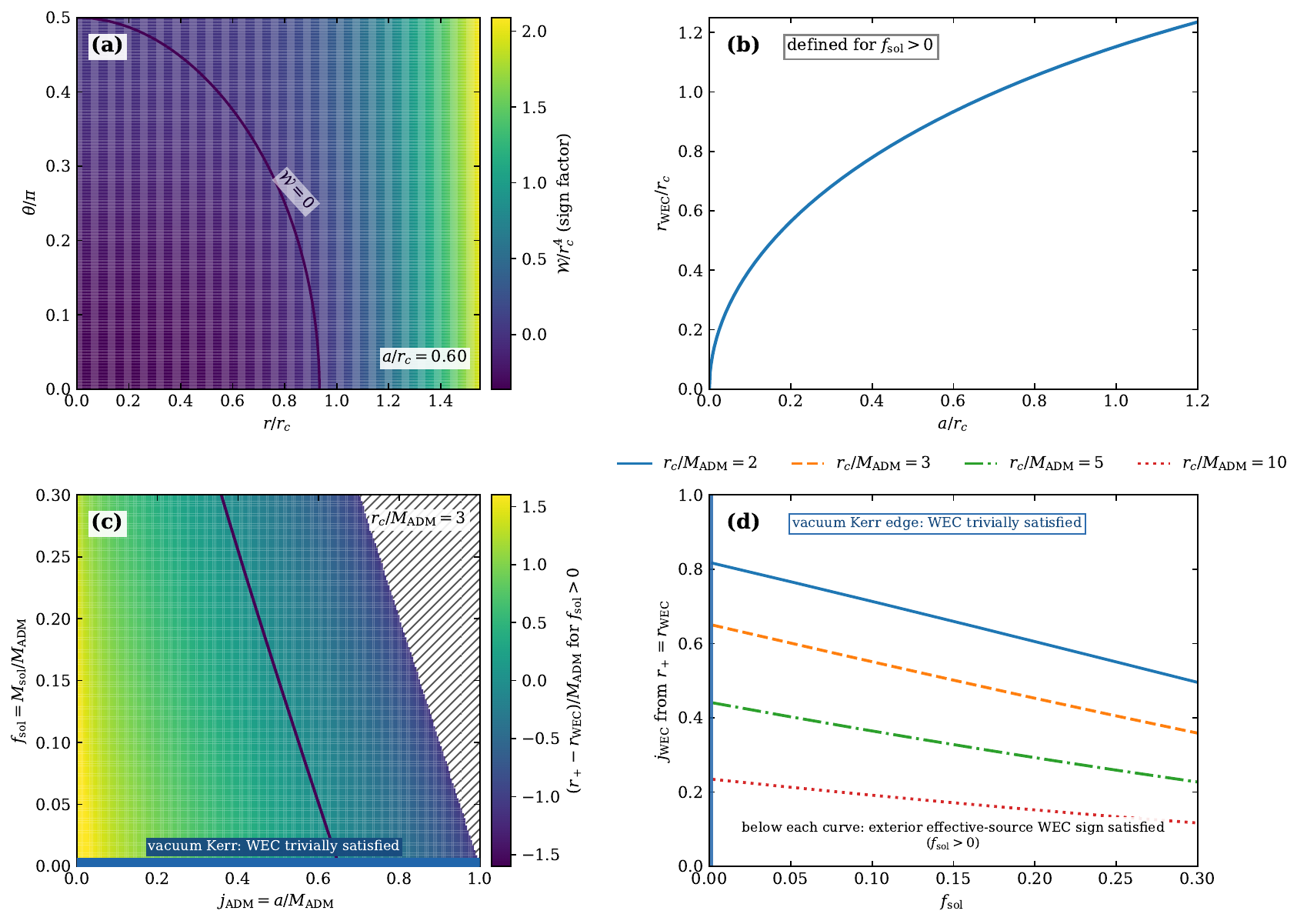}
\caption{
Weak-energy-condition structure of the rotating effective source.
Panel (a) shows the normalized angular factor
${\cal W}/r_{c}^{4}$ for \(a/r_{c}=0.60\); its sign diagnoses the source
only for \(f_{\rm sol}>0\).
The contour \({\cal W}=0\) separates the WEC-satisfying and
WEC-violating regions.
Panel (b) displays the closed-form sign-factor root
\(r_{\rm WEC}/r_{c}\) as a function of \(a/r_{c}\).
Panel (c) shows the exterior-WEC margin
\((r_{+}-r_{\rm WEC})/M_{\rm ADM}\) in the
\((j_{\rm ADM},f_{\rm sol})\) plane for
\(r_{c}/M_{\rm ADM}=3\).
For \(f_{\rm sol}>0\), positive values correspond to complete exterior
effective-Einstein-source WEC satisfaction,
negative values indicate that a WEC-violating polar region extends
outside the horizon, and the hatched region is horizonless.
Panel (d) presents the boundary \(r_{+}=r_{\rm WEC}\) for several core
scales.  Parameter values below each curve have an effective source whose
WEC sign is non-negative throughout the black-hole exterior.  The
\(f_{\rm sol}=0\) edge is marked separately as vacuum Kerr, where the WEC
is trivially satisfied; none of these labels is a fundamental-matter
viability claim.
}
\label{fig:rotating_wec_domain}
\end{figure*}

Figure~\ref{fig:rotating_wec_domain} summarizes the local and exterior
WEC structure.

Panel~(a) of Fig.~\ref{fig:rotating_wec_domain} illustrates the angular
structure of the closed sign factor
Eq.~\eqref{eq:sec4_wec_factor}.  The WEC is automatically satisfied on
the equatorial plane, whereas the violating region first appears near
the rotation axis and at small radius.  The contour
\({\cal W}=0\) closes onto the axis at \(r=r_{\rm WEC}\).

Panel~(b) confirms that \(r_{\rm WEC}\) increases continuously with the
rotation parameter.  Rotation therefore enlarges the inner region in
which the effective source fails the transverse WEC.

Panel~(c) combines the local WEC radius with the numerically determined
outer horizon.  At fixed core scale, increasing \(j_{\rm ADM}\)
decreases the exterior-WEC margin.  Increasing \(f_{\rm sol}\) also
tends to reduce the margin because a larger fraction of the ADM mass is
distributed outside the strong-field region, thereby changing the
location of the outer horizon.

The hatched region in panel~(c) contains no positive outer horizon and
must not be interpreted as a black-hole exterior.  The boundary between
the colored and hatched regions will be analyzed more systematically in
Sec.~\ref{sec:horizon_extremality}.

Panel~(d) shows that the effective-Einstein-source WEC-sign threshold
decreases as the core scale
increases.  A more extended profile places less environmental mass
inside the near-horizon region at fixed \(f_{\rm sol}\), and the outer
horizon is consequently less effective at hiding the axial WEC-violating
domain.

\subsection{Other energy conditions and model scope}
\label{subsec:other_energy_conditions}

The WEC is the only energy condition used in the parameter
classification. For completeness, \(p_r=-\rho_{\rm rot}\) reduces the
strong-energy condition to \(p_\perp\geq0\), which fails in an inner
region. The dominant-energy condition additionally requires
\(p_\perp/\rho_{\rm rot}\leq1\), equivalent here to
\(r\leq r_c/\sqrt{3\alpha}\); because
\(p_\perp/\rho_{\rm rot}\to7\), it cannot hold throughout the
asymptotically flat exterior. These local diagnostics are not used as
admissibility priors.

Consistently with the scope statement in
Sec.~\ref{sec:static_geometry}, the exterior-WEC condition
\begin{equation}
r_{+}\geq r_{\rm WEC}
\end{equation}
is an internal sign diagnostic only for \(f_{\rm sol}>0\).
Configurations that satisfy it possess a non-negative
effective energy density and satisfy the null and weak energy conditions
throughout the domain outside the outer horizon.  Configurations that
violate it remain mathematically well-defined effective geometries, but
their exterior source contains a polar region with
\(\rho_{\rm rot}+p_{\perp}<0\).  At \(f_{\rm sol}=0\), the separate vacuum
Kerr classification applies and the WEC is trivially satisfied.

In the subsequent analysis, the WEC boundary will be shown together
with the horizon and extremality boundaries rather than used silently
as a prior.  This makes it possible to distinguish clearly among

\begin{enumerate}
\item vacuum Kerr or nonzero-source black-hole geometries whose
effective-Einstein-source WEC sign is satisfied throughout the exterior;

\item black-hole geometries with an exterior effective-source
WEC-sign-violating region; and

\item horizonless configurations.
\end{enumerate}

The complete horizon and extremality structure is developed in
Sec.~\ref{sec:horizon_extremality}.

\section{Convexity criterion for general radial mass functions}
\label{sec:horizon_extremality}

We now analyze the horizon structure of the rotating effective geometry
constructed in Sec.~\ref{sec:rotating_geometry}.  The location and
multiplicity of the horizons are controlled by the generalized radial
function
\begin{equation}
\Delta(r)
=
r^{2}-2r\,m(r)+a^{2},
\label{eq:sec5_delta}
\end{equation}
where
\begin{equation}
m(r)
=
M_{\bullet}+M_{\rm FDM}(r).
\label{eq:sec5_mass_function}
\end{equation}
Unlike the Kerr case, the mass entering
Eq.~\eqref{eq:sec5_delta} is a radial function rather than a constant.
Consequently, the horizon positions and the selected completion's
double-root locus depend not
only on the total ADM mass, but also on the distribution of that mass
through \(m(r)\) and \(m'(r)\).

The present section first proves a sufficient theorem for a class of
monotone radial mass functions.  It then establishes the adopted profile
as a strict example, introduces the dimensionless horizon problem, and
combines the extremality boundary with the exterior
weak-energy-condition criterion obtained in
Sec.~\ref{sec:rotating_effective_source}.

\subsection{General theorem and extremal branch}
\label{subsec:horizon_equation}

The inverse radial metric component is
\begin{equation}
g^{rr}
=
\frac{\Delta(r)}{\Sigma},
\qquad
\Sigma
=
r^{2}+a^{2}\cos^{2}\theta.
\label{eq:sec5_inverse_radial_metric}
\end{equation}
Since \(\Sigma>0\) away from the ring singularity, a Killing horizon is
located at a positive zero of
\begin{equation}
{
\Delta(r)=0
}.
\label{eq:sec5_horizon_condition}
\end{equation}
Equivalently, the horizon radii satisfy
\begin{equation}
r^{2}
-
2r
\left[
M_{\bullet}+M_{\rm FDM}(r)
\right]
+
a^{2}
=
0.
\label{eq:sec5_explicit_horizon_equation}
\end{equation}

Let \(r_{+}\) denote the largest positive root of
Eq.~\eqref{eq:sec5_horizon_condition}.  When a second positive root is
present, it is denoted by \(r_{-}\).  In the parameter range analyzed
below and for \(a\ne0\), the solutions fall into three classes:
\begin{align}
r_{-}<r_{+}
&:
&
&\text{subextremal black hole},
\label{eq:sec5_subextremal_class}
\\
r_{-}=r_{+}=r_{e}
&:
&
&\text{extremal black hole},
\label{eq:sec5_extremal_class}
\\
\text{no positive zero of }\Delta
&:
&
&\text{horizonless rotating geometry}.
\label{eq:sec5_horizonless_class}
\end{align}
On the static edge \(a=0\), the coordinate zero at \(r=0\) is excluded
and the black-hole sector contains exactly one simple positive root, as
proved below.
The last class does not belong to the black-hole sector of the effective
model, since the central Kerr-like singular region is no longer hidden
behind an outer Killing horizon.

At large radius,
\begin{equation}
m(r)
=
M_{\rm ADM}
+
{\cal O}(r^{-13}),
\label{eq:sec5_mass_asymptotic}
\end{equation}
and therefore
\begin{equation}
\Delta(r)
=
r^{2}
-
2M_{\rm ADM}r
+
a^{2}
+
{\cal O}(r^{-12}).
\label{eq:sec5_delta_asymptotic}
\end{equation}
The asymptotic form is Kerr-like, but the near-horizon roots generally
differ from the Kerr values because
\begin{equation}
m(r_{+})
\neq
M_{\rm ADM}.
\label{eq:sec5_local_mass_difference}
\end{equation}

\paragraph{General one-minimum theorem.}
The following statement applies to a class of mass functions and is not
specific to the FDM-inspired example.  Let
\(m\in C^2(0,\infty)\) obey
\begin{equation}
m(r)>0,\qquad m'(r)\geq0,
\label{eq:sec5_theorem_mass_conditions}
\end{equation}
together with the boundary conditions
\begin{equation}
\begin{aligned}
\lim_{r\to0^+}\left[m(r)+rm'(r)\right]&=m_0>0,\\
\lim_{r\to\infty}m(r)&=M_{\rm ADM}<\infty,\\
\lim_{r\to\infty}rm'(r)&=0.
\end{aligned}
\label{eq:sec5_theorem_boundary_conditions}
\end{equation}
Suppose further that the pointwise inequality
\begin{equation}
2m'(r)+rm''(r)<1
\qquad\text{for every }r>0.
\label{eq:sec5_general_convexity_condition}
\end{equation}
Define
\begin{equation}
H(r)=r-m(r)-rm'(r),
\qquad
\Delta'(r)=2H(r).
\label{eq:sec5_H_definition}
\end{equation}
Then
\begin{equation}
H'(r)=1-2m'(r)-rm''(r)>0.
\label{eq:sec5_H_monotonicity}
\end{equation}
Because \(m\in C^2(0,\infty)\), \(H\) is continuously differentiable on
every compact subinterval of the positive half-line.  Its one-sided
central limit does not require an extension of \(m\) to \(r=0\): it
follows directly from the first condition in
Eq.~\eqref{eq:sec5_theorem_boundary_conditions}.  Thus
\(H(0^+)=-m_0<0\), while the two asymptotic conditions give
\(H(r)\to r-M_{\rm ADM}>0\).  Hence \(H\) has exactly one positive
zero \(r_m\).  Equivalently,
\begin{equation}
\Delta''(r)=2H'(r)>0,
\label{eq:sec5_strict_convexity}
\end{equation}
so \(\Delta\) is strictly convex and \(r_m\) is its unique global
positive-radius minimum.

For \(a\ne0\), \(\Delta(0^+)=a^2>0\) and
\(\Delta(r)\to+\infty\), so strict convexity gives
\begin{equation}
\begin{array}{ccl}
\Delta(r_m)<0&\Longrightarrow&\text{two simple positive roots},\\
\Delta(r_m)=0&\Longrightarrow&\text{one positive double root},\\
\Delta(r_m)>0&\Longrightarrow&\text{no positive root}.
\end{array}
\label{eq:sec5_general_root_classification}
\end{equation}
The static case is different.  When \(a=0\), one may write
\(\Delta=r^2f\), with \(f=1-2m(r)/r\).  The algebraic zero
\(\Delta(0)=0\) is therefore supplied by the prefactor \(r^2\), not by
the horizon equation \(f=0\).  For the central-mass geometries used here,
\(m(r)\to M_\bullet=m_0>0\), so the origin is the singular inner boundary
of the selected static geometry rather than a regular Killing horizon.
In the general theorem it is in any case outside the open domain
\(r>0\).  Moreover,
\(\Delta'(0^+)=-2m_0<0\).  The function therefore decreases from the
excluded factor root to its negative minimum and then increases to
\(+\infty\).  It has exactly one simple positive root in addition to
\(r=0\), not two.  Thus the theorem excludes three- or four-root
structures for \(a\ne0\) and excludes more than one physical positive
root for \(a=0\).  There is no second extremal branch.  When
\(\Delta(r_m)=0\), the unique double root \(r_e=r_m\) satisfies
\begin{equation}
r_e=m(r_e)+r_em'(r_e),\qquad
a_{\rm ext}^2=r_e^2[1-2m'(r_e)],
\label{eq:sec5_general_unique_double_root}
\end{equation}
provided the second expression is nonnegative.  If
\(1-2m'(r_e)<0\), the stationary equation for \(r_e\) still has a formal
solution, but it would require \(a_{\rm ext}^2<0\) and hence no real
rotation parameter; it is not a physical extremal branch.  Under the
full hypotheses stated above this negative case cannot occur:
\(\Delta_{a=0}\) decreases below zero immediately outside the origin, so
its unique minimum obeys
\(\Delta_{a=0}(r_m)=-a_{\rm ext}^{2}<0\).  The negative-\(a_{\rm ext}^2\)
warning applies when the stationary-radius formula is used outside the
theorem's central-limit or convexity domain.  Conditions
\eqref{eq:sec5_theorem_mass_conditions}--%
\eqref{eq:sec5_general_convexity_condition} are sufficient rather than
necessary; profiles outside this class may still have two roots, but the
theorem no longer excludes additional extrema.

\paragraph{What the criterion adds and where it applies.}
The differentiation leading to \(H'\) is elementary and is not presented
as a new identity.  The useful step is the global statement obtained only
after combining that local sign with the explicit central and ADM
derivative limits: it gives a checkable sufficient domain in which a
profile-by-profile root search is complete, excludes missed secondary
horizon branches, and distinguishes a stationary-radius solution from a
real extremal rotation parameter.  This addresses a practical ambiguity
in generalized Newman--Janis and rotating regular-black-hole studies,
where \(\Delta=0\) is commonly solved for individual mass functions and
additional roots can otherwise be hidden by a sign-change scan
\cite{BambiModesto2013,AzregAinou2014,NevesSaa2014,Toshmatov2014,
Toshmatov2017,SimpsonVisser2022}.

The criterion is portable to any off-shell Kerr or Kerr--Schild
completion whose inverse radial metric has
\(g^{rr}=\Delta(r)/\Sigma\) with the same one-variable mass function,
independently of the pressure closure, Newman--Janis interpretation, or
Carter separability \cite{GursesGursey1975,DrakeSzekeres2000,
BeltracchiGondolo2021,BeltracchiGondolo2021II}.  It does not apply without
modification to metrics with a \(\theta\)-dependent mass function, an
independent radial redshift factor in \(g^{rr}\), a nonfinite ADM mass, or
a source closure whose horizon equation is not of the displayed
\(\Delta(r)\) form.  Nor is the inequality necessary.  Its value is
therefore classificatory rather than dynamical: passing it certifies the
one-minimum horizon topology, whereas failing it correctly leaves room
for the secondary horizons found in sufficiently compact source-derived
environments \cite{Fonseca2026}.

Strictly, the theorem counts positive zeros of the radial horizon
candidate \(\Delta\) within this circular ansatz.  Identifying the outer
Killing horizon with a global event horizon additionally requires the
relevant circularity, regularity, and causal hypotheses; these issues for
parametrically deformed metrics are analyzed in
Ref.~\cite{HeumannPsaltis2023}.  We therefore retain ``outer Killing
horizon'' as the geometric term used in the calculations below.

\begin{table*}[t]
\caption{Scope of the convexity criterion relative to representative
off-shell and source-derived constructions.  ``Applicable'' means that
the horizon-candidate equation reduces to the stated one-variable
\(\Delta(r)\) and that all boundary and derivative hypotheses hold; it
does not identify a global event horizon without the additional
circularity and regularity assumptions discussed in the text.}
\label{tab:geometry_scope_comparison}
\begin{ruledtabular}
\begin{tabular}{p{0.18\textwidth}p{0.21\textwidth}p{0.16\textwidth}
p{0.19\textwidth}p{0.20\textwidth}}
Construction & Independent metric/source data & Carter separability &
Criterion status & Relation to this work\\
\hline
One-function radial-\(\Delta\) family &
\(m(r)\) plus a selected completion &
built in &
directly applicable &
the theorem's precise domain\\
Newman--Janis/Kerr--Schild regular black holes &
profile and complexification prescription &
often retained for radial \(\Delta\) &
applicable only when all hypotheses hold &
prior roots are commonly profile specific\\
Static source-derived environment &
mass, lapse/redshift, and pressure closure &
spherical integrability &
tests the radial horizon factor only when it matches \(\Delta\) &
supplies physical closures absent here\\
Finite-spin source-derived anisotropic fluid &
several metric functions and matter rotation &
not guaranteed &
generally outside the one-function family &
robustness benchmark for finite-spin curves\\
Ultracompact source-derived environment &
coupled metric and matter functions &
model dependent &
may fail the derivative or horizon-form hypotheses &
can exhibit extra light rings or horizons\\
\end{tabular}
\end{ruledtabular}
\end{table*}

The stationary-limit surface is determined separately by
\begin{equation}
g_{tt}=0,
\label{eq:sec5_stationary_limit_definition}
\end{equation}
or
\begin{equation}
r^{2}
+
a^{2}\cos^{2}\theta
-
2r\,m(r)
=
0.
\label{eq:sec5_stationary_limit_equation}
\end{equation}
On the rotation axis this equation coincides with the horizon equation.
Away from the axis, the stationary-limit surface lies outside the outer
horizon and defines the boundary of the ergoregion.

An extremal horizon is a double positive root of the radial function.
It therefore satisfies
\begin{equation}
\Delta(r_{e})=0,
\qquad
\Delta'(r_{e})=0.
\label{eq:sec5_double_root_conditions}
\end{equation}
Differentiating Eq.~\eqref{eq:sec5_delta} gives
\begin{equation}
\Delta'(r)
=
2r
-
2m(r)
-
2r\,m'(r).
\label{eq:sec5_delta_derivative}
\end{equation}
The derivative condition can therefore be written as
\begin{equation}
{
r_{e}
=
m(r_{e})
+
r_{e}m'(r_{e})
}.
\label{eq:sec5_extremal_radius_equation}
\end{equation}
This equation determines the extremal radius independently of \(a\).
Once \(r_{e}\) is known, the extremal rotation parameter follows from
\(\Delta(r_{e})=0\):
\begin{equation}
a_{\rm ext}^{2}
=
2r_{e}m(r_{e})-r_{e}^{2}.
\label{eq:sec5_extremal_spin_first_form}
\end{equation}
Using Eq.~\eqref{eq:sec5_extremal_radius_equation}, this becomes
\begin{equation}
{
a_{\rm ext}^{2}
=
r_{e}^{2}
\left[
1-2m'(r_{e})
\right]
}.
\label{eq:sec5_extremal_spin_derivative_form}
\end{equation}

Equations~\eqref{eq:sec5_extremal_radius_equation} and
\eqref{eq:sec5_extremal_spin_derivative_form} are the double-root
conditions for the present Kerr-like geometry.  They show that the
selected completion's double-root threshold depends on two local
quantities:
\begin{equation}
m(r_{e}),
\qquad
m'(r_{e}).
\end{equation}
The enclosed mass fixes the location of the double root, while its
radial derivative changes the double-root rotation parameter within the
selected completion.

A real extremal rotation parameter requires
\begin{equation}
m'(r_{e})
\leq
\frac{1}{2}.
\label{eq:sec5_extremal_reality_condition}
\end{equation}
For the adopted positive-density profile,
\begin{equation}
m'(r)
=
4\pi r^{2}\rho_{\rm FDM}(r)
\geq0,
\label{eq:sec5_positive_mass_derivative}
\end{equation}
so the environmental mass gradient generally reduces
\(a_{\rm ext}/r_{e}\) relative to the vacuum Kerr value.

The Kerr limit is recovered when
\begin{equation}
M_{\rm FDM}(r)\rightarrow0,
\qquad
m'(r)\rightarrow0.
\end{equation}
Equations~\eqref{eq:sec5_extremal_radius_equation} and
\eqref{eq:sec5_extremal_spin_derivative_form} then give
\begin{equation}
r_{e}=M_{\bullet},
\qquad
a_{\rm ext}=M_{\bullet}.
\label{eq:sec5_kerr_extremal_limit}
\end{equation}

To compare configurations with different masses, we normalize all
lengths by the total ADM mass:
\begin{equation}
x
=
\frac{r}{M_{\rm ADM}},
\qquad
\widehat r_{c}
=
\frac{r_{c}}{M_{\rm ADM}},
\qquad
j_{\rm ADM}
=
\frac{a}{M_{\rm ADM}}.
\label{eq:sec5_dimensionless_variables}
\end{equation}
The total soliton mass fraction is
\begin{equation}
f_{\rm sol}
=
\frac{M_{\rm sol}}{M_{\rm ADM}},
\label{eq:sec5_soliton_fraction}
\end{equation}
so that
\begin{equation}
\frac{M_{\bullet}}{M_{\rm ADM}}
=
1-f_{\rm sol}.
\label{eq:sec5_central_mass_fraction}
\end{equation}

Define the normalized radial mass function
\begin{equation}
\mu(x)
\equiv
\frac{m(r)}{M_{\rm ADM}}
=
1-f_{\rm sol}
+
f_{\rm sol}
{\cal F}
\left(
\frac{x}{\widehat r_{c}}
\right),
\label{eq:sec5_normalized_mass_function}
\end{equation}
where
\begin{equation}
{\cal F}(y)
=
\frac{M_{\rm FDM}(r)}{M_{\rm sol}}
=
I_{z(y)}
\left(
\frac{3}{2},
\frac{13}{2}
\right),
\label{eq:sec5_normalized_soliton_mass}
\end{equation}
and
\begin{equation}
z(y)
=
\frac{\alpha y^{2}}
{1+\alpha y^{2}}.
\label{eq:sec5_beta_argument}
\end{equation}
The derivative of the normalized enclosed-mass function is
\begin{equation}
{\cal F}'(y)
=
\frac{4\pi}{\mu_{\infty}}
\frac{y^{2}}
{\left(1+\alpha y^{2}\right)^{8}},
\label{eq:sec5_mass_fraction_derivative}
\end{equation}
where
\begin{equation}
\mu_{\infty}
=
\frac{M_{\rm sol}}
{\rho_{c}r_{c}^{3}}
=
\frac{33\pi^{2}}
{1024\alpha^{3/2}}.
\label{eq:sec5_total_mass_coefficient}
\end{equation}
It follows that
\begin{equation}
\mu'(x)
=
\frac{f_{\rm sol}}{\widehat r_{c}}
{\cal F}'
\left(
\frac{x}{\widehat r_{c}}
\right).
\label{eq:sec5_dimensionless_mass_derivative}
\end{equation}

\paragraph{FDM-inspired corollary.}
For the normalized profile, define
\begin{equation}
G(x)=x-\mu(x)-x\mu'(x),
\qquad
\widehat\Delta'(x)=2G(x).
\label{eq:sec5_G_definition}
\end{equation}
Writing \(y=x/\widehat r_c\), direct differentiation gives
\begin{equation}
\begin{aligned}
2\mu'(x)+x\mu''(x)
&=\frac{f_{\rm sol}}{\widehat r_c}{\cal Q}(y),\\
{\cal Q}(y)
&=\frac{4C\,y^2(1-3\alpha y^2)}
{(1+\alpha y^2)^9},
\qquad C=\frac{4\pi}{\mu_\infty}.
\end{aligned}
\label{eq:sec5_G_derivative}
\end{equation}
The positive maximum is
\begin{equation}
\max_{y\geq0}{\cal Q}(y)=1.446282
\quad\text{at}\quad y=0.94810.
\label{eq:sec5_Q_bound}
\end{equation}
Therefore the hypotheses of the general theorem hold for
\begin{equation}
0\leq f_{\rm sol}<1,\qquad
\frac{f_{\rm sol}}{\widehat r_c}
<\frac{1}{1.446282}=0.691428,
\label{eq:sec5_fdm_theorem_domain}
\end{equation}
independently of \(j_{\rm ADM}\).  This is a strict analytic domain for
the profile, not merely the box used in the figures.  The production box
is
\begin{equation}
0\leq f_{\rm sol}\leq0.30,\qquad
2\leq\widehat r_c\leq10,\qquad
0\leq j_{\rm ADM}\leq1.10,
\label{eq:sec5_scan_box}
\end{equation}
where \(f_{\rm sol}/\widehat r_c\leq0.15\) and hence
\begin{equation}
G'(x)
=1-\frac{f_{\rm sol}}{\widehat r_c}{\cal Q}(y)
\geq1-\frac{0.30}{2}(1.446282)>0.783.
\label{eq:sec5_G_monotonic_bound}
\end{equation}
The all-root scan in Appendix~\ref{app:numerics} is consequently a
verification and plotting tool; exclusion of additional branches follows
from the theorem and this corollary.

\paragraph{Applications beyond the FDM-inspired profile.}
Table~\ref{tab:profile_theorem_tests} shows that the criterion is neither
profile specific nor automatic.  For the finite-mass comparison profiles,
\(M_0>0\) denotes the central mass, \(M_h\) the environmental mass, and
\(b>0\) its scale.  Direct differentiation gives the displayed global
maxima.

\begin{table*}[t]
\caption{Tests of the sufficient convexity theorem for representative
monotone mass functions.  The quantity
\({\cal S}(r)=2m'(r)+rm''(r)\) must have
\(\max_{r>0}{\cal S}<1\), in addition to the boundary hypotheses.
The entries classify the off-shell radial function, not the distinct
matter closures of the cited source-based models.}
\label{tab:profile_theorem_tests}
\begin{ruledtabular}
\begin{tabular}{p{0.19\textwidth}p{0.29\textwidth}p{0.17\textwidth}
p{0.13\textwidth}p{0.15\textwidth}}
Profile & Mass function & \(\max{\cal S}\) & Theorem domain &
Root implication \\
\hline
FDM-inspired &
\(M_{\rm ADM}[1-f_{\rm sol}+f_{\rm sol}{\cal F}(r/r_c)]\) &
\(1.446282\,f_{\rm sol}/\widehat r_c\) &
\(f_{\rm sol}/\widehat r_c<0.691428\) &
\(\leq2\) for \(a\ne0\); one for \(a=0\) \\
BH-compressed hydrogenic shape &
\(M_0+M_h[1-e^{-2y}(1+2y+2y^2)]\), \(y=r/R\) &
\(1.257713\,M_h/R\) &
\(M_h/R<0.795094\) &
same classification \\
Hernquist-like &
\(M_0+M_h r^2/(r+b)^2\) &
\(81M_h/(128b)\) &
\(M_h/b<128/81\) &
same classification \\
Jaffe-like &
\(M_0+M_h r/(r+b)\) &
\(2M_h/b\) &
\(M_h/b<1/2\) &
same classification \\
Untruncated NFW &
\(M_0+4\pi\rho_s b^3[\ln(1+r/b)-r/(r+b)]\) &
not sufficient &
fails finite-ADM limit &
no conclusion \\
Regular Hayward &
\(Mr^3/(r^3+b^3)\) &
profile dependent &
fails \(m_0>0\) &
no conclusion \\
Thin smooth shell &
\(1+2[1+\tanh((r-5)/0.2)]\) &
\(201.062\) &
violated &
three extrema; four roots possible \\
\end{tabular}
\end{ruledtabular}
\end{table*}

The hydrogenic row is a limiting-shape benchmark motivated by the
black-hole-dominated Schr\"odinger--Poisson profile
\(\rho_H\propto e^{-2r/R}\), not a relativistic backreacted scalar
solution \cite{DaviesMocz2020}.  Direct differentiation gives
\begin{equation}
2m_H'(r)+rm_H''(r)
=\frac{M_h}{R}\,8y^2(2-y)e^{-2y},
\label{eq:sec5_hydrogenic_convexity_factor}
\end{equation}
whose positive maximum is \(1.257713\,M_h/R\) at
\(y=(7-\sqrt{17})/4\).  It supplies a second compact analytic example,
while in the controlled weak-coupling regime
\(R/M_\bullet\sim\alpha_g^{-2}\gg1\) its photon-region gradient remains
small.

The shell row is an explicit counterexample satisfying \(m>0\),
\(m'\geq0\), finite ADM mass, and the derivative limits, but not
Eq.~\eqref{eq:sec5_general_convexity_condition}.  It has zeros of \(H\)
at
\begin{equation}
r/M_0=1.00000,\quad4.60841,\quad5.59121,
\end{equation}
and for \(a/M_0=0.1\) the four positive roots are
\begin{equation}
r/M_0=0.005013,\quad1.994987,\quad4.946143,\quad9.999000.
\label{eq:sec5_shell_four_roots}
\end{equation}
Thus monotonicity of \(m\) alone does not control horizon multiplicity.

The source-based Einstein-cluster models of Fonseca \emph{et al.}
\cite{Fonseca2026} provide a physical counterpart.  Their reported
secondary-horizon thresholds occur only in the high-compactness regime.
For the zero-inner-cutoff representatives in
Table~\ref{tab:profile_theorem_tests}, the Hernquist and Jaffe theorem
bounds are \(M_h/b<1.580\) and \(M_h/b<0.5\), respectively, whereas their
secondary horizons occur at compactnesses about \(2.00\) and \(0.56\).
Their precise inner-cutoff profiles and redshift closure differ, but the
logic is general: multiple horizon extrema require \(H\) to cease being
strictly increasing, so \({\cal S}\geq1\) somewhere, or another boundary
hypothesis must fail.  The untruncated NFW mass instead lies outside the
theorem because it has no finite ADM limit.

The dimensionless horizon function is
\begin{equation}
\widehat\Delta(x)
\equiv
\frac{\Delta(r)}{M_{\rm ADM}^{2}}
=
x^{2}
-
2x\mu(x)
+
j_{\rm ADM}^{2}.
\label{eq:sec5_dimensionless_delta}
\end{equation}
The outer horizon is the largest positive solution of
\begin{equation}
\widehat\Delta(x_{+})=0.
\label{eq:sec5_dimensionless_horizon_equation}
\end{equation}
The extremal equations become
\begin{equation}
{
x_{e}
=
\mu(x_{e})
+
x_{e}\mu'(x_{e})
}
\label{eq:sec5_dimensionless_extremal_radius}
\end{equation}
and
\begin{equation}
{
j_{\rm ext}^{2}
=
x_{e}^{2}
\left[
1-2\mu'(x_{e})
\right]
}.
\label{eq:sec5_dimensionless_extremal_spin}
\end{equation}

For \(f_{\rm sol}=0\),
\begin{equation}
\mu(x)=1,
\qquad
\mu'(x)=0,
\end{equation}
and hence
\begin{equation}
x_{e}=1,
\qquad
j_{\rm ext}=1.
\label{eq:sec5_dimensionless_kerr_limit}
\end{equation}
For nonzero \(f_{\rm sol}\), however, a fraction of the ADM mass lies
outside the near-horizon region.  The extremal radius and spin are then
determined by the local mass profile rather than by the total mass alone.

\subsection{Kinematics and phase diagram}
\label{subsec:horizon_kinematics}

The outer horizon is generated by the Killing vector
\begin{equation}
\chi
=
\partial_{t}
+
\Omega_{H}\partial_{\phi}.
\label{eq:sec5_horizon_generator}
\end{equation}
The horizon angular velocity retains the Kerr-like form
\begin{equation}
{
\Omega_{H}
=
\frac{a}
{r_{+}^{2}+a^{2}}
}.
\label{eq:sec5_horizon_angular_velocity}
\end{equation}
The area of a spatial horizon section is
\begin{equation}
{
A_{H}
=
4\pi
\left(
r_{+}^{2}+a^{2}
\right)
}.
\label{eq:sec5_horizon_area}
\end{equation}
Although the functional forms are identical to those of Kerr, the
horizon radius \(r_{+}\) is determined by the full radial mass function.

The surface gravity is
\begin{equation}
{
\kappa_{H}
=
\frac{
\Delta'(r_{+})
}{
2\left(r_{+}^{2}+a^{2}\right)
}
}
\label{eq:sec5_surface_gravity}
\end{equation}
or explicitly
\begin{equation}
\kappa_{H}
=
\frac{
r_{+}
-
m(r_{+})
-
r_{+}m'(r_{+})
}{
r_{+}^{2}+a^{2}
}.
\label{eq:sec5_surface_gravity_explicit}
\end{equation}
The corresponding Hawking temperature of the effective geometry is
\begin{equation}
T_{H}
=
\frac{\kappa_{H}}{2\pi}.
\label{eq:sec5_hawking_temperature}
\end{equation}
At extremality,
\begin{equation}
\Delta'(r_{e})=0,
\end{equation}
and therefore
\begin{equation}
\kappa_{H}=0,
\qquad
T_{H}=0.
\label{eq:sec5_extremal_zero_temperature}
\end{equation}

Within Einstein gravity, the geometric horizon entropy is
\begin{equation}
S_{H}
=
\frac{A_{H}}{4}
=
\pi
\left(
r_{+}^{2}+a^{2}
\right).
\label{eq:sec5_horizon_entropy}
\end{equation}
These thermodynamic quantities are properties of the effective metric.
They do not imply that the underlying FDM-inspired source represents a
fundamental equilibrium scalar configuration.  In the absence of a
specified fundamental matter action, integrable charge variation, and
first-law analysis, they are used only as local geometric horizon
quantities and not as a complete FDM black-hole thermodynamics.

The horizon and extremality equations are solved numerically in terms of
\((f_{\rm sol},\widehat r_{c},j_{\rm ADM})\).  For each pair
\((f_{\rm sol},\widehat r_{c})\), we first solve
Eq.~\eqref{eq:sec5_dimensionless_extremal_radius} for the outer positive
double-root radius \(x_{e}\).  The corresponding extremal spin is then
obtained from Eq.~\eqref{eq:sec5_dimensionless_extremal_spin}.
For subextremal configurations, the largest positive zero of
\(\widehat\Delta(x)\) is identified as \(x_{+}\).

\begin{figure*}[t]
\centering
\includegraphics[
width=0.98\textwidth
]{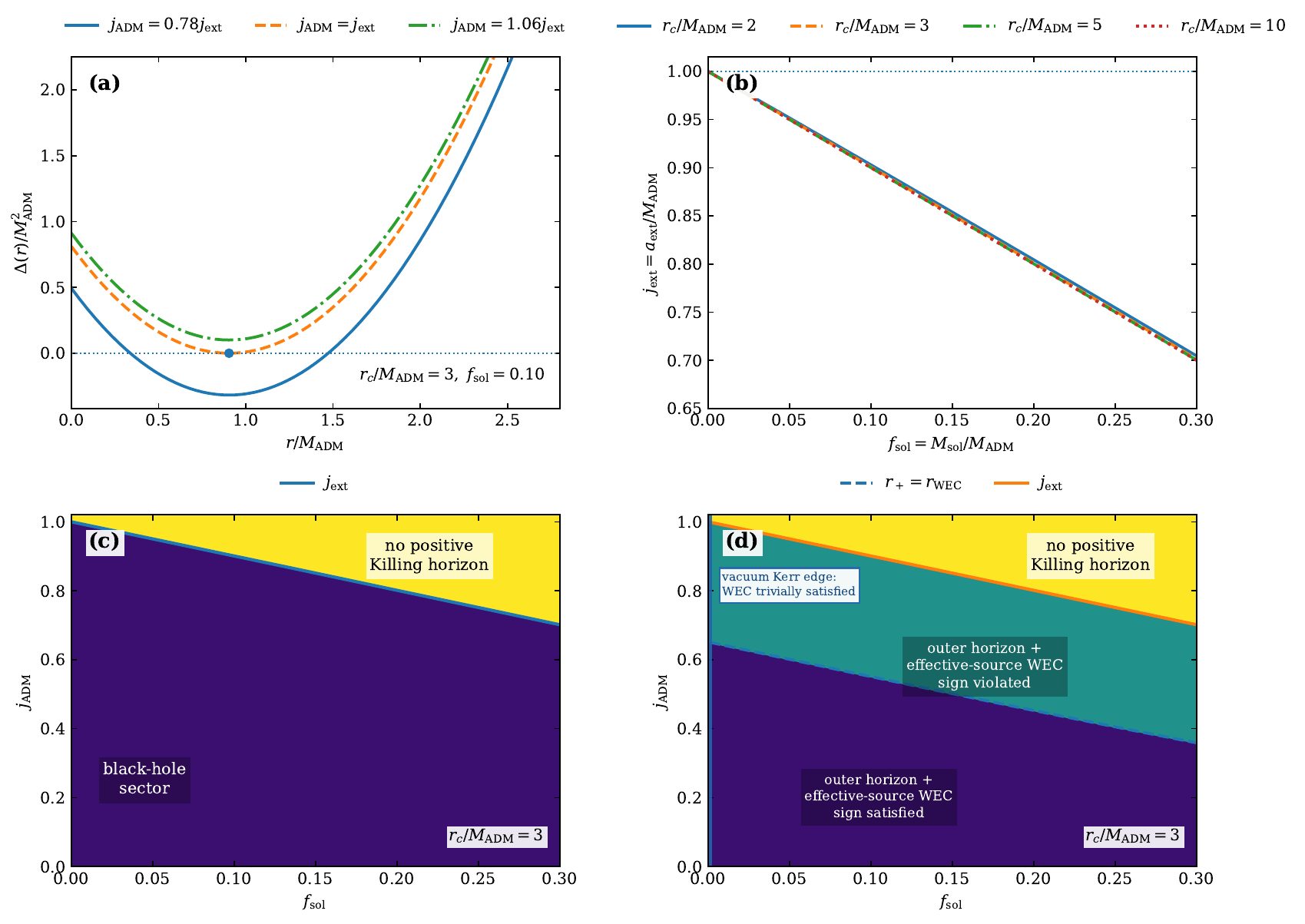}
\caption{
Horizon structure, extremality, and the combined sign-based
effective-source consistency domain of the FDM-inspired Kerr-like geometry.
Panel (a) shows the radial horizon function
$\Delta(r)/M_{\rm ADM}^{2}$ for subextremal, extremal, and
superextremal configurations at
$r_{c}/M_{\rm ADM}=3$ and $f_{\rm sol}=0.10$.
The black point marks the extremal double root.
Panel (b) displays the extremal spin
$j_{\rm ext}=a_{\rm ext}/M_{\rm ADM}$ as a function of the total
soliton mass fraction for several core radii.  This is the double-root
locus of the selected Newman--Janis completion, not a modification of a
universal Kerr angular-momentum bound.
Panel (c) presents the horizon phase diagram in the
$(f_{\rm sol},j_{\rm ADM})$ plane for
$r_{c}/M_{\rm ADM}=3$; the extremality curve separates the
black-hole sector from the horizonless region.  Interior points with
\(j_{\rm ADM}>0\) have two positive roots; the static
\(j_{\rm ADM}=0\) edge has one positive root after excluding \(r=0\).
Panel (d) combines the extremality boundary with the exterior-WEC
boundary $r_{+}=r_{\rm WEC}$.  The three regions are labeled
effective-Einstein-source WEC sign satisfied,
effective-Einstein-source WEC sign violated, and no
positive Killing horizon.  The labels classify the ansatz and are not
fundamental-matter viability statements.  The \(f_{\rm sol}=0\) edge is
shown separately as vacuum Kerr, where \(T_{\mu\nu}^{\rm eff}=0\) and the
WEC is trivially satisfied.
}
\label{fig:horizon_extremality}
\end{figure*}

Figure~\ref{fig:horizon_extremality} summarizes the resulting phase
structure.

For the representative parameters used in panel~(a),
\begin{equation}
\frac{r_{c}}{M_{\rm ADM}}=3,
\qquad
f_{\rm sol}=0.10,
\label{eq:sec5_representative_parameters}
\end{equation}
the extremal solution is
\begin{equation}
\frac{r_{e}}{M_{\rm ADM}}
\simeq
0.90370,
\qquad
j_{\rm ext}
\simeq
0.90094.
\label{eq:sec5_representative_extremal_solution}
\end{equation}
For \(0<j_{\rm ADM}<j_{\rm ext}\), the radial function crosses zero twice
and the geometry possesses inner and outer horizons.  At
\(j_{\rm ADM}=j_{\rm ext}\), the two roots merge.  For
\(j_{\rm ADM}>j_{\rm ext}\), the minimum of \(\Delta\) lies above zero
and the geometry is horizonless.  At \(j_{\rm ADM}=0\), only the outer
positive root remains after the origin is excluded.

Panel~(b) shows that the extremal spin decreases as the environmental
mass fraction increases.  At fixed \(M_{\rm ADM}\), increasing
\(f_{\rm sol}\) reduces the central mass fraction and places a larger
fraction of the total mass outside the near-horizon region.  The
effective spacetime therefore reaches extremality at a smaller value of
\(a/M_{\rm ADM}\).

The core scale controls how rapidly the environmental mass becomes
enclosed.  A compact core approaches its asymptotic mass at smaller
radius and produces a less pronounced reduction of the extremal spin.
A more extended core leaves more mass outside the extremal radius and
therefore lowers \(j_{\rm ext}\) more strongly.

Panel~(c) displays the extremality boundary for
\(\widehat r_{c}=3\).  The black-hole region is defined by
\begin{equation}
0
\leq
j_{\rm ADM}
<
j_{\rm ext}
\left(
f_{\rm sol},
\widehat r_{c}
\right),
\label{eq:sec5_black_hole_parameter_region}
\end{equation}
while the extremal curve satisfies
\begin{equation}
j_{\rm ADM}
=
j_{\rm ext}
\left(
f_{\rm sol},
\widehat r_{c}
\right).
\label{eq:sec5_extremal_parameter_curve}
\end{equation}
Values above this curve do not possess a positive outer horizon.

\paragraph{Effective-source WEC-sign boundary and model scope.}
\label{subsec:combined_admissible_domain}

The extremality boundary determines whether the geometry contains an
outer Killing horizon, while the WEC boundary derived in
Sec.~\ref{sec:rotating_effective_source} determines whether the
effective source satisfies the WEC throughout the exterior.

For \(f_{\rm sol}>0\), the exterior-WEC boundary is defined by
\begin{equation}
r_{+}
=
r_{\rm WEC},
\label{eq:sec5_wec_boundary_equation}
\end{equation}
where
\begin{equation}
r_{\rm WEC}^{2}
=
\frac{
-3\alpha a^{2}
+
\sqrt{
9\alpha^{2}a^{4}
+
16\alpha a^{2}r_{c}^{2}
}
}{
8\alpha
}.
\label{eq:sec5_wec_radius_repeated}
\end{equation}
Let
\begin{equation}
j_{\rm WEC}
\left(
f_{\rm sol},\widehat r_{c}
\right)
\label{eq:sec5_wec_spin_boundary}
\end{equation}
denote, only for \(f_{\rm sol}>0\), the spin at which
Eq.~\eqref{eq:sec5_wec_boundary_equation} is satisfied.

The nonzero-source parameter plane can then be divided into three regions:
\(0\leq j_{\rm ADM}\leq j_{\rm WEC}\) gives a black hole with exterior
effective-Einstein-source WEC-sign satisfaction;
\(j_{\rm WEC}<j_{\rm ADM}<j_{\rm ext}\) gives a black
hole with exterior effective-source WEC-sign violation; and
\(j_{\rm ADM}>j_{\rm ext}\) gives a
horizonless geometry.
At the extremal and WEC boundaries, the corresponding inequalities are
understood in the limiting sense.  The edge \(f_{\rm sol}=0\) is not the
limit of this sign-factor classification: it is exactly vacuum Kerr and
satisfies the WEC trivially for \(j_{\rm ADM}\leq1\).

Panel~(d) of Fig.~\ref{fig:horizon_extremality} shows these regions for
\begin{equation}
\widehat r_{c}=3.
\end{equation}
The exterior-WEC boundary lies below the extremality curve over a
substantial part of the displayed parameter domain.  Consequently, the
existence of an outer Killing horizon alone does not guarantee that the
effective source satisfies the WEC everywhere outside it.

The corresponding sign-based effective-source consistency domain is
\begin{equation}
\begin{aligned}
{\cal D}_{\rm sign}
={}&\bigl\{(f_{\rm sol},\widehat r_{c},j_{\rm ADM})\ \big|\
f_{\rm sol}>0,\
\Delta(r_{+})=0,
\\[-2pt]
&\hspace{24mm}r_{+}>0,\ r_{+}\geq r_{\rm WEC}\bigr\}.
\end{aligned}
\label{eq:sec5_admissible_domain}
\end{equation}
We use this domain only as an explicit model-consistency classification.
It is not called a physical allowed region: horizon existence, the sign of
the WEC combinations, their absolute amplitude, and realizability by a
fundamental scalar theory are four distinct questions.  The WEC is not
silently imposed as a prior in subsequent calculations.

The value \(j_{\rm ext}<1\) is only the ADM-normalized coordinate of the
double-root locus in this selected off-shell ansatz.  It is neither a
correction to nor a violation of the Kerr bound.  The Kerr relation
\begin{equation}
|J|\leq M^{2}
\end{equation}
assumes a vacuum Kerr geometry characterized by a single constant mass.
Here the near-horizon geometry is instead controlled by a radial mass
function, and the extremality condition is
\begin{equation}
a_{\rm ext}^{2}
=
r_{e}^{2}
\left[
1-2m'(r_{e})
\right].
\end{equation}
The relevant bound is therefore the generalized double-root condition,
not the vacuum relation \(a=M_{\rm ADM}\).

Similarly, the absence of a positive root of \(\Delta\) means only that
the chosen effective geometry is outside its black-hole sector.  It
does not demonstrate that such a parameter combination can be realized
as a regular horizonless scalar configuration.  In the present metric,
a horizonless rotating solution would expose the central Kerr-like
singular region and is therefore excluded from the benchmark black-hole
sector.

The results of this section establish the parameter region in which the
subsequent photon dynamics are to be interpreted as black-hole optics.
In Sec.~\ref{sec:photon_dynamics}, we derive the separable null
Hamilton--Jacobi equations and analyze the spherical photon orbits and
shadow observables within this horizon-classified parameter space.

\section{Separable photon dynamics and black-hole shadow}
\label{sec:photon_dynamics}

We now study the null geodesics and shadow of the rotating effective
geometry. The analysis is restricted to parameter configurations for
which the generalized radial function
\begin{equation}
\Delta(r)
=
r^{2}-2r\,m(r)+a^{2}
\label{eq:sec6_delta}
\end{equation}
possesses a positive outer root. The corresponding horizon
classification was established in Sec.~\ref{sec:horizon_extremality}.

An important property of the Newman--Janis prescription adopted in
Sec.~\ref{sec:rotating_geometry} is that \(\Delta\) depends only on the
radial coordinate. The inverse metric therefore retains the algebraic
structure required for Hamilton--Jacobi separability. This permits the
photon region and the critical curve on the observer's screen to be
obtained analytically in terms of \(\Delta(r)\) and its radial
derivatives, without integrating individual null rays.

The comparison with Kerr is always performed at fixed total ADM mass,
fixed ADM angular momentum, and fixed observer inclination:
\begin{equation}
M_{\rm ADM}^{\rm FDM}
=
M_{\rm ADM}^{\rm Kerr},
\qquad
J_{\rm ADM}^{\rm FDM}
=
J_{\rm ADM}^{\rm Kerr},
\qquad
\iota_{\rm FDM}
=
\iota_{\rm Kerr}.
\label{eq:sec6_kerr_comparison_definition}
\end{equation}
This normalization prevents an artificial shadow difference produced
only by comparing spacetimes with different asymptotic charges.

\subsection{Separated equations and critical curve}
\label{subsec:hj_separation}

The Hamilton--Jacobi equation for a null geodesic is
\begin{equation}
g^{\mu\nu}
\frac{\partial S}{\partial x^{\mu}}
\frac{\partial S}{\partial x^{\nu}}
=
0.
\label{eq:sec6_hamilton_jacobi}
\end{equation}
Stationarity and axisymmetry imply two conserved quantities,
\begin{equation}
E
=
-p_{t},
\qquad
L_{z}
=
p_{\phi},
\label{eq:sec6_energy_angular_momentum}
\end{equation}
where \(E\) is the photon energy and \(L_{z}\) is its axial angular
momentum. We use the separated ansatz
\begin{equation}
S
=
-Et
+
L_{z}\phi
+
S_{r}(r)
+
S_{\theta}(\theta).
\label{eq:sec6_hj_ansatz}
\end{equation}
Substitution into Eq.~\eqref{eq:sec6_hamilton_jacobi} gives
\begin{equation}
\Delta
\left(
\frac{dS_{r}}{dr}
\right)^{2}
-
\frac{
\left[
E(r^{2}+a^{2})-aL_{z}
\right]^{2}
}{\Delta}
+
(L_{z}-aE)^{2}
+
{\cal Q}
=
0
\label{eq:sec6_radial_separation}
\end{equation}
and
\begin{equation}
\left(
\frac{dS_{\theta}}{d\theta}
\right)^{2}
-
a^{2}E^{2}\cos^{2}\theta
+
L_{z}^{2}\cot^{2}\theta
-
{\cal Q}
=
0,
\label{eq:sec6_angular_separation}
\end{equation}
where \({\cal Q}\) is the Carter-like separation constant.

The first-order null geodesic equations can therefore be written as
\begin{align}
\Sigma^{2}
\left(
\frac{dr}{d\lambda}
\right)^{2}
&=
{\cal R}(r),
\label{eq:sec6_radial_geodesic}
\\
\Sigma^{2}
\left(
\frac{d\theta}{d\lambda}
\right)^{2}
&=
\Theta(\theta),
\label{eq:sec6_polar_geodesic}
\end{align}
where \(\lambda\) is an affine parameter and
\begin{equation}
{\cal R}(r)
=
\left[
E(r^{2}+a^{2})-aL_{z}
\right]^{2}
-
\Delta(r)
\left[
(L_{z}-aE)^{2}
+
{\cal Q}
\right],
\label{eq:sec6_radial_potential_dimensional}
\end{equation}
while
\begin{equation}
\Theta(\theta)
=
{\cal Q}
+
a^{2}E^{2}\cos^{2}\theta
-
L_{z}^{2}\cot^{2}\theta.
\label{eq:sec6_polar_potential_dimensional}
\end{equation}

It is convenient to introduce the energy-normalized impact parameters
\begin{equation}
\xi
\equiv
\frac{L_{z}}{E},
\qquad
\eta
\equiv
\frac{{\cal Q}}{E^{2}}.
\label{eq:sec6_impact_parameter_definitions}
\end{equation}
The normalized radial and angular potentials become
\begin{equation}
\frac{{\cal R}(r)}{E^{2}}
=
\left[
r^{2}+a^{2}-a\xi
\right]^{2}
-
\Delta(r)
\left[
(\xi-a)^{2}+\eta
\right]
\label{eq:sec6_radial_potential}
\end{equation}
and
\begin{equation}
\frac{\Theta(\theta)}{E^{2}}
=
\eta
+
a^{2}\cos^{2}\theta
-
\xi^{2}\cot^{2}\theta.
\label{eq:sec6_angular_potential}
\end{equation}
The only modification relative to Kerr is the replacement of the
vacuum radial function by
\begin{equation}
\Delta(r)
=
r^{2}-2r\,m(r)+a^{2}.
\end{equation}
The polar potential retains its Kerr form.

A spherical photon orbit remains at a constant Boyer--Lindquist radius
\(r=r_{p}\), although it may oscillate in the polar direction. It
satisfies
\begin{equation}
{\cal R}(r_{p})=0,
\qquad
\left.
\frac{d{\cal R}}{dr}
\right|_{r=r_{p}}
=
0.
\label{eq:sec6_spherical_conditions}
\end{equation}
These are the standard Carter--Bardeen spherical-orbit conditions
\cite{Carter1968,Bardeen1973,Chandrasekhar1983}.
Solving these two equations for the critical impact parameters gives
\begin{equation}
{
\xi_{c}(r_{p})
=
\frac{
\left(r_{p}^{2}+a^{2}\right)\Delta'(r_{p})
-
4r_{p}\Delta(r_{p})
}{
a\Delta'(r_{p})
}
}
\label{eq:sec6_xi_critical}
\end{equation}
and
\begin{equation}
{
\eta_{c}(r_{p})
=
\frac{
16r_{p}^{2}\Delta(r_{p})
}{
\left[\Delta'(r_{p})\right]^{2}
}
-
\left[
\xi_{c}(r_{p})-a
\right]^{2}
}.
\label{eq:sec6_eta_critical}
\end{equation}
With \(\xi=L_z/E\) and \(\eta={\cal Q}/E^2\), the dimensions are
\begin{equation}
[\xi]=\mathrm{length},\qquad [\eta]=\mathrm{length}^{2}.
\label{eq:sec6_impact_dimensions}
\end{equation}
Dimensionless plots display \(\xi/M_{\rm ADM}\) and
\(\eta/M_{\rm ADM}^{2}\).
For the present mass profile,
\begin{equation}
\Delta'(r)
=
2r
-
2m(r)
-
2r\,m'(r),
\label{eq:sec6_delta_prime}
\end{equation}
and
\begin{equation}
\Delta''(r)
=
2
-
4m'(r)
-
2r\,m''(r).
\label{eq:sec6_delta_second}
\end{equation}
The mass gradient therefore enters the critical impact parameters
directly through \(\Delta'\).

The second derivative of the radial potential, evaluated at fixed
\(\xi_{c}\) and \(\eta_{c}\), is
\begin{align}
\frac{1}{E^{2}}
\left.
\frac{d^{2}{\cal R}}{dr^{2}}
\right|_{r=r_{p}}
={}&
4
\left[
r_{p}^{2}+a^{2}-a\xi_{c}
\right]
+
8r_{p}^{2}
\nonumber\\
&-
\Delta''(r_{p})
\left[
(\xi_{c}-a)^{2}
+
\eta_{c}
\right].
\label{eq:sec6_radial_second_derivative}
\end{align}
The unstable branch that forms the shadow boundary satisfies
\begin{equation}
\left.
\frac{d^{2}{\cal R}}{dr^{2}}
\right|_{r=r_{p}}
>
0.
\label{eq:sec6_instability_condition}
\end{equation}

Equatorial circular photon orbits are obtained by setting
\begin{equation}
\eta_{c}=0.
\label{eq:sec6_equatorial_eta_zero}
\end{equation}
The smaller unstable root corresponds to the prograde orbit, while the
larger unstable root corresponds to the retrograde orbit:
\begin{equation}
r_{\rm ph}^{-}
<
r_{\rm ph}^{+}.
\label{eq:sec6_prograde_retrograde_order}
\end{equation}
Here the minus sign denotes the prograde branch and the plus sign denotes
the retrograde branch.

In the static limit, the photon region collapses to a spherical photon
surface. For
\begin{equation}
f(r)
=
1-\frac{2m(r)}{r},
\end{equation}
the circular null-orbit equation
\begin{equation}
r f'(r)-2f(r)=0
\label{eq:sec6_static_photon_condition_initial}
\end{equation}
reduces to
\begin{equation}
{
r_{\rm ph}
-
3m(r_{\rm ph})
+
r_{\rm ph}m'(r_{\rm ph})
=
0
}.
\label{eq:sec6_static_photon_sphere}
\end{equation}
The critical impact parameter in the static geometry is
\begin{equation}
b_{c}^{2}
=
\frac{
r_{\rm ph}^{2}
}{
f(r_{\rm ph})
}
=
\frac{
r_{\rm ph}^{3}
}{
r_{\rm ph}-2m(r_{\rm ph})
}.
\label{eq:sec6_static_critical_impact}
\end{equation}

Equation~\eqref{eq:sec6_static_photon_sphere} shows that the photon
sphere cannot in general be obtained by replacing the radial mass
function with either its asymptotic value or its local value while
neglecting \(m'(r)\). The mass gradient contributes at the same order as
the enclosed mass.

\begin{figure*}[t]
\centering
\includegraphics[
width=0.98\textwidth
]{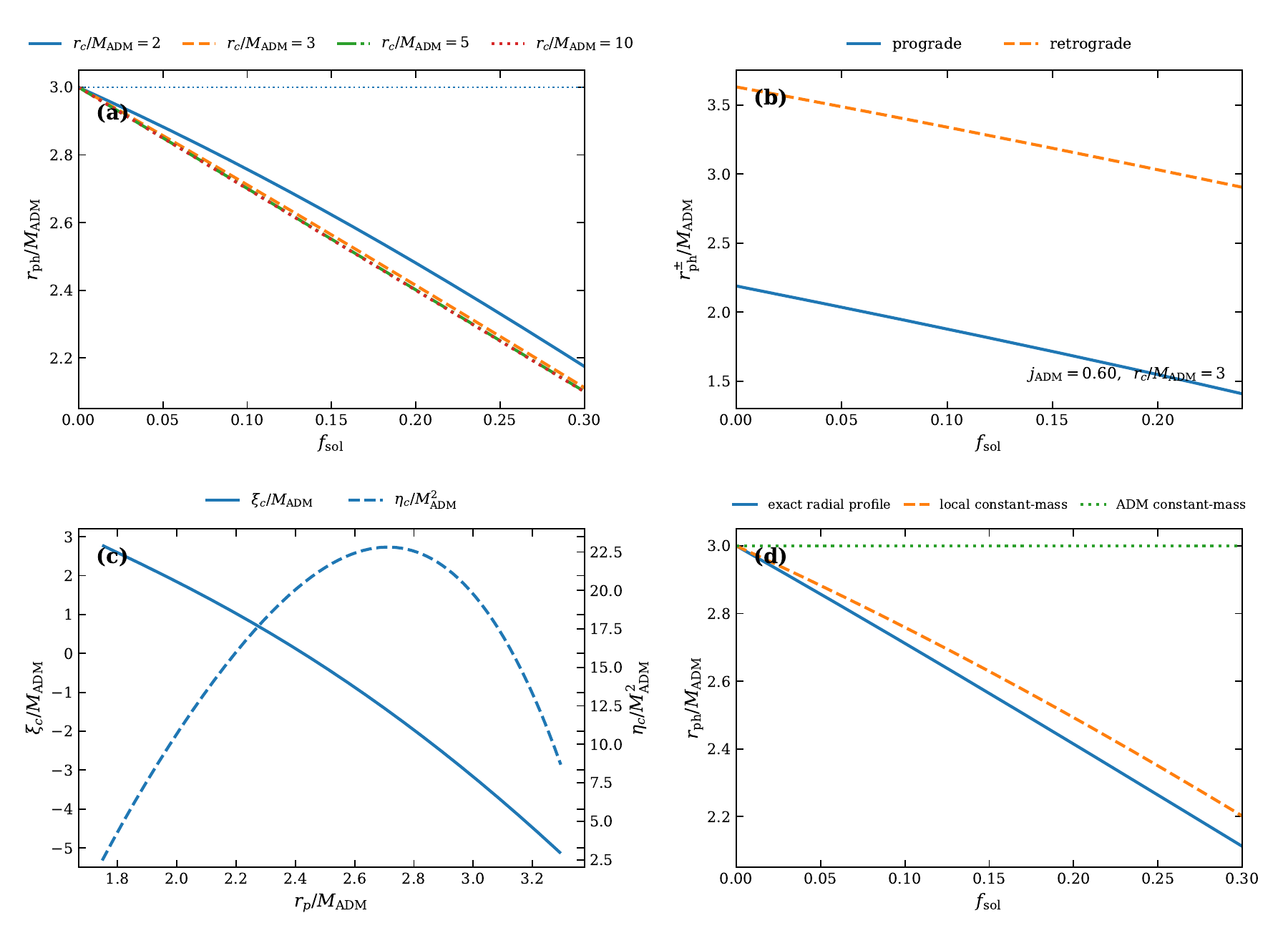}
\caption{
Photon spheres, unstable spherical photon orbits, and critical impact
parameters within the selected Newman--Janis completion of the
FDM-inspired radial profile.
Panel (a) shows the static photon-sphere radius as a function of the
total soliton fraction for several core scales. The horizontal line
marks the Schwarzschild value \(r_{\rm ph}=3M_{\rm ADM}\).
Panel (b) displays the prograde and retrograde equatorial photon-orbit
radii for \(j_{\rm ADM}=0.60\) and
\(r_c/M_{\rm ADM}=3\).
Panel (c) shows the critical impact parameters
\(\xi_c(r_p)\) and \(\eta_c(r_p)\) along the visible unstable
spherical-orbit branch for
\(j_{\rm ADM}=0.70\), \(f_{\rm sol}=0.10\), and
\(r_c/M_{\rm ADM}=3\).
Panel (d) compares the full radial-profile static photon-sphere equation with a local
constant-mass approximation and with an ADM constant-mass
approximation. The separation between the curves demonstrates the
importance of the radial mass-gradient term \(r m'(r)\).
}
\label{fig:photon_orbits_impact}
\end{figure*}

Figure~\ref{fig:photon_orbits_impact} summarizes the photon-region
structure.

Panel~(a) of Fig.~\ref{fig:photon_orbits_impact} demonstrates that the
photon-sphere radius depends on both the total environmental fraction
and the core scale. At fixed \(M_{\rm ADM}\), a more extended core
places a larger fraction of its mass outside the photon region. The
local geometry sampled by the photon orbit is then controlled by an
enclosed mass smaller than the ADM mass.

Panel~(b) shows that the environmental mass profile shifts both
equatorial branches. The separation between the prograde and retrograde
radii remains primarily controlled by rotation, whereas their common
displacement relative to Kerr is sensitive to the radial distribution
of the environmental mass.

Panel~(c) illustrates the parametric critical curve
\begin{equation}
r_{p}
\longmapsto
\left[
\xi_{c}(r_{p}),
\eta_{c}(r_{p})
\right].
\label{eq:sec6_parametric_critical_curve}
\end{equation}
Only the unstable portion satisfying
Eq.~\eqref{eq:sec6_instability_condition} and visible to the chosen
observer contributes to the shadow boundary.  The calculation does not
assume that this set is connected: all exterior zeros of the visibility
and instability functions are enumerated first.  The 2976-point
all-branch audit in Appendix~\ref{app:numerics} finds exactly one visible
unstable interval and two simple visibility endpoints throughout the
audited production box, including \(j_{\rm ADM}/j_{\rm ext}=0.999\).

Panel~(d) demonstrates that neither the replacement
\begin{equation}
m(r)\rightarrow M_{\rm ADM}
\end{equation}
nor the local relation
\begin{equation}
r_{\rm ph}=3m(r_{\rm ph})
\end{equation}
reproduces the full result within the selected ansatz. The discrepancy originates from the
explicit \(r_{\rm ph}m'(r_{\rm ph})\) term in
Eq.~\eqref{eq:sec6_static_photon_sphere}.

\subsection{Screen observables and limitations}
\label{subsec:shadow_coordinates}

Consider a distant observer located at inclination
\begin{equation}
\theta_{o}=\iota
\end{equation}
relative to the rotation axis. For an asymptotically flat spacetime, the
celestial coordinates of a critical photon are
\begin{equation}
{
X
=
-\frac{\xi_{c}}{\sin\iota}
}
\label{eq:sec6_celestial_x}
\end{equation}
and
\begin{equation}
{
Y
=
\pm
\sqrt{
\eta_{c}
+
a^{2}\cos^{2}\iota
-
\xi_{c}^{2}\cot^{2}\iota
}
}.
\label{eq:sec6_celestial_y}
\end{equation}
The shadow boundary is generated parametrically by varying \(r_{p}\)
over the visible unstable spherical-orbit branch:
\begin{equation}
{\cal C}_{\rm sh}
=
\left\{
\left[
X(r_{p}),Y(r_{p})
\right]
\right\}.
\label{eq:sec6_shadow_curve_definition}
\end{equation}
Reality of the screen coordinate requires
\begin{equation}
\eta_{c}
+
a^{2}\cos^{2}\iota
-
\xi_{c}^{2}\cot^{2}\iota
\geq0.
\label{eq:sec6_visibility_condition}
\end{equation}

To quantify the shadow size, we define its area by
\begin{equation}
A_{\rm sh}
=
\frac{1}{2}
\left|
\oint_{{\cal C}_{\rm sh}}
\left(
X\,dY-Y\,dX
\right)
\right|.
\label{eq:sec6_shadow_area}
\end{equation}
The corresponding area-equivalent radius is
\begin{equation}
{
R_{A}
=
\sqrt{
\frac{A_{\rm sh}}{\pi}
}
}.
\label{eq:sec6_area_radius}
\end{equation}
The horizontal geometric center is defined as
\begin{equation}
X_{c}
=
\frac{
X_{\max}+X_{\min}
}{2},
\label{eq:sec6_horizontal_center}
\end{equation}
and the shadow displacement is
\begin{equation}
{
D_{\rm sh}
=
|X_{c}|
}.
\label{eq:sec6_shadow_displacement}
\end{equation}

After shifting the curve by \(X_c\), its radial profile can be written as
\begin{equation}
R(\psi)
=
\sqrt{
\left[X(\psi)-X_c\right]^{2}
+
Y(\psi)^{2}
},
\label{eq:sec6_radial_shadow_profile}
\end{equation}
where
\begin{equation}
\psi
=
\operatorname{atan2}
\left[
Y,
X-X_c
\right].
\label{eq:sec6_shadow_polar_angle}
\end{equation}
The angular residual relative to Kerr is
\begin{equation}
{
\Delta R(\psi)
=
R_{\rm FDM}(\psi)
-
R_{\rm Kerr}(\psi)
}.
\label{eq:sec6_radial_residual}
\end{equation}
Because both curves are first recentered, \(\Delta R(\psi)\) measures a
genuine change of size and shape rather than a simple displacement.

\begin{figure*}[t]
\centering
\includegraphics[
width=0.98\textwidth
]{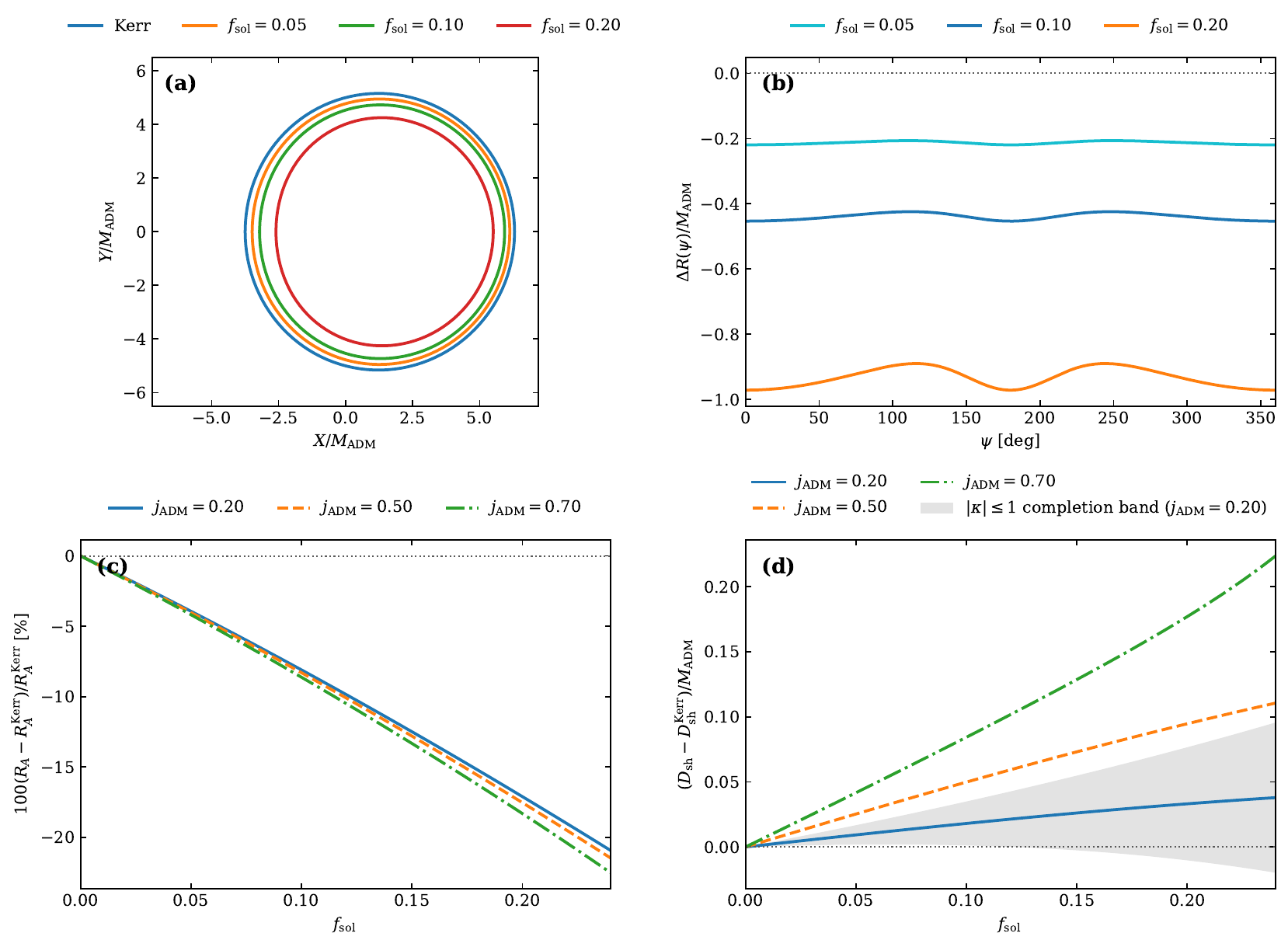}
\caption{
Modification of the black-hole shadow relative to a Kerr spacetime
with the same ADM mass and ADM angular momentum.  All finite-spin curves
are illustrations within the selected Newman--Janis ansatz, not universal
predictions of a radial mass function alone.
Panel (a) compares the critical shadow contours for
\(j_{\rm ADM}=0.70\), \(r_c/M_{\rm ADM}=3\),
\(\iota=60^{\circ}\), and several soliton mass fractions.
Panel (b) shows the recentered radial residual
\(\Delta R(\psi)=R_{\rm FDM}(\psi)-R_{\rm Kerr}(\psi)\).
Panel (c) presents the fractional change of the area-equivalent shadow
radius for several spins.
Panel (d) shows the change of the horizontal shadow displacement
relative to Kerr.  The shaded low-spin band is generated by
Eq.~\eqref{eq:sec3_explicit_alternative_completion} with
\(|\kappa|\leq1\) at \(j_{\rm ADM}=0.20\); it quantifies representative
completion uncertainty rather than numerical error. Only parameter points possessing a positive outer
horizon are retained.  Every contour uses Brent-refined visibility
endpoints; panels (c,d) use 49 uniformly spaced values of \(f_{\rm sol}\)
and plot the refined extrema directly, without grid-maximum extraction or
smoothing.
}
\label{fig:shadow_relative_kerr}
\end{figure*}

Figure~\ref{fig:shadow_relative_kerr} compares the FDM-inspired shadows
with Kerr shadows having the same ADM charges.

Panel~(a) of Fig.~\ref{fig:shadow_relative_kerr} shows that the
environmental deformation changes both the characteristic scale and the
left--right asymmetry of the shadow. Since the asymptotic mass and
angular momentum are held fixed, these changes originate from the
radial redistribution of mass in the photon region rather than from a
different normalization of the spacetime.

The nonzero and angle-dependent residuals in panel~(b) show that the
environmental correction cannot generally be absorbed into a single
rescaling of the shadow diameter. A pure mass rescaling would produce an
approximately constant radial residual after recentering, whereas the
calculated \(\Delta R(\psi)\) retains a nontrivial angular structure.

Panel~(c) isolates the modification of the global shadow scale through
the area-equivalent radius. Panel~(d) shows that the same mass profile
also modifies the spin-induced horizontal displacement.  The shaded band
shows that a completion change of the same environmental scale can be
comparable to, or larger than, the Newman--Janis profile-induced shift
even at \(j_{\rm ADM}=0.20\).  Thus \(R_A\) primarily probes the radial
mass scale, whereas \(D_{\rm sh}\) is conditional on the rotating
completion and is not promoted here as a robust FDM observable.  The
\(j_{\rm ADM}=0.50\) and \(0.70\) curves are retained only as results of
the selected Newman--Janis ansatz; the slow-rotation band is not
extrapolated to them.

The curves in Fig.~\ref{fig:shadow_relative_kerr} are geometric
predictions of the effective metric. They are not observational
posteriors and do not include uncertainties associated with plasma
emission, interferometric sampling, calibration, or image
reconstruction.
This separation between the critical curve, lensing/photon rings, and a
measured intensity image follows the modern shadow literature
\cite{Grenzebach2014,CunhaHerdeiro2018,GrallaHolzWald2019,
PerlickTsupko2022,XuHouWang2018,HouXuWang2018,Haroon2019,
Jusufi2019,Konoplya2019,EHT2019M87I,EHT2019M87VI,
EHT2022SgrAI,EHT2022SgrAVI}.

No phenomenological crescent or beam-convolved intensity image is retained:
the manuscript reports only observables fixed by the effective metric and
its null geodesics.

The shadow calculations in this section follow directly from the
effective metric and require no phenomenological modification of the
null geodesic equations. The separability of the Hamilton--Jacobi
equation and the closed expressions
Eqs.~\eqref{eq:sec6_xi_critical} and
\eqref{eq:sec6_eta_critical} make the photon region reproducible for any
specified radial mass function.

The shadow boundary is a property of geometric optics and does not
determine the brightness distribution of an accretion flow. Plasma
refraction, absorption, scattering, and frequency-dependent
radiative-transfer effects have not been included; the geometric model
limitations are centralized in the Model scope paragraph of
Sec.~\ref{sec:static_geometry}.

The metric-defined observables are the critical curve, photon-orbit
radii, impact parameters, area-equivalent shadow radius, and displacement
relative to Kerr at the same asymptotic charges.  Only the static and
profile-controlled parts are completion robust; the spin-odd displacement
inherits the band in Fig.~\ref{fig:shadow_relative_kerr}. A direct comparison with observational
data would additionally require a self-consistent emission model,
relativistic radiative transfer, realistic baseline sampling, thermal
noise, calibration uncertainties, and image reconstruction.

Within these qualifications, the principal conclusion is that a finite
FDM-inspired radial mass distribution modifies both the location of the
unstable photon region and the mapping of that region to the observer's
screen. The resulting shadow deformation is controlled not only by the
total environmental mass, but also by its core scale and by the radial
mass gradient in the strong-field region.

\section{General radial-profile response and approximation error}
\label{sec:approximation_hierarchy}

The preceding sections treated one radial mass function without expanding
in its amplitude.  We now derive the first-order response for an arbitrary
smooth profile deformation and only afterward specialize it to the
FDM-inspired example.  This separates the general strong-field response
from the numerical benchmark and quantifies the information lost when a
distributed mass is replaced by a constant.

The qualification ``within the selected effective ansatz'' follows the
centralized Model scope statement in Sec.~\ref{sec:static_geometry}.

We work with the dimensionless variables
\begin{equation}
x
=
\frac{r}{M_{\rm ADM}},
\qquad
\widehat r_c
=
\frac{r_c}{M_{\rm ADM}},
\qquad
j_{\rm ADM}
=
\frac{a}{M_{\rm ADM}},
\label{eq:sec7_dimensionless_variables}
\end{equation}
and introduce a formal deformation parameter \(\varepsilon\).  Let
\begin{equation}
\mu(x;\varepsilon)
\equiv\frac{m(r)}{M_{\rm ADM}}
=1+\varepsilon h(x),
\qquad
\lim_{x\to\infty}h(x)=0,
\label{eq:sec7_epsilon_definition}
\end{equation}
where \(h\in C^2\) is otherwise arbitrary on the radial domain of
interest.  The ADM normalization fixes the asymptotic condition above;
the response formulas below do not require the FDM functional form.  For
the example studied numerically,
\begin{equation}
\varepsilon=f_{\rm sol},\qquad
h(x)
\equiv
{\cal F}
\left(
\frac{x}{\widehat r_c}
\right)
-1
\label{eq:sec7_h_definition}
\end{equation}
and
\begin{equation}
{\cal F}(y)
=
I_{\alpha y^{2}/(1+\alpha y^{2})}
\left(
\frac{3}{2},
\frac{13}{2}
\right).
\label{eq:sec7_mass_fraction}
\end{equation}
Since
\begin{equation}
0\leq{\cal F}(y)\leq1,
\end{equation}
one has
\begin{equation}
-1\leq h(x)\leq0.
\label{eq:sec7_h_bounds}
\end{equation}
so \(h(x)\) measures how much of the asymptotic environmental mass remains
outside a radius \(x\).  The bounds in
Eq.~\eqref{eq:sec7_h_bounds} are profile specific, whereas the response
formulas are general.

The normalized radial function is
\begin{equation}
\widehat\Delta(x;\varepsilon)
=
x^{2}
-
2x
\left[
1+\varepsilon h(x)
\right]
+
j_{\rm ADM}^{2}.
\label{eq:sec7_dimensionless_delta}
\end{equation}
It is useful to separate this into
\begin{equation}
\widehat\Delta
=
\widehat\Delta_{0}
+
\varepsilon\,\delta\widehat\Delta,
\label{eq:sec7_delta_decomposition}
\end{equation}
where
\begin{equation}
\widehat\Delta_{0}(x)
=
x^{2}-2x+j_{\rm ADM}^{2}
\label{eq:sec7_kerr_delta}
\end{equation}
is the Kerr radial function and
\begin{equation}
\delta\widehat\Delta(x)
=
-2x h(x).
\label{eq:sec7_delta_perturbation}
\end{equation}
Its derivative is
\begin{equation}
\delta\widehat\Delta'(x)
=
-2
\left[
h(x)+x h'(x)
\right].
\label{eq:sec7_delta_prime_perturbation}
\end{equation}

\paragraph{Unexpanded reference.}
The unexpanded calculation uses the horizon and double-root equations in
Eqs.~\eqref{eq:sec5_dimensionless_horizon_equation}--%
\eqref{eq:sec5_dimensionless_extremal_spin}, the static photon equation
\eqref{eq:sec6_static_photon_sphere}, and the critical-curve construction
of Sec.~\ref{sec:photon_dynamics}, with the complete function
\({\cal F}(x/\widehat r_c)\). These relations are not repeated here; this
section is concerned only with their expansion and numerical error.

\subsection{First-order response of a general profile}
\label{subsec:sec7_perturbation}

For sufficiently small environmental fraction, an observable
\({\cal O}\) can be expanded as
\begin{equation}
{\cal O}(\varepsilon)
=
{\cal O}^{(0)}
+
\varepsilon{\cal O}^{(1)}
+
{\cal O}(\varepsilon^{2}),
\label{eq:sec7_generic_expansion}
\end{equation}
where \({\cal O}^{(0)}\) is the Kerr or Schwarzschild value at fixed
ADM mass.

For the outer horizon, write
\begin{equation}
x_+
=
x_+^{(0)}
+
\varepsilon x_+^{(1)}
+
{\cal O}(\varepsilon^{2}).
\label{eq:sec7_horizon_expansion}
\end{equation}
The zeroth-order radius is
\begin{equation}
x_+^{(0)}
=
1+\sqrt{1-j_{\rm ADM}^{2}}.
\label{eq:sec7_kerr_outer_horizon}
\end{equation}
Expanding
\(\widehat\Delta(x_+;\varepsilon)=0\) gives
\begin{equation}
\widehat\Delta_0'
\left(
x_+^{(0)}
\right)
x_+^{(1)}
+
\delta\widehat\Delta
\left(
x_+^{(0)}
\right)
=
0.
\label{eq:sec7_horizon_linear_equation}
\end{equation}
Since
\begin{equation}
\widehat\Delta_0'
\left(
x_+^{(0)}
\right)
=
2
\left[
x_+^{(0)}-1
\right],
\end{equation}
the first-order correction is
\begin{equation}
{
x_+^{(1)}
=
\frac{
x_+^{(0)}
h\!\left(x_+^{(0)}\right)
}{
x_+^{(0)}-1
}
}.
\label{eq:sec7_horizon_first_order}
\end{equation}
Because \(h\leq0\), the correction is negative. At fixed ADM mass and
spin, distributing part of the total mass outside the near-horizon
region moves the outer horizon inward.

Equation~\eqref{eq:sec7_horizon_first_order} also reveals that the
expansion becomes nonuniform near the Kerr extremal limit:
\begin{equation}
x_+^{(0)}-1
=
\sqrt{1-j_{\rm ADM}^{2}}
\longrightarrow0.
\label{eq:sec7_near_extremal_denominator}
\end{equation}
A small \(\varepsilon\) is therefore not by itself sufficient near
\(j_{\rm ADM}=1\). A stronger consistency condition is
\begin{equation}
\left|
\varepsilon
\frac{
x_+^{(0)}h(x_+^{(0)})
}{
x_+^{(0)}-1
}
\right|
\ll
x_+^{(0)}.
\label{eq:sec7_horizon_perturbative_condition}
\end{equation}

For the extremal configuration, let
\begin{equation}
x_e
=
1+\varepsilon x_e^{(1)}
+
{\cal O}(\varepsilon^{2}).
\label{eq:sec7_extremal_radius_expansion}
\end{equation}
Expansion of
Eq.~\eqref{eq:sec5_dimensionless_extremal_radius} yields
\begin{equation}
{
x_e^{(1)}
=
h(1)+h'(1)
}.
\label{eq:sec7_extremal_radius_first_order}
\end{equation}
Substitution into
Eq.~\eqref{eq:sec5_dimensionless_extremal_spin} gives
\begin{equation}
j_{\rm ext}^{2}
=
1+2\varepsilon h(1)
+
{\cal O}(\varepsilon^{2}),
\label{eq:sec7_extremal_spin_squared_expansion}
\end{equation}
and hence
\begin{equation}
{
j_{\rm ext}
=
1+\varepsilon h(1)
+
{\cal O}(\varepsilon^{2})
}.
\label{eq:sec7_extremal_spin_first_order}
\end{equation}
The terms involving \(h'(1)\) cancel from the first-order double-root spin,
although they remain present in the double-root radius. Since \(h(1)<0\),
the selected ansatz places this double-root locus below unity; this is not
a modified Kerr angular-momentum bound.

For the static photon sphere, write
\begin{equation}
x_{\rm ph}
=
3
+
\varepsilon x_{\rm ph}^{(1)}
+
{\cal O}(\varepsilon^{2}).
\label{eq:sec7_photon_sphere_expansion}
\end{equation}
Expansion of
Eq.~\eqref{eq:sec6_static_photon_sphere} gives
\begin{equation}
{
x_{\rm ph}^{(1)}
=
3
\left[
h(3)-h'(3)
\right]
}.
\label{eq:sec7_photon_sphere_first_order}
\end{equation}
The two contributions have distinct origins. The term \(3h(3)\)
describes the reduction of the enclosed mass relative to
\(M_{\rm ADM}\), while the term \(-3h'(3)\) is the explicit
mass-gradient correction.

The perturbation of the rotating critical curve follows from
\begin{equation}
\widehat\Delta
=
\widehat\Delta_0
+
\varepsilon\delta\widehat\Delta
\end{equation}
and
\begin{equation}
\xi_c
=
\xi_c^{(0)}
+
\varepsilon\xi_c^{(1)}
+
{\cal O}(\varepsilon^{2}).
\label{eq:sec7_xi_expansion}
\end{equation}
At fixed spherical-orbit parameter \(x_p\), one obtains
\begin{equation}
{
\xi_c^{(1)}
=
\frac{
4x_p
\left[
\widehat\Delta_0
\delta\widehat\Delta'
-
\delta\widehat\Delta
\widehat\Delta_0'
\right]
}{
j_{\rm ADM}
\left(
\widehat\Delta_0'
\right)^{2}
}
}.
\label{eq:sec7_xi_first_order}
\end{equation}
Similarly,
\begin{align}
\eta_c^{(1)}
={}&
16x_p^{2}
\left[
\frac{
\delta\widehat\Delta
}{
(\widehat\Delta_0')^{2}
}
-
\frac{
2\widehat\Delta_0
\delta\widehat\Delta'
}{
(\widehat\Delta_0')^{3}
}
\right]
\nonumber\\
&-
2
\left(
\xi_c^{(0)}-j_{\rm ADM}
\right)
\xi_c^{(1)}.
\label{eq:sec7_eta_first_order}
\end{align}
The screen-coordinate corrections are
\begin{equation}
X^{(1)}
=
-\frac{\xi_c^{(1)}}{\sin\iota}
\label{eq:sec7_screen_x_first_order}
\end{equation}
and
\begin{equation}
Y^{(1)}
=
\frac{
\eta_c^{(1)}
-
2\xi_c^{(0)}
\xi_c^{(1)}
\cot^{2}\iota
}{
2Y^{(0)}
}.
\label{eq:sec7_screen_y_first_order}
\end{equation}

Equation~\eqref{eq:sec7_screen_y_first_order} is a \emph{local} expansion
at fixed \(x_p\); it is not valid pointwise at the visible-branch
endpoints, where \(Y^{(0)}=0\).  Let
\begin{equation}
{\cal B}(x_p,\varepsilon)
=\eta_c+j_{\rm ADM}^{2}\cos^2\iota
-\xi_c^2\cot^2\iota
\label{eq:sec7_visibility_function}
\end{equation}
and write the two simple endpoint roots as
\(x_\pm=x_\pm^{(0)}+\varepsilon x_\pm^{(1)}+\cdots\).  Their displacement is
\begin{equation}
x_\pm^{(1)}
=-\left.
\frac{\partial_\varepsilon{\cal B}}
{\partial_{x_p}{\cal B}}
\right|_{(x_\pm^{(0)},0)}.
\label{eq:sec7_endpoint_shift}
\end{equation}
Although \(Y^{(1)}\sim|x_p-x_\pm^{(0)}|^{-1/2}\) in the fixed-\(x_p\)
description, the singularity is integrable.  In the line-integral area
variation the explicit moving-endpoint contribution vanishes because the
unperturbed upper and lower branches meet at \(Y^{(0)}=0\).  A pointwise
endpoint value of Eq.~\eqref{eq:sec7_screen_y_first_order} is nevertheless
undefined and is never used numerically.

For the global contour we instead use a fixed screen angle.  After
recentring by \(X_c(\varepsilon)\), define
\begin{equation}
X-X_c=R(\psi,\varepsilon)\cos\psi,\qquad
Y=R(\psi,\varepsilon)\sin\psi,
\label{eq:sec7_screen_angle_parameterization}
\end{equation}
with \(0\leq\psi<2\pi\) and
\(R=R^{(0)}+\varepsilon R^{(1)}+\cdots\).  Then
\begin{equation}
A_{\rm sh}^{(1)}
=\int_0^{2\pi}R^{(0)}(\psi)R^{(1)}(\psi)\,d\psi,
\label{eq:sec7_area_fixed_angle}
\end{equation}
which is regular at the left and right screen points and automatically
includes the endpoint motion in
Eq.~\eqref{eq:sec7_endpoint_shift}.  The fixed-\(x_p\) formulas for
\(\xi_c^{(1)}\) and \(\eta_c^{(1)}\) are therefore analytic local response
coefficients; the global shadow-area coefficient reported below is obtained
by endpoint-refined numerical differentiation in this fixed-\(\psi\)
representation, not claimed as a closed analytic area variation.

Expanding the shadow area as
\begin{equation}
A_{\rm sh}
=
A_{\rm sh}^{(0)}
+
\varepsilon A_{\rm sh}^{(1)}
+
{\cal O}(\varepsilon^{2}),
\label{eq:sec7_area_expansion}
\end{equation}
the first-order area variation is
\begin{align}
A_{\rm sh}^{(1)}
=
\frac{1}{2}
\oint
\big[
&
X^{(1)}dY^{(0)}
+
X^{(0)}dY^{(1)}
\nonumber\\
&
-
Y^{(1)}dX^{(0)}
-
Y^{(0)}dX^{(1)}
\big].
\label{eq:sec7_area_first_order}
\end{align}
The area-equivalent radius then satisfies
\begin{equation}
{
R_A
=
R_A^{(0)}
+
\varepsilon
\frac{
A_{\rm sh}^{(1)}
}{
2\pi R_A^{(0)}
}
+
{\cal O}(\varepsilon^{2})
}.
\label{eq:sec7_shadow_radius_first_order}
\end{equation}

For the representative parameters
\begin{equation}
j_{\rm ADM}=0.70,
\qquad
\widehat r_c=3,
\qquad
\iota=60^{\circ},
\label{eq:sec7_representative_parameters}
\end{equation}
the endpoint-refined Kerr value and one-sided first-order coefficient are
\begin{equation}
\frac{R_A^{(0)}}{M_{\rm ADM}}
=
5.09440904
\label{eq:sec7_kerr_shadow_radius_numeric}
\end{equation}
and
\begin{equation}
\left.
\frac{\partial(R_A/M_{\rm ADM})}
{\partial f_{\rm sol}}
\right|_{f_{\rm sol}=0^+}
=
-4.132227\pm0.0000005.
\label{eq:sec7_shadow_linear_coefficient}
\end{equation}
Table~\ref{tab:app_shadow_fit_stability} shows the direct one-sided
differences and Richardson extrapolants.  The uncertainty is dominated by
the residual \(f_{\rm sol}\)-window truncation; quadrature and endpoint
errors are reported separately in Appendix~\ref{app:numerics}.  This is a
numerical derivative of the regularized contour, not a closed analytic
evaluation of Eq.~\eqref{eq:sec7_area_first_order}. Thus
\begin{equation}
\frac{R_A}{M_{\rm ADM}}
=
5.09440904
-
4.132227 f_{\rm sol}
+
{\cal O}(f_{\rm sol}^{2})
\label{eq:sec7_shadow_numeric_expansion}
\end{equation}
for this specific parameter choice.

\subsection{Constant-mass approximation and numerical validity}
\label{subsec:sec7_comparison}

The constant-ADM-mass approximation replaces the complete radial mass
function by
\begin{equation}
m(r)
\longrightarrow
M_{\rm ADM},
\label{eq:sec7_constant_mass_replacement}
\end{equation}
or equivalently,
\begin{equation}
\mu(x)
\longrightarrow1,
\qquad
h(x)\longrightarrow0,
\qquad
h'(x)\longrightarrow0.
\label{eq:sec7_constant_mass_h_zero}
\end{equation}
The radial function then becomes
\begin{equation}
\widehat\Delta_{\rm const}
=
x^{2}-2x+j_{\rm ADM}^{2},
\label{eq:sec7_constant_mass_delta}
\end{equation}
which is exactly the Kerr radial function. Consequently,
\begin{equation}
x_+^{\rm const}
=
1+\sqrt{1-j_{\rm ADM}^{2}},
\label{eq:sec7_constant_horizon}
\end{equation}
\begin{equation}
j_{\rm ext}^{\rm const}
=
1,
\label{eq:sec7_constant_extremal_spin}
\end{equation}
\begin{equation}
x_{\rm ph}^{\rm const}
=
3,
\label{eq:sec7_constant_photon_sphere}
\end{equation}
and
\begin{equation}
R_A^{\rm const}
=
R_A^{\rm Kerr}(j_{\rm ADM},\iota).
\label{eq:sec7_constant_shadow_radius}
\end{equation}
This approximation is therefore independent of \(f_{\rm sol}\).
It cannot represent any response caused by the radial distribution of
the environmental mass.

\begin{figure*}[t]
\centering
\includegraphics[
width=0.98\textwidth
]{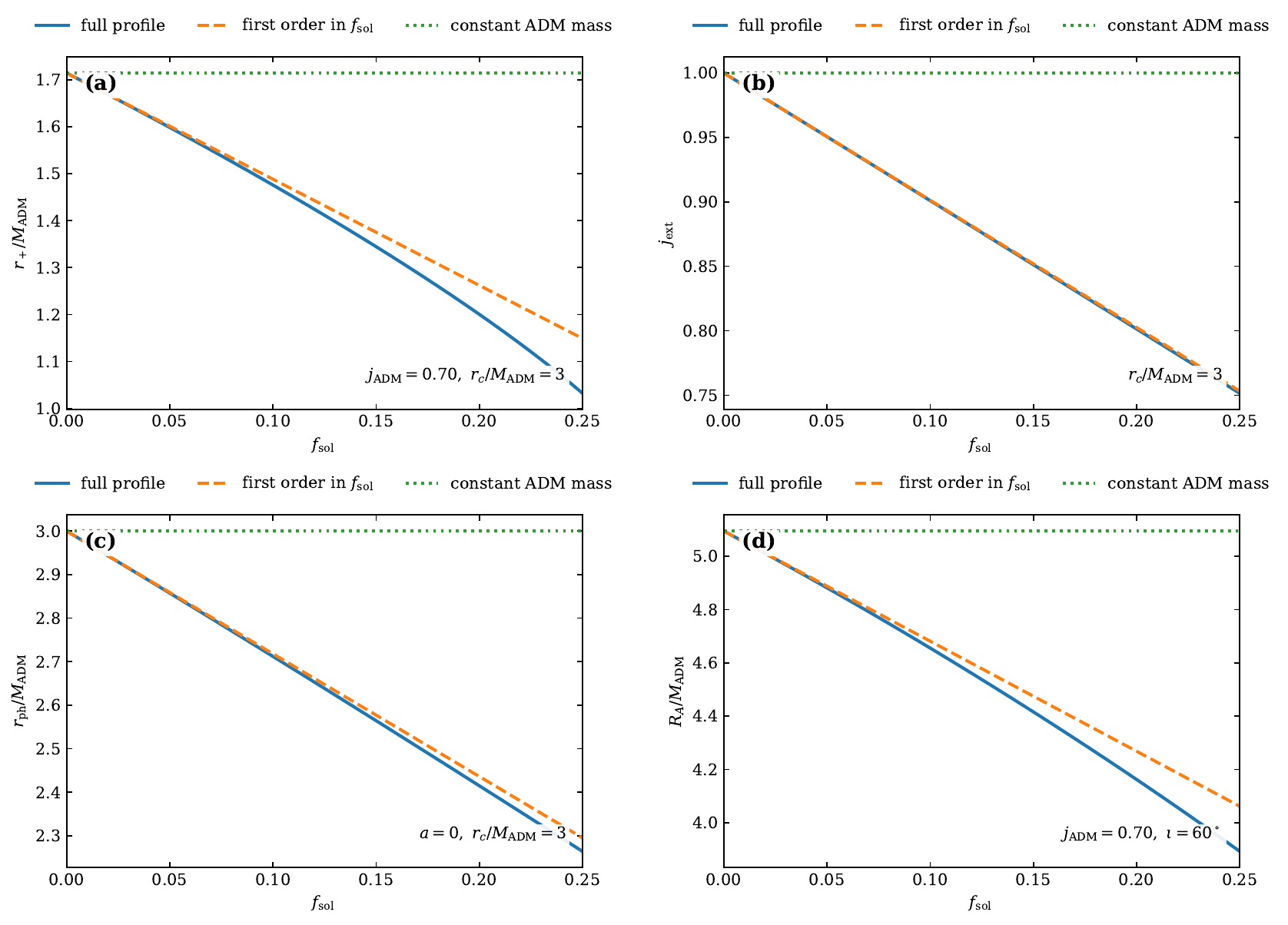}
\caption{
Comparison among the full radial-profile calculation within the selected
effective ansatz, the first-order
perturbative expansion in the soliton mass fraction, and the
constant-ADM-mass approximation.
Panel (a) shows the outer-horizon radius for
\(j_{\rm ADM}=0.70\) and \(r_c/M_{\rm ADM}=3\).
Panel (b) displays the generalized extremal spin as a function of
\(f_{\rm sol}\).
Panel (c) compares the static photon-sphere radius obtained from the
full radial-profile equation
\(r_{\rm ph}-3m(r_{\rm ph})+r_{\rm ph}m'(r_{\rm ph})=0\)
with its first-order expansion and the Schwarzschild constant-mass
value.
Panel (d) presents the area-equivalent shadow radius for
\(j_{\rm ADM}=0.70\), \(r_c/M_{\rm ADM}=3\), and
\(\iota=60^{\circ}\).
The perturbative approximation follows the full-profile result at small
\(f_{\rm sol}\) but progressively departs from it as the environmental
fraction increases, whereas the constant-mass approximation removes the
radial-profile response entirely.
}
\label{fig:exact_perturbative_constant_mass}
\end{figure*}

Figure~\ref{fig:exact_perturbative_constant_mass} compares the three
descriptions.

Panel~(a) of Fig.~\ref{fig:exact_perturbative_constant_mass} shows that
the full-profile outer horizon moves inward as \(f_{\rm sol}\) increases. The
first-order expression captures the initial slope but eventually
underestimates the nonlinear curvature of the full result within the ansatz. This
observable is particularly sensitive to higher-order terms because the
root of \(\widehat\Delta\) shifts into a region where both \(h(x)\) and
its derivatives differ from their values at the unperturbed Kerr
horizon.

Panel~(b) demonstrates that the generalized extremal spin decreases
approximately linearly with the soliton fraction over the displayed
range. The first-order approximation is especially accurate for this
quantity. This behavior follows from the cancellation of the
\(h'(1)\) terms in
Eq.~\eqref{eq:sec7_extremal_spin_first_order}. The constant-mass
approximation instead predicts \(j_{\rm ext}=1\) for every
\(f_{\rm sol}\) and therefore misses the environmental correction.

Panel~(c) shows that the first-order photon-sphere formula remains close
to the radial-profile result over a broader interval than the first-order
horizon formula. The constant-mass approximation retains the fixed
Schwarzschild value \(x_{\rm ph}=3\) and fails to capture both the
enclosed-mass correction and the explicit mass-gradient contribution.

Panel~(d) exhibits the same approximation hierarchy for the rotating
shadow. At small \(f_{\rm sol}\), the first-order contour deformation
reproduces the full-profile area-equivalent radius. At larger fractions,
nonlinear changes in the spherical photon-orbit interval and in the
critical impact parameters become important. The constant-mass curve
remains at the Kerr value and therefore increasingly overestimates the
full-profile shadow size within the selected ansatz.

To quantify these statements, define the absolute relative approximation
errors
\begin{equation}
{\cal E}_{\rm pert}^{({\cal O})}
=
\left|
\frac{
{\cal O}_{\rm exact}
-
{\cal O}_{\rm pert}
}{
{\cal O}_{\rm exact}
}
\right|
\label{eq:sec7_perturbative_error}
\end{equation}
and
\begin{equation}
{\cal E}_{\rm const}^{({\cal O})}
=
\left|
\frac{
{\cal O}_{\rm exact}
-
{\cal O}_{\rm const}
}{
{\cal O}_{\rm exact}
}
\right|.
\label{eq:sec7_constant_error}
\end{equation}
For the parameters used in
Fig.~\ref{fig:exact_perturbative_constant_mass}, the first-order errors
at
\begin{equation}
f_{\rm sol}=0.10
\end{equation}
are approximately
\begin{equation}
{\cal E}_{\rm pert}^{(r_+)}
\simeq0.8\%,
\qquad
{\cal E}_{\rm pert}^{(j_{\rm ext})}
\simeq0.04\%,
\label{eq:sec7_errors_f01_first}
\end{equation}
\begin{equation}
{\cal E}_{\rm pert}^{(r_{\rm ph})}
\simeq0.2\%,
\qquad
{\cal E}_{\rm pert}^{(R_A)}
\simeq0.5\%.
\label{eq:sec7_errors_f01_second}
\end{equation}
Thus the linearized description is accurate at approximately the
percent level for this representative configuration when
\(f_{\rm sol}\lesssim0.1\).

At
\begin{equation}
f_{\rm sol}=0.20,
\end{equation}
the corresponding errors are approximately
\begin{equation}
{\cal E}_{\rm pert}^{(r_+)}
\simeq5.1\%,
\qquad
{\cal E}_{\rm pert}^{(j_{\rm ext})}
\simeq0.15\%,
\label{eq:sec7_errors_f02_first}
\end{equation}
\begin{equation}
{\cal E}_{\rm pert}^{(r_{\rm ph})}
\simeq0.9\%,
\qquad
{\cal E}_{\rm pert}^{(R_A)}
\simeq2.5\%.
\label{eq:sec7_errors_f02_second}
\end{equation}
The perturbative description therefore fails first for the horizon
location, whereas the extremal spin and static photon-sphere radius
remain comparatively well approximated.

\begin{table*}[t]
\caption{Endpoint-refined values and absolute relative errors for
\(\widehat r_c=3\), \(j_{\rm ADM}=0.70\), and
\(\iota=60^\circ\).  Errors are defined by
Eqs.~\eqref{eq:sec7_perturbative_error} and
\eqref{eq:sec7_constant_error}.}
\label{tab:sec7_error_values}
\begin{ruledtabular}
\begin{tabular}{lcccccc}
& \multicolumn{3}{c}{\(f_{\rm sol}=0.10\)}
& \multicolumn{3}{c}{\(f_{\rm sol}=0.20\)}\\
Observable & profile result & first order [\%] & constant mass [\%]
& profile result & first order [\%] & constant mass [\%]\\
\hline
\(r_+/M_{\rm ADM}\) & 1.47563794 & 0.844 & 16.163
& 1.20037099 & 5.138 & 42.801\\
\(j_{\rm ext}\) & 0.90093833 & 0.036 & 10.995
& 0.80133492 & 0.149 & 24.792\\
\(r_{\rm ph}/M_{\rm ADM}\) & 2.71195922 & 0.223 & 10.621
& 2.41464836 & 0.885 & 24.242\\
\(R_A/M_{\rm ADM}\) & 4.65524223 & 0.557 & 9.434
& 4.16189081 & 2.549 & 22.406\\
\end{tabular}
\end{ruledtabular}
\end{table*}

The constant-mass approximation is substantially less accurate; the
observable-by-observable values are reported in
Table~\ref{tab:sec7_error_values}, rather than only as ranges.  These
deviations are not caused by a failure of a particular truncation order.
They occur because the replacement
\(m(r)\rightarrow M_{\rm ADM}\) removes the radial-profile information
by construction.

\begin{figure*}[t]
\centering
\includegraphics[width=0.99\textwidth]
{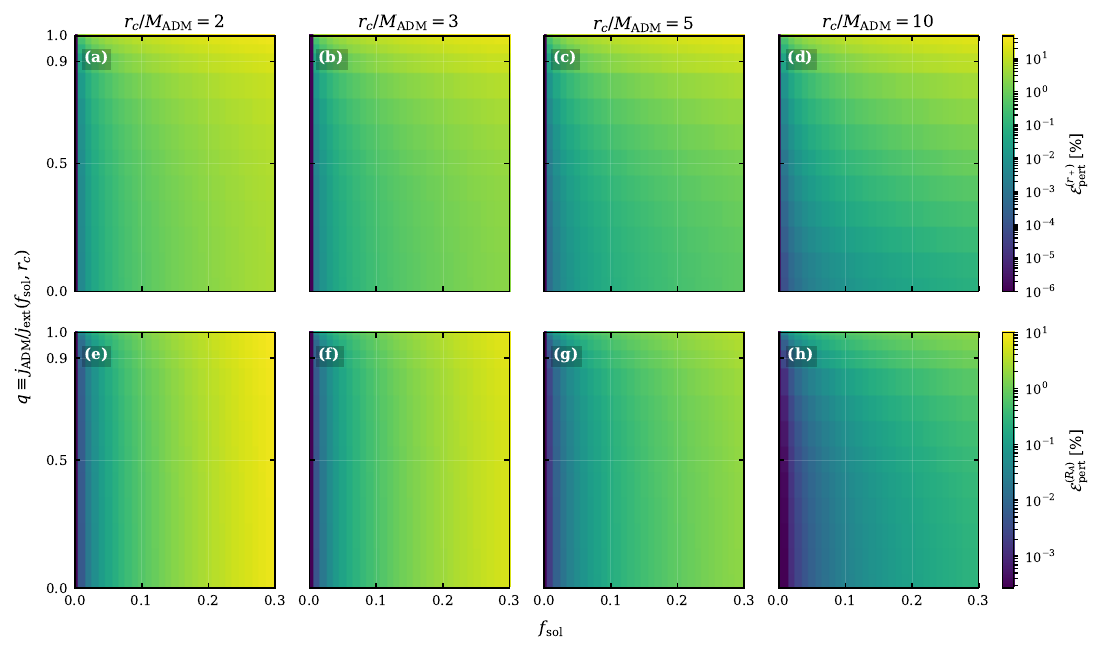}
\caption{Absolute relative first-order errors over 1,736 audited
configurations at \(\iota=60^\circ\).  Columns correspond to
\(\widehat r_c=2,3,5,10\); the upper row shows
\({\cal E}_{\rm pert}^{(r_+)}\), and the lower row shows
\({\cal E}_{\rm pert}^{(R_A)}\).  The grid contains 31 values of
\(0\leq f_{\rm sol}\leq0.30\) and 14 values of
\(q=j_{\rm ADM}/j_{\rm ext}\), including
\(q=0.95,0.975,0.99,0.999\).  Shadow areas use all refined visible
intervals, an endpoint-regularizing substitution, and adaptive
quadrature.  Exact zeros at \(f_{\rm sol}=0\) are clipped only for display
on the logarithmic color scale; the archived CSV retains zero.  These are
errors within the selected off-shell completion, not completion
uncertainties.}
\label{fig:error_surfaces}
\end{figure*}

Figure~\ref{fig:error_surfaces} tests the first-order approximation beyond
the single representative configuration.  It uses
\[
q\equiv j_{\rm ADM}/j_{\rm ext}(f_{\rm sol},\widehat r_c)
\]
to compare equal relative distances from the selected completion's
double-root curve.  At each grid point the perturbative formula is
evaluated at the same fixed physical \(j_{\rm ADM}\), rather than
differentiated along a constant-\(q\) path.

For \(f_{\rm sol}\leq0.10\) and \(q\leq0.70\), the maxima over all four
core scales are \(0.9163\%\) for \(r_+\) and \(0.6640\%\) for \(R_A\).
Extending to \(q\leq0.90\) raises them to \(2.5689\%\) and \(0.8014\%\).
If \(q=0.999\) is included, the corresponding horizon maximum is already
\(19.0718\%\), although the shadow-radius maximum remains \(1.4708\%\).
Over the full audited grid the maxima are \(47.4957\%\) and \(10.0445\%\),
both at \(\widehat r_c=2\), \(f_{\rm sol}=0.30\), and \(q=0.999\).
Thus the horizon expansion becomes strongly nonuniform near the
double-root curve, while the area radius is appreciably more robust.

The one-dimensional quantitative values above apply to
\(\widehat r_c=3\), \(j_{\rm ADM}=0.70\), and
\(\iota=60^{\circ}\). They are not universal bounds. The perturbative
domain also depends on the core scale and spin. In particular, the
horizon expansion becomes increasingly restrictive near extremality
because of the factor
\begin{equation}
\frac{1}{\sqrt{1-j_{\rm ADM}^{2}}}
\end{equation}
in Eq.~\eqref{eq:sec7_horizon_first_order}. A suitable general
criterion is therefore not merely
\(\varepsilon\ll1\), but
\begin{equation}
\varepsilon
\left|
\frac{
{\cal O}^{(1)}
}{
{\cal O}^{(0)}
}
\right|
\ll1
\label{eq:sec7_observable_validity_condition}
\end{equation}
for every observable used in a given calculation.

The comparison establishes the following hierarchy. For sufficiently
small \(f_{\rm sol}\) and away from the near-extremal region, the
first-order expansion provides a controlled and transparent
approximation to the full radial-profile model within the selected ansatz.
For moderate environmental fractions, the nonperturbative mass function is
required, particularly for horizon-sensitive quantities. The
constant-ADM-mass approximation is useful only as a Kerr reference
baseline; it is not an approximation capable of retaining the
geometrical effects of the FDM-inspired mass distribution.

This distinction is central to the interpretation of the preceding
shadow results. The changes in the photon region and critical curve are
not determined solely by the total ADM mass. They depend on the enclosed
mass \(m(r)\), the mass gradient \(m'(r)\), and, for stability
properties, the second derivative \(m''(r)\). Replacing this structure
by a constant removes precisely the information that the effective
model is designed to probe.

\section{Discussion and conclusions}
\label{sec:discussion_conclusions}

The central result is a profile-portable sufficient one-minimum
criterion within the one-function radial-\(\Delta\) family.  For a
positive, nondecreasing \(C^2\) radial mass with the stated central and
asymptotic derivative limits, the bound \(2m'+rm''<1\) makes
\(\Delta=r^2-2rm+a^2\) strictly convex on the positive half-line.
For \(a\ne0\), the sign of its unique minimum gives two simple positive
roots, one double root, or none.  For \(a=0\), \(r=0\) is an excluded
factor root at the singular inner boundary and exactly one simple positive
root remains.  There is
at most one extremal double-root branch, and it is physical only when the
derived \(a_{\rm ext}^2\) is nonnegative.
The adopted FDM-inspired profile satisfies the theorem whenever
\(f_{\rm sol}/(r_c/M_{\rm ADM})<0.691428\), a domain much wider than the
production box.  Hernquist-like and Jaffe-like examples give independent
analytic applications, while a smooth monotone shell explicitly produces
three stationary points and four positive roots after the convexity bound
is violated.  This complements the source-based ultracompact branches of
Fonseca \emph{et al.} \cite{Fonseca2026}.  The large all-root scan verifies
the implementation; it is not the proof.

The second general result is the response theory for
\(\mu=1+\varepsilon h\).  The outer-horizon, extremal-branch,
photon-sphere, and fixed-angle critical-curve shifts are determined by
local combinations of \(h\) and \(h'\), with endpoint motion included in
the global shadow functional.  The FDM-inspired mass integral and WEC
factor are analytic examples of these general relations.  Replacing
\(m(r)\) by \(M_{\rm ADM}\) returns Kerr by definition and is retained only
as a zero-profile reference.

Two numerical audits delimit these formulas.  The all-branch search over
2976 configurations, extending to \(j_{\rm ADM}/j_{\rm ext}=0.999\),
finds two visibility roots, no instability root, and one connected
visible unstable interval at every point; the near-extremal cutoff test is
stable at the \(10^{-14}\) level.  The independent 1736-point error map
shows that \(f_{\rm sol}\leq0.1\) gives subpercent errors in both \(r_+\)
and \(R_A\) only for moderate relative spin
\(j_{\rm ADM}/j_{\rm ext}\leq0.7\).  The horizon expansion becomes
nonuniform near the double-root curve, reaching a \(19.1\%\) error at
\(q=0.999\) even in that small-fraction domain, whereas the corresponding
area-radius error remains below \(1.5\%\).  No production-box-wide
``percent accuracy'' claim is made.

Rotational nonuniqueness is now separated by a slow-rotation comparison.
The static mass profile fixes the \(a=0\) horizon and photon response, and
variations of the frame-dragging completion do not move the horizon at
linear order in \(a\).  They do change the spin-odd shadow displacement at
that order, while the area-equivalent radius first acquires completion
uncertainty at order \(a^2\).  The explicit same-charge family
\(\omega_\kappa\) produces displacement half-widths \(0.01687M_{\rm ADM}\)
and \(0.04343M_{\rm ADM}\) at \(f_{\rm sol}=0.1\) and \(0.2\) for the
low-spin benchmark, showing that this uncertainty is not negligible.
Unlike the source-based equation of Datta and Singha
\cite{DattaSingha2026}, this band is a completion diagnostic rather than a
matter solution.  Source-derived finite-spin environments can require
additional metric functions and can lose Carter integrability altogether
\cite{FernandesCardoso2025,DestounisFernandes2026}; accordingly, the
\(j_{\rm ADM}=0.5\) and \(0.7\) curves are illustrations within the
selected Newman--Janis ansatz.  A slow-rotation construction cannot validate
the near-extremal branch; the reported \(j_{\rm ext}\) is therefore the
double-root locus of the selected completion, not a correction to a
universal Kerr bound.

As a subsidiary consistency check, the static seed also admits a
positive-potential test-field statement.
When \(m-rm'\geq0\), the massless-scalar radial operator is nonnegative and
has no exponentially growing modes.  The FDM-inspired profile obeys this
condition throughout the production box with the uniform dimensionless
margin \(\mu-x\mu'>0.5956\).  The explicit energy identity covers
\(\ell=0\) and excludes a normalizable zero mode.  This is mode stability
only, not boundedness, decay, or rotating gravitational and matter-sector
stability.

The physical interpretation remains sharply delimited.  A weak-field
mapping with \(m_\phi=10^{-19}\,\mathrm{eV}\),
\(f_{\rm sol}=10^{-2}\), and a \(4\times10^6M_\odot\) central black hole
has \(\alpha_g\simeq3.0\times10^{-3}\) and
\(r_c/M_{\rm ADM}\simeq2.9\times10^7\).  Its genuine strong-field
profile-gradient effect is below \(10^{-20}\); percent-level shifts at
fixed total ADM mass are charge-normalization effects.  Conversely, the
compact production box is not a self-consistent FDM core.  A relativistic
black-hole--scalar solution, a rotating mode analysis, and a source-based
determination of the frame-dragging function remain necessary before the
benchmark can be promoted to an astrophysical FDM prediction.

\appendix

\section{Static Einstein equations and closure nonuniqueness}
\label{app:static_closure}

The soliton density profile determines the enclosed mass uniquely, but it
does not by itself determine a unique static spacetime. To make this point
explicit, consider the general static and spherically symmetric line
element
\begin{equation}
ds^{2}
=
-e^{2\Phi(r)}dt^{2}
+
\frac{dr^{2}}{1-2m(r)/r}
+
r^{2}
\left(
d\theta^{2}
+
\sin^{2}\theta\,d\phi^{2}
\right),
\label{eq:app_general_static_metric}
\end{equation}
with an anisotropic effective stress tensor
\begin{equation}
T^{\mu}{}_{\nu}
=
\operatorname{diag}
\left(
-\rho,
p_{r},
p_{\perp},
p_{\perp}
\right).
\label{eq:app_static_stress_tensor}
\end{equation}
The independent Einstein equations are
\begin{equation}
m'(r)
=
4\pi r^{2}\rho(r),
\label{eq:app_mass_equation}
\end{equation}
\begin{equation}
\Phi'(r)
=
\frac{
m(r)+4\pi r^{3}p_{r}(r)
}{
r\left[r-2m(r)\right]
},
\label{eq:app_redshift_equation}
\end{equation}
and the anisotropic conservation equation
\begin{equation}
p_{r}'(r)
=
-
\left[
\rho(r)+p_{r}(r)
\right]
\Phi'(r)
+
\frac{2}{r}
\left[
p_{\perp}(r)-p_{r}(r)
\right].
\label{eq:app_anisotropic_tov}
\end{equation}

Equation~\eqref{eq:app_mass_equation} shows that the density fixes
\(m(r)\), up to the central integration constant. It does not fix the
redshift function \(\Phi(r)\). A further relation, such as an equation of
state, a pressure anisotropy prescription, or a direct condition on
\(\Phi(r)\), is required to close the system.

The static seed used in the main text adopts the Schwarzschild gauge
\begin{equation}
e^{2\Phi(r)}
=
1-\frac{2m(r)}{r}
\equiv
f(r),
\label{eq:app_schwarzschild_gauge}
\end{equation}
so that
\begin{equation}
ds^{2}
=
-f(r)\,dt^{2}
+
\frac{dr^{2}}{f(r)}
+
r^{2}d\Omega^{2}.
\label{eq:app_static_seed}
\end{equation}
Substitution of Eq.~\eqref{eq:app_schwarzschild_gauge} into
Eq.~\eqref{eq:app_redshift_equation} imposes
\begin{equation}
{
p_{r}(r)
=
-\rho(r)
}.
\label{eq:app_radial_closure}
\end{equation}
The tangential pressure is then fixed by
Eq.~\eqref{eq:app_anisotropic_tov}, or equivalently by the angular
Einstein equation:
\begin{equation}
{
p_{\perp}(r)
=
-\frac{m''(r)}{8\pi r}
=
-\rho(r)
-
\frac{r}{2}\rho'(r)
}.
\label{eq:app_tangential_pressure}
\end{equation}

For the adopted FDM profile,
\begin{equation}
\rho_{\rm FDM}(r)
=
\rho_{c}
\left(
1+\alpha\frac{r^{2}}{r_{c}^{2}}
\right)^{-8},
\label{eq:app_fdm_density}
\end{equation}
one has
\begin{equation}
\frac{\rho_{\rm FDM}'(r)}
{\rho_{\rm FDM}(r)}
=
-\frac{
16\alpha r
}{
r_{c}^{2}+\alpha r^{2}
}.
\label{eq:app_density_log_derivative}
\end{equation}
The tangential pressure therefore becomes
\begin{equation}
p_{\perp}(r)
=
\rho_{\rm FDM}(r)
\frac{
7\alpha r^{2}-r_{c}^{2}
}{
r_{c}^{2}+\alpha r^{2}
}.
\label{eq:app_fdm_tangential_pressure}
\end{equation}
The radial null-energy combination is saturated,
\begin{equation}
\rho_{\rm FDM}+p_{r}=0,
\label{eq:app_static_radial_nec}
\end{equation}
whereas the tangential combination is
\begin{equation}
\rho_{\rm FDM}+p_{\perp}
=
\frac{
8\alpha r^{2}
}{
r_{c}^{2}+\alpha r^{2}
}
\rho_{\rm FDM}
\geq0.
\label{eq:app_static_tangential_nec}
\end{equation}
Thus the static seed satisfies the null and weak energy conditions.

The relation \(p_{r}=-\rho\), however, is not implied by the FDM density
profile itself. It follows from the additional gauge and closure choice
\begin{equation}
g_{tt}g_{rr}=-1.
\end{equation}
For the same prescribed density, one could instead impose isotropic
pressure,
\begin{equation}
p_{r}=p_{\perp},
\end{equation}
choose a barotropic equation of state,
\begin{equation}
p_{r}=p_{r}(\rho),
\end{equation}
or specify an independent anisotropy function
\begin{equation}
\Pi(r)
\equiv
p_{\perp}(r)-p_{r}(r).
\end{equation}
Each choice generally produces a different \(\Phi(r)\), pressure
profile, and static metric while retaining the same density and enclosed
mass.

The Newman--Janis construction in this work is therefore conditional on
the particular seed
Eq.~\eqref{eq:app_static_seed}. The density profile uniquely determines
the radial mass function, but the complete static geometry is fixed only
after the Schwarzschild-gauge closure has been imposed.

\section{Algebraic identities and Newman--Janis consistency checks}
\label{app:nja_checks}

The static seed is defined by
\begin{equation}
f(r)
=
1-\frac{2m(r)}{r}.
\label{eq:app_seed_function}
\end{equation}
The fixed complexification rule used in this work leaves the radial mass
function real and uncomplexified:
\begin{equation}
m(r)
\longrightarrow
m(r),
\label{eq:app_mass_complexification}
\end{equation}
while
\begin{equation}
r^{2}
\longrightarrow
\Sigma
=
r^{2}+a^{2}\cos^{2}\theta,
\qquad
\frac{2m(r)}{r}
\longrightarrow
\frac{2r\,m(r)}{\Sigma}.
\label{eq:app_fixed_complexification}
\end{equation}
This prescription yields
\begin{equation}
\Delta(r)
=
r^{2}-2r\,m(r)+a^{2},
\label{eq:app_delta_definition}
\end{equation}
which remains a function of \(r\) alone.

After transforming to Boyer--Lindquist-type coordinates, the rotating
metric is
\begin{align}
ds^{2}
={}&
-\left(
1-\frac{2r\,m(r)}{\Sigma}
\right)dt^{2}
-\frac{
4ar\,m(r)\sin^{2}\theta
}{
\Sigma
}
dt\,d\phi
\nonumber\\
&+
\frac{\Sigma}{\Delta}\,dr^{2}
+
\Sigma\,d\theta^{2}
\nonumber\\
&+
\left[
r^{2}+a^{2}
+
\frac{
2a^{2}r\,m(r)\sin^{2}\theta
}{
\Sigma
}
\right]
\sin^{2}\theta\,d\phi^{2}.
\label{eq:app_rotating_metric}
\end{align}
The coordinate transformation from the rotating null coordinates
\((u,r,\theta,\widetilde\phi)\) is
\begin{equation}
du
=
dt
-
\frac{
r^{2}+a^{2}
}{
\Delta(r)
}
dr,
\label{eq:app_bl_u_transform}
\end{equation}
\begin{equation}
d\widetilde\phi
=
d\phi
-
\frac{a}{\Delta(r)}dr.
\label{eq:app_bl_phi_transform}
\end{equation}
Because \(\Delta\) depends only on \(r\), this transformation removes
both \(g_{tr}\) and \(g_{r\phi}\). A complexification producing an
explicitly \(\theta\)-dependent \(\Delta\) would not admit this
Boyer--Lindquist transformation in the same form.

The nonvanishing inverse-metric components are
\begin{equation}
g^{tt}
=
-
\frac{
(r^{2}+a^{2})^{2}
-
a^{2}\Delta\sin^{2}\theta
}{
\Sigma\Delta
},
\label{eq:app_inverse_tt}
\end{equation}
\begin{equation}
g^{t\phi}
=
-
\frac{
2ar\,m(r)
}{
\Sigma\Delta
},
\label{eq:app_inverse_tphi}
\end{equation}
\begin{equation}
g^{\phi\phi}
=
\frac{
\Delta-a^{2}\sin^{2}\theta
}{
\Sigma\Delta\sin^{2}\theta
},
\label{eq:app_inverse_phiphi}
\end{equation}
together with
\begin{equation}
g^{rr}
=
\frac{\Delta}{\Sigma},
\qquad
g^{\theta\theta}
=
\frac{1}{\Sigma}.
\label{eq:app_inverse_radial_angular}
\end{equation}
Direct multiplication gives
\begin{equation}
g_{\mu\alpha}g^{\alpha\nu}
=
\delta_{\mu}^{\nu}.
\label{eq:app_inverse_identity}
\end{equation}

The determinant is
\begin{equation}
{
\det(g_{\mu\nu})
=
-\Sigma^{2}\sin^{2}\theta
}.
\label{eq:app_metric_determinant}
\end{equation}
In particular,
\begin{equation}
\sqrt{-g}
=
\Sigma\sin\theta.
\label{eq:app_volume_element}
\end{equation}

Several limiting cases provide immediate algebraic checks. In the
nonrotating limit,
\begin{equation}
a\longrightarrow0,
\end{equation}
one obtains
\begin{equation}
\Sigma\longrightarrow r^{2},
\qquad
\Delta\longrightarrow r^{2}-2r\,m(r),
\end{equation}
and Eq.~\eqref{eq:app_rotating_metric} reduces to the static seed
Eq.~\eqref{eq:app_static_seed}.

When the mass function is constant,
\begin{equation}
m(r)
\longrightarrow M,
\end{equation}
the metric becomes the Kerr metric with
\begin{equation}
\Delta_{\rm Kerr}
=
r^{2}-2Mr+a^{2}.
\label{eq:app_kerr_limit_delta}
\end{equation}
When both
\begin{equation}
a\longrightarrow0,
\qquad
m(r)\longrightarrow M,
\end{equation}
the Schwarzschild metric is recovered.

For the finite-mass FDM profile,
\begin{equation}
m(r)
=
M_{\rm ADM}
+
{\cal O}(r^{-13})
\end{equation}
at large radius. The leading asymptotic metric components are
\begin{equation}
g_{tt}
=
-1
+
\frac{2M_{\rm ADM}}{r}
+
{\cal O}(r^{-2}),
\label{eq:app_asymptotic_gtt}
\end{equation}
and
\begin{equation}
g_{t\phi}
=
-
\frac{
2aM_{\rm ADM}\sin^{2}\theta
}{
r
}
+
{\cal O}(r^{-2}).
\label{eq:app_asymptotic_gtphi}
\end{equation}
The corresponding asymptotic charges are therefore
\begin{equation}
M_{\rm ADM}
=
M_{\bullet}+M_{\rm sol},
\qquad
J_{\rm ADM}
=
aM_{\rm ADM}.
\label{eq:app_asymptotic_charges}
\end{equation}

The inverse metric also provides a direct separability check. For the
Hamilton--Jacobi ansatz
\begin{equation}
S
=
-Et
+
L_{z}\phi
+
S_{r}(r)
+
S_{\theta}(\theta),
\end{equation}
the null Hamilton--Jacobi equation reduces to
\begin{align}
0
={}&
\Delta
\left(
\frac{dS_{r}}{dr}
\right)^{2}
-
\frac{
\left[
E(r^{2}+a^{2})-aL_{z}
\right]^{2}
}{
\Delta
}
\nonumber\\
&+
\left(
\frac{dS_{\theta}}{d\theta}
\right)^{2}
+
L_{z}^{2}\cot^{2}\theta
-
a^{2}E^{2}\cos^{2}\theta
+
(L_{z}-aE)^{2}.
\label{eq:app_hj_separability_identity}
\end{align}
The \(r\)- and \(\theta\)-dependent terms can be separated by a single
constant. Thus the fixed Newman--Janis prescription retains the
off-shell Kerr separability structure.

The symbolic verification targets used in the calculation are
\begin{align}
g_{\mu\alpha}g^{\alpha\nu}
&=
\delta_{\mu}^{\nu},
\label{eq:app_check_inverse}
\\
\det(g_{\mu\nu})
&=
-\Sigma^{2}\sin^{2}\theta,
\label{eq:app_check_determinant}
\\
g_{tr}
&=
g_{r\phi}
=
0,
\label{eq:app_check_bl}
\\
\left.g_{\mu\nu}\right|_{a=0}
&=
g_{\mu\nu}^{\rm static},
\label{eq:app_check_static}
\\
\left.g_{\mu\nu}\right|_{m(r)=M}
&=
g_{\mu\nu}^{\rm Kerr}.
\label{eq:app_check_kerr}
\end{align}
The same symbolic pass computes \(G_{\mu\nu}\), forms
\(T_{\mu\nu}=G_{\mu\nu}/8\pi\), and simplifies each component of the
covariant divergence to
\begin{equation}
\left\{\nabla_\mu T^{\mu t},\nabla_\mu T^{\mu r},
\nabla_\mu T^{\mu\theta},\nabla_\mu T^{\mu\phi}\right\}
=\{0,0,0,0\}.
\label{eq:app_stress_conservation}
\end{equation}
Diagonalizing \(T^{(a)}{}_{(b)}\) gives the two doubly degenerate
eigenvalues stated in Sec.~\ref{subsec:carter_effective_source}.  Independent
contractions reproduce Eqs.~\eqref{eq:sec4_rotating_ricci} and
\eqref{eq:sec4_rotating_ricci_squared}.  After putting the Kretschmann
scalar over a common denominator and canceling common factors, its only
geometric denominator is \(\Sigma^6\); neither \(\Delta\) nor
\(\Delta'\) remains.  Therefore simple and double Killing horizons are
regular curvature loci, whereas \(\Sigma=0\) is the only possible
curvature singularity for a smooth \(m(r)\).
The distributed script \texttt{kretschmann\_cas.wls} constructs the metric,
connection, Riemann tensor, and contraction from first principles and
exports the uncancelled scalar, the complete \(\Sigma^6\) numerator, its
denominator, and the constant-\(m\) Kerr limit; no hand-entered curvature
formula is used.

The closed-timelike-curve check does not require a numerical scan.  The
azimuthal component can be rearranged exactly as
\begin{equation}
\Sigma\,g_{\phi\phi}/\sin^2\theta
=(r^2+a^2)\Sigma+2a^2r\,m(r)\sin^2\theta,
\label{eq:app_gphiphi_positive}
\end{equation}
which is positive for the complete \(r>0\) benchmark domain.  Finally,
the Carter coframe contains \(\sqrt{\Delta/\Sigma}\) and is used only in
stationary blocks with \(\Delta>0\); tensor invariants and
horizon-penetrating coordinates, rather than that frame, are used on the
horizon itself.
These identities test the internal algebra of the selected construction.
They do not remove the intrinsic nonuniqueness of the Newman--Janis
complexification itself.

\section{Closed mass integral and energy-condition reduction}
\label{app:mass_energy}

The enclosed FDM mass is
\begin{equation}
M_{\rm FDM}(r)
=
4\pi\rho_{c}
\int_{0}^{r}
s^{2}
\left(
1+\alpha\frac{s^{2}}{r_{c}^{2}}
\right)^{-8}
ds.
\label{eq:app_mass_integral}
\end{equation}
Introducing
\begin{equation}
u
=
\alpha\frac{s^{2}}{r_{c}^{2}},
\qquad
z
=
\frac{u}{1+u},
\label{eq:app_mass_substitutions}
\end{equation}
gives the incomplete-beta representation
\begin{equation}
{
M_{\rm FDM}(r)
=
\frac{
2\pi\rho_{c}r_{c}^{3}
}{
\alpha^{3/2}
}
B_{z(r)}
\left(
\frac{3}{2},
\frac{13}{2}
\right)
},
\label{eq:app_mass_beta}
\end{equation}
where
\begin{equation}
z(r)
=
\frac{
\alpha r^{2}
}{
r_{c}^{2}+\alpha r^{2}
}.
\label{eq:app_z_definition}
\end{equation}
An equivalent hypergeometric form is
\begin{equation}
{
M_{\rm FDM}(r)
=
\frac{4\pi}{3}
\rho_{c}r^{3}
\,{}_2F_{1}
\left(
\frac{3}{2},
8;
\frac{5}{2};
-\alpha\frac{r^{2}}{r_{c}^{2}}
\right)
}.
\label{eq:app_mass_hypergeometric}
\end{equation}

For numerical and symbolic cross-checks, the integral also has an
elementary representation. Defining
\begin{equation}
y
=
\sqrt{\alpha}\frac{r}{r_{c}},
\label{eq:app_y_definition}
\end{equation}
one finds
\begin{equation}
M_{\rm FDM}(r)
=
\frac{
4\pi\rho_{c}r_{c}^{3}
}{
\alpha^{3/2}
}
{\cal I}(y),
\label{eq:app_mass_elementary}
\end{equation}
where
\begin{equation}
{\cal I}(y)
=
\frac{33}{2048}\arctan y
+
\frac{
y\,P(y^{2})
}{
215040(1+y^{2})^{7}
},
\label{eq:app_elementary_I}
\end{equation}
with
\begin{align}
P(u)
={}&
-3465
+
48580u
+
92323u^{2}
+
101376u^{3}
\nonumber\\
&+
65373u^{4}
+
23100u^{5}
+
3465u^{6}.
\label{eq:app_elementary_polynomial}
\end{align}

Differentiation of any of
Eqs.~\eqref{eq:app_mass_beta},
\eqref{eq:app_mass_hypergeometric}, or
\eqref{eq:app_mass_elementary} gives
\begin{equation}
{
M_{\rm FDM}'(r)
=
4\pi r^{2}\rho_{\rm FDM}(r)
}.
\label{eq:app_mass_derivative_check}
\end{equation}
The total soliton mass is finite:
\begin{align}
M_{\rm sol}
&=
M_{\rm FDM}(\infty)
\nonumber\\
&=
\frac{
2\pi\rho_{c}r_{c}^{3}
}{
\alpha^{3/2}
}
B
\left(
\frac{3}{2},
\frac{13}{2}
\right)
\nonumber\\
&=
{
\frac{
33\pi^{2}
}{
1024\alpha^{3/2}
}
\rho_{c}r_{c}^{3}
}.
\label{eq:app_total_soliton_mass}
\end{align}
The normalized enclosed-mass fraction is therefore
\begin{equation}
{\cal F}(x)
\equiv
\frac{
M_{\rm FDM}(r)
}{
M_{\rm sol}
}
=
I_{\frac{\alpha x^{2}}{1+\alpha x^{2}}}
\left(
\frac{3}{2},
\frac{13}{2}
\right),
\qquad
x=\frac{r}{r_{c}}.
\label{eq:app_normalized_mass_fraction}
\end{equation}

Near the center,
\begin{align}
M_{\rm FDM}(r)
=
\frac{4\pi}{3}\rho_{c}r^{3}
\bigg[
1
&-
\frac{24\alpha}{5}
\frac{r^{2}}{r_{c}^{2}}
+
\frac{108\alpha^{2}}{7}
\frac{r^{4}}{r_{c}^{4}}
\nonumber\\
&+
{\cal O}
\left(
\frac{r^{6}}{r_{c}^{6}}
\right)
\bigg].
\label{eq:app_central_mass_expansion}
\end{align}
At large radius,
\begin{equation}
M_{\rm sol}
-
M_{\rm FDM}(r)
=
\frac{
4\pi\rho_{c}r_{c}^{16}
}{
13\alpha^{8}r^{13}
}
\left[
1+
{\cal O}
\left(
\frac{r_{c}^{2}}{r^{2}}
\right)
\right].
\label{eq:app_mass_asymptotic_tail}
\end{equation}

For the rotating geometry, the eigenvalues of the effective stress
tensor in the natural orthonormal frame can be written as
\begin{equation}
8\pi\varrho
=
\frac{
2r^{2}m'(r)
}{
\Sigma^{2}
},
\label{eq:app_rotating_density}
\end{equation}
\begin{equation}
p_{r}
=
-\varrho,
\label{eq:app_rotating_radial_pressure}
\end{equation}
and
\begin{equation}
8\pi p_{\perp}
=
-
\frac{
r\Sigma m''(r)
+
2a^{2}\cos^{2}\theta\,m'(r)
}{
\Sigma^{2}
}.
\label{eq:app_rotating_tangential_pressure}
\end{equation}
Since
\begin{equation}
m'(r)
=
4\pi r^{2}\rho_{\rm FDM}(r)
\geq0,
\end{equation}
the effective density is nonnegative:
\begin{equation}
\varrho
=
\frac{
r^{4}\rho_{\rm FDM}(r)
}{
\Sigma^{2}
}
\geq0.
\label{eq:app_effective_density_positive}
\end{equation}
The radial NEC is saturated,
\begin{equation}
\varrho+p_{r}=0.
\label{eq:app_rotating_radial_nec}
\end{equation}

The tangential NEC combination is
\begin{equation}
8\pi
\left(
\varrho+p_{\perp}
\right)
=
\frac{
2(r^{2}-a^{2}\cos^{2}\theta)m'
-
r\Sigma m''
}{
\Sigma^{2}
}.
\label{eq:app_tangential_nec_general}
\end{equation}
Using the closed FDM density derivative, this reduces to
\begin{equation}
{
\varrho+p_{\perp}
=
\frac{
2r^{2}\rho_{\rm FDM}(r)
}{
\Sigma^{2}
\left(
r_{c}^{2}+\alpha r^{2}
\right)
}
{\cal Q}(r,\theta)
},
\label{eq:app_tangential_nec_reduced}
\end{equation}
where
\begin{align}
{\cal Q}(r,\theta)
&=
4\alpha r^{2}\Sigma
-
a^{2}\cos^{2}\theta
\left(
r_{c}^{2}+\alpha r^{2}
\right)
\nonumber\\
&=
4\alpha r^{4}
+
a^{2}\cos^{2}\theta
\left(
3\alpha r^{2}-r_{c}^{2}
\right).
\label{eq:app_Q_factor}
\end{align}

For fixed \(r\), \({\cal Q}\) is linear in
\begin{equation}
u=\cos^{2}\theta.
\end{equation}
When
\begin{equation}
3\alpha r^{2}<r_{c}^{2},
\end{equation}
its minimum occurs on the rotation axis, \(u=1\). When
\begin{equation}
3\alpha r^{2}\geq r_{c}^{2},
\end{equation}
its minimum occurs at the equator, where
\begin{equation}
{\cal Q}(r,\pi/2)
=
4\alpha r^{4}>0.
\end{equation}
Consequently, any tangential effective-Einstein-source NEC- or WEC-sign
violation first appears on the rotation axis.

The all-angle exterior condition therefore reduces to
\begin{equation}
{\cal Q}(r,0)
=
4\alpha r^{4}
+
3\alpha a^{2}r^{2}
-
a^{2}r_{c}^{2}
\geq0.
\label{eq:app_axis_wec_condition}
\end{equation}
For \(f_{\rm sol}>0\), the positive root defines the critical WEC
sign-factor radius:
\begin{equation}
{
r_{\rm WEC}^{2}
=
\frac{
-3\alpha a^{2}
+
\sqrt{
9\alpha^{2}a^{4}
+
16\alpha a^{2}r_{c}^{2}
}
}{
8\alpha
}
}.
\label{eq:app_wec_radius}
\end{equation}
For \(f_{\rm sol}>0\), the WEC is satisfied throughout the black-hole
exterior if and only if
\begin{equation}
r_{+}\geq r_{\rm WEC}.
\label{eq:app_exterior_wec_condition}
\end{equation}
For \(f_{\rm sol}=0\), the effective matter tensor vanishes identically;
in that vacuum limit the WEC is trivially satisfied and
\(r_{\rm WEC}\) should not be interpreted as a physical matter
boundary.

\section{Numerical algorithms and convergence tests}
\label{app:numerics}

All numerical calculations are performed in units
\begin{equation}
M_{\rm ADM}=1.
\end{equation}
The dimensionless mass function is
\begin{equation}
\mu(x)
=
1-f_{\rm sol}
+
f_{\rm sol}
{\cal F}
\left(
\frac{x}{\widehat r_{c}}
\right),
\label{eq:app_numerical_mass_function}
\end{equation}
where
\begin{equation}
x=\frac{r}{M_{\rm ADM}},
\qquad
\widehat r_{c}
=
\frac{r_{c}}{M_{\rm ADM}}.
\end{equation}
The incomplete-beta representation is used for the production
calculations, while the hypergeometric and elementary expressions are
used as independent cross-checks.

The derivative is evaluated analytically:
\begin{equation}
\mu'(x)
=
\frac{
f_{\rm sol}
}{
\widehat r_{c}
}
{\cal F}'
\left(
\frac{x}{\widehat r_{c}}
\right),
\label{eq:app_numerical_mass_derivative}
\end{equation}
with
\begin{equation}
{\cal F}'(y)
=
\frac{
4\pi
}{
\mu_{\infty}
}
\frac{
y^{2}
}{
(1+\alpha y^{2})^{8}
},
\qquad
\mu_{\infty}
=
\frac{
33\pi^{2}
}{
1024\alpha^{3/2}
}.
\label{eq:app_F_derivative}
\end{equation}
Analytical derivatives are used instead of finite differences in the
horizon, extremality, and photon-orbit equations.

\subsection{Horizon and extremality algorithms}

The dimensionless horizon function is
\begin{equation}
\widehat\Delta(x)
=
x^{2}-2x\mu(x)+j_{\rm ADM}^{2}.
\label{eq:app_numerical_delta}
\end{equation}
Positive roots are first bracketed on a combined logarithmic and linear
radial grid. Each sign-changing interval is then refined with Brent's
method. Duplicate roots closer than the root tolerance are discarded, and
\emph{all} positive roots are retained before the largest is identified as
\(x_{+}\).  Because a sign-change search alone can miss an even root, all
zeros of \(\widehat\Delta'\) are independently bracketed and refined; the
value of \(\widehat\Delta\) at every stationary point is tested against the
double-root tolerance.  The calculation also samples
\(\widehat\Delta''\) on the same intervals.  Thus the extremal curve is not
inferred from a failed sign change of \(\widehat\Delta\).

The classification grid covers Eq.~\eqref{eq:sec5_scan_box} with 4001
uniform values of \(f_{\rm sol}\), 1601 spin values, and four fixed-core
slices, for \(25\,622\,404\) spin--amplitude points.  The radial work is
not repeated at every spin.  Because
\(\widehat\Delta'=2G(x;f_{\rm sol},\widehat r_c)\) is independent of
\(j_{\rm ADM}\), the profile arrays are precomputed once per core scale on
the hybrid mesh
\(10^{-8}\leq x\leq50\) (20000 linear plus 4000 logarithmic points, 23999
unique values).  \(G\) and \(G'\) are then evaluated for all amplitudes in
vectorized blocks of 128.  After the unique stationary radius and
Brent-refined \(j_{\rm ext}\) are known, all 1601 spins are classified by
comparison with that boundary.  Thus the radial stage contains
\(4\times4001\times23999\) vectorized values, rather than an implied
\(6.15\times10^{11}\) repeated evaluations.

This complete two-stage audit required \(4.62\) s wall time on the
reference machine specified below.  It found exactly one positive zero of
\(\widehat\Delta'\) for every \((f_{\rm sol},\widehat r_c)\); the minimum
sampled \(G'\) was \(0.78305768\), \(0.85537179\), \(0.91322307\), and
\(0.95661153\) for \(\widehat r_c=2,3,5,10\), respectively.  All plotted
boundaries are subsequently refined by bracketed scalar root finding.
The result agrees with the general theorem in
Eqs.~\eqref{eq:sec5_general_convexity_condition}--%
\eqref{eq:sec5_general_root_classification} and its profile-specific
margin Eq.~\eqref{eq:sec5_G_monotonic_bound}; no three-root, four-root, or
second-double-root case was found.  The analytic theorem, not the grid, is
the exclusion argument.

The extremal radius is obtained by solving
\begin{equation}
{\cal E}(x)
\equiv
x-\mu(x)-x\mu'(x)
=
0.
\label{eq:app_extremal_root_function}
\end{equation}
The extremal spin is subsequently evaluated from
\begin{equation}
j_{\rm ext}
=
x_{e}
\sqrt{
1-2\mu'(x_{e})
}.
\label{eq:app_numerical_extremal_spin}
\end{equation}
This two-step procedure is more stable than solving
\(\widehat\Delta=\widehat\Delta'=0\) simultaneously.

The numerical solution is accepted only when
\begin{equation}
\left|
\widehat\Delta(x_{e})
\right|
<
\epsilon_{\rm root},
\qquad
\left|
\widehat\Delta'(x_{e})
\right|
<
\epsilon_{\rm root}.
\label{eq:app_extremal_residual_test}
\end{equation}
The calculations reported in the main text use
\begin{equation}
\epsilon_{\rm root}
=
10^{-11}
\end{equation}
or smaller.

For \(f_{\rm sol}>0\), the exterior effective-source WEC-sign boundary is
found by solving
\begin{equation}
{\cal W}(j_{\rm ADM})
\equiv
x_{+}
\left(
j_{\rm ADM},f_{\rm sol},\widehat r_{c}
\right)
-
x_{\rm WEC}
\left(
j_{\rm ADM},\widehat r_{c}
\right)
=
0
\label{eq:app_wec_boundary_function}
\end{equation}
within the subextremal interval
\begin{equation}
0<j_{\rm ADM}<j_{\rm ext}.
\end{equation}

\subsection{Photon-region and shadow algorithms}

For each subextremal configuration, the spherical-orbit impact
parameters are evaluated from
\begin{equation}
\xi_{c}(x_{p})
=
\frac{
(x_{p}^{2}+j_{\rm ADM}^{2})\widehat\Delta'
-
4x_{p}\widehat\Delta
}{
j_{\rm ADM}\widehat\Delta'
},
\label{eq:app_numerical_xi}
\end{equation}
\begin{equation}
\eta_{c}(x_{p})
=
\frac{
16x_{p}^{2}\widehat\Delta
}{
(\widehat\Delta')^{2}
}
-
(\xi_{c}-j_{\rm ADM})^{2}.
\label{eq:app_numerical_eta}
\end{equation}
The initial radial scan begins immediately outside the outer horizon,
\begin{equation}
x_{p,\min}
=
x_{+}(1+\epsilon_{h}),
\end{equation}
with
\begin{equation}
\epsilon_{h}
=
10^{-6}.
\end{equation}

The visible branch is selected by the simultaneous conditions
\begin{equation}
{\cal B}(x_{p})
\equiv
\eta_{c}
+
j_{\rm ADM}^{2}\cos^{2}\iota
-
\xi_{c}^{2}\cot^{2}\iota
\geq0
\label{eq:app_visibility_function}
\end{equation}
and
\begin{equation}
{\cal R}''(x_{p})>0.
\label{eq:app_instability_test}
\end{equation}
The exterior scan enumerates all roots of both \({\cal B}=0\) and
\({\cal R}''=0\), including independent tests for non-sign-changing
near-zero minima.  Their ordered union partitions the exterior radial
domain; a subinterval is retained only when its midpoint satisfies both
Eqs.~\eqref{eq:app_visibility_function} and
\eqref{eq:app_instability_test}.  The endpoints of every retained interval
are refined by solving
\begin{equation}
{\cal B}(x_{p})=0
\label{eq:app_shadow_endpoint_equation}
\end{equation}
with Brent's method when the boundary is a visibility zero.  The audited
production box contains exactly one retained interval with two such
endpoints.  The single-component area routine checks this count and fails
explicitly rather than joining nonadjacent endpoints.  If a future profile
produces several intervals, each must be mapped separately and the
screen-space union or outer envelope constructed before an area is
assigned.

The upper shadow branch is
\begin{equation}
X(x_{p})
=
-\frac{\xi_{c}(x_{p})}{\sin\iota},
\qquad
Y(x_{p})
=
\sqrt{{\cal B}(x_{p})},
\label{eq:app_upper_shadow_branch}
\end{equation}
and the lower branch follows from reflection:
\begin{equation}
Y\longrightarrow-Y.
\end{equation}

For a discretized closed contour
\begin{equation}
\left\{
X_{k},Y_{k}
\right\}_{k=1}^{N},
\end{equation}
the shadow area is evaluated with the polygon formula
\begin{equation}
A_{\rm sh}^{(N)}
=
\frac{1}{2}
\left|
\sum_{k=1}^{N}
\left(
X_{k}Y_{k+1}
-
Y_{k}X_{k+1}
\right)
\right|,
\label{eq:app_polygon_area}
\end{equation}
where
\begin{equation}
X_{N+1}=X_{1},
\qquad
Y_{N+1}=Y_{1}.
\end{equation}
The area-equivalent radius is
\begin{equation}
R_{A}^{(N)}
=
\sqrt{
\frac{
A_{\rm sh}^{(N)}
}{\pi}
}.
\label{eq:app_numerical_area_radius}
\end{equation}

For the \(f_{\rm sol}\to0^+\) derivative, the production calculation avoids
subtracting two polygonal contours.  It evaluates the continuous upper
branch as
\begin{equation}
A_{\rm sh}
=2\left|\int_{x_{p,-}}^{x_{p,+}}
Y(x_p)\frac{dX}{dx_p}\,dx_p\right|
\label{eq:app_continuous_area}
\end{equation}
and uses
\begin{equation}
x_p=\frac{x_{p,+}+x_{p,-}}{2}
+\frac{x_{p,+}-x_{p,-}}{2}\sin u
\label{eq:app_endpoint_regularization}
\end{equation}
to regularize both square-root endpoints.  The derivative \(dX/dx_p\) is
evaluated analytically and the remaining integral is adaptively
quadratured.

The horizontal displacement is calculated from
\begin{equation}
D_{\rm sh}
=
\left|
\frac{
X_{\max}+X_{\min}
}{2}
\right|.
\label{eq:app_numerical_displacement}
\end{equation}
For the angular residual, the curve is recentered and interpolated on a
uniform grid in
\begin{equation}
\psi
=
\operatorname{atan2}
\left(
Y,
X-X_{c}
\right).
\end{equation}

\paragraph{All-branch and near-extremal audit.}
To remove the poles in the impact-parameter representation without
changing any exterior root, the independent audit uses the smooth
numerators
\begin{equation}
Q_{\cal B}
=j_{\rm ADM}^{2}(\widehat\Delta')^{2}{\cal B},
\qquad
Q_{{\cal R}''}
=\frac{(\widehat\Delta')^{2}}{8x_p}{\cal R}''.
\label{eq:app_smooth_photon_root_functions}
\end{equation}
Outside the outer horizon \(\widehat\Delta'>0\), so these functions have
the same signs and zeros as \({\cal B}\) and \({\cal R}''\).
Sign-changing roots are bracketed on a log-dense horizon-adapted mesh from
\(x_+(1+10^{-12})\) to \(x=10^6\); local minima of the normalized
absolute numerators are tested independently for tangent roots.  The
remaining exterior is controlled by
\(Q_{\cal B}\sim-4\csc^2\!\iota\,x^6<0\) and
\(Q_{{\cal R}''}\sim4x^3>0\), since the profile tail decays as
\(\rho\sim r^{-16}\).

The completeness grid contains
\[
\widehat r_c=2,3,5,10,\qquad
f_{\rm sol}=0,0.01,\ldots,0.30,
\]
and 24 relative spins
\[
\begin{split}
q={}&0.05,0.10,\ldots,0.90,\\
&0.925,0.95,0.975,0.99,0.995,0.999
\end{split}
\]
at \(\iota=60^\circ\), for 2976 configurations.  Every point has exactly
two exterior zeros of \({\cal B}\), no exterior zero of
\({\cal R}''\), and one connected visible unstable interval.  No
non-sign-changing root candidate is found.  The maximum normalized
\({\cal B}\)-root residual is \(4.40\times10^{-13}\), and the minimum
sampled \(Q_{{\cal R}''}/S_{{\cal R}''}\) is \(0.09835\).  The static
edge is audited separately at all \(4\times31=124\) profile points:
each has one exterior photon circle, all are unstable, and no tangent
root is found.

\begin{figure*}[t]
\centering
\includegraphics[width=0.96\textwidth]
{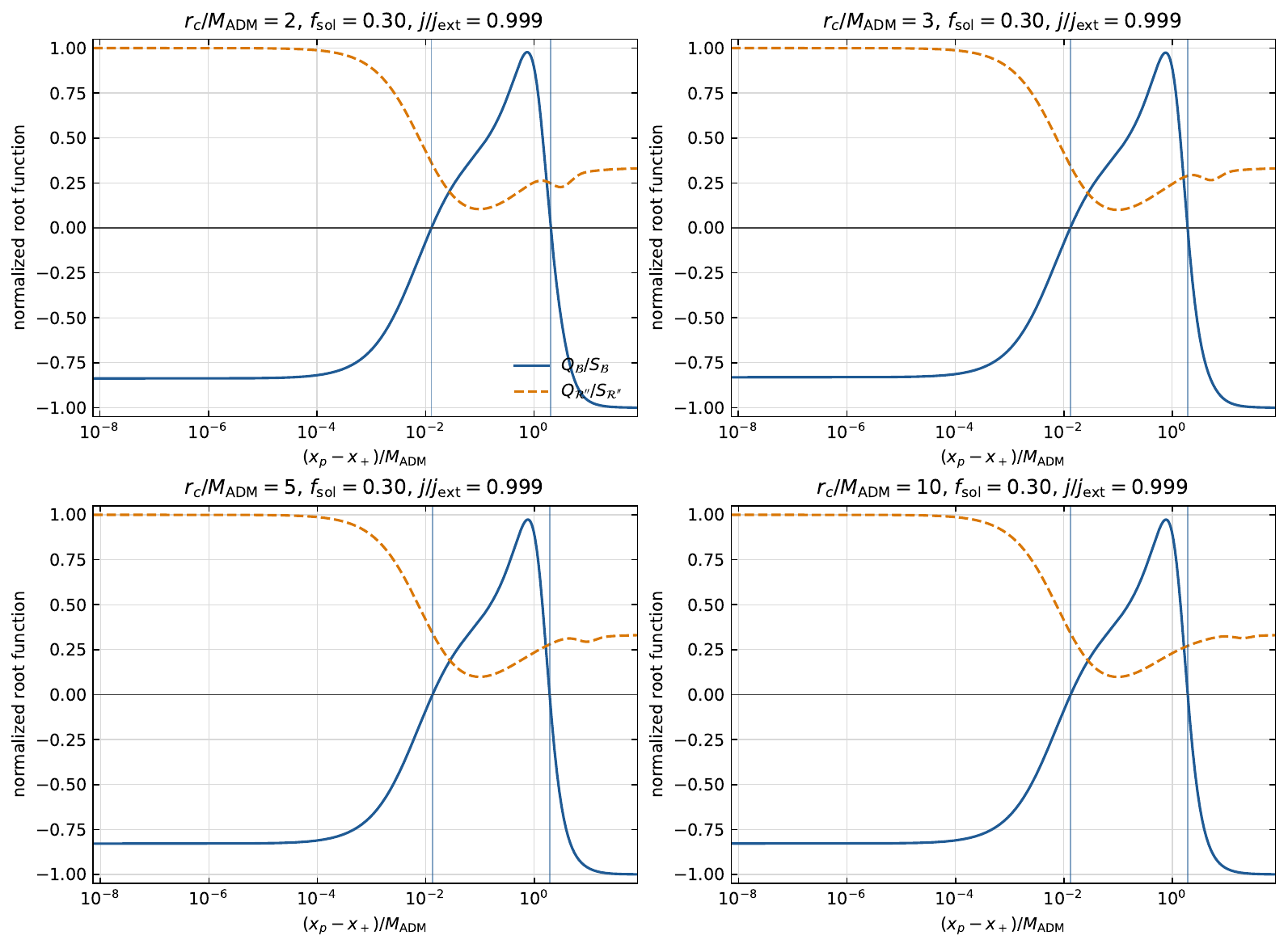}
\caption{Representative exterior root functions for the most near-extremal
audited slice, \(f_{\rm sol}=0.30\) and \(q=0.999\), at the four displayed
core scales.  The horizontal coordinate is the distance from the outer
horizon.  The plotted quantities are
\(Q_{\cal B}/S_{\cal B}\) and
\(Q_{{\cal R}''}/S_{{\cal R}''}\), where each positive scale \(S\) is
the sum of the absolute algebraic terms in the corresponding smooth
numerator.  The blue vertical lines are the two visibility roots; the
instability function has no exterior zero.  The visible unstable set is
the single interval between the blue lines.  The numerical search extends
to \(x=10^6\), beyond the plotted range, and is closed by the analytic
asymptotic signs stated in the text.}
\label{fig:app_photon_branch_audit}
\end{figure*}

The smallest separation found between the prograde visible endpoint and
the horizon is
\[
\frac{x_{p,-}-x_+}{x_+}=1.70428\times10^{-2};
\]
the smallest equatorial prograde-orbit separation is
\(6.85486\times10^{-3}\).  Both occur at
\((\widehat r_c,f_{\rm sol},q)=(2,0.30,0.999)\), so even the formerly used
\(\epsilon_h=2\times10^{-4}\) lies well inside the empty near-horizon
gap.

Table~\ref{tab:app_near_extremal_cutoff} gives the explicit convergence
test.  It covers \(q=0.95,0.99,0.999\), seven values
\(f_{\rm sol}=0,0.05,\ldots,0.30\), and all four core scales: 84
configurations per cutoff.  Differences are measured against the
\(10^{-12}\) reference search.
The completeness, static-edge, and cutoff stages together required
\(29.916\) s on the reference machine.

\begin{table*}[t]
\caption{Near-extremal sensitivity to the horizon-relative starting
offset.  No configuration failed.  Entries are maximum absolute
differences over the 84 tested configurations; \(x_{p,-}\) and
\(x_{p,+}\) are the prograde and retrograde visible endpoints, and
\(x_{\rm ph}^{\rm pro}\) is the equatorial prograde orbit.}
\label{tab:app_near_extremal_cutoff}
\begin{ruledtabular}
\begin{tabular}{ccccccc}
\(\epsilon_h\) & failures &
\(\max|\delta x_{p,-}|\) & \(\max|\delta x_{p,+}|\) &
\(\max|\delta x_{\rm ph}^{\rm pro}|\) &
\(\max|\delta R_A|\) & \(\max|\delta D_{\rm sh}|\)\\
\hline
\(10^{-3}\) & \(0/84\) & \(1.02\times10^{-14}\) & \(0\) &
\(1.42\times10^{-14}\) & \(8.88\times10^{-16}\) & \(1.11\times10^{-14}\)\\
\(2\times10^{-4}\) & \(0/84\) & \(1.51\times10^{-14}\) & \(0\) &
\(2.02\times10^{-14}\) & \(1.78\times10^{-15}\) & \(1.60\times10^{-14}\)\\
\(10^{-4}\) & \(0/84\) & \(1.53\times10^{-14}\) & \(0\) &
\(1.48\times10^{-14}\) & \(1.78\times10^{-15}\) & \(1.87\times10^{-14}\)\\
\(10^{-5}\) & \(0/84\) & \(1.51\times10^{-14}\) & \(0\) &
\(1.51\times10^{-14}\) & \(1.78\times10^{-15}\) & \(1.80\times10^{-14}\)\\
\(10^{-6}\) & \(0/84\) & \(1.83\times10^{-14}\) & \(0\) &
\(1.44\times10^{-14}\) & \(1.78\times10^{-15}\) & \(1.67\times10^{-14}\)\\
\end{tabular}
\end{ruledtabular}
\end{table*}

Thus the adopted \(10^{-6}\) production offset neither removes a
near-horizon branch nor biases the reported endpoints, area radius, or
horizontal displacement on the audited domain.  This is a finite-grid
verification for the selected Newman--Janis completion, not an analytic
light-ring theorem for arbitrary matter geometries.

\subsection{Perturbative coefficients}

The first-order corrections to the horizon, extremal spin, and static
photon sphere are evaluated from the analytical expressions
\begin{equation}
x_{+}^{(1)}
=
\frac{
x_{+}^{(0)}h(x_{+}^{(0)})
}{
x_{+}^{(0)}-1
},
\label{eq:app_horizon_linear_coefficient}
\end{equation}
\begin{equation}
j_{\rm ext}^{(1)}
=
h(1),
\label{eq:app_extremal_linear_coefficient}
\end{equation}
and
\begin{equation}
x_{\rm ph}^{(1)}
=
3
\left[
h(3)-h'(3)
\right].
\label{eq:app_photon_linear_coefficient}
\end{equation}
For the shadow area radius, let \(s>0\) denote a numerical step in
\(f_{\rm sol}\), distinct from the profile function \(h(x)\), and define
the one-sided quotient
\begin{equation}
D(s)=\frac{R_A(s)-R_A(0)}{s}
\label{eq:app_shadow_one_sided_quotient}
\end{equation}
and its first Richardson extrapolant
\begin{equation}
D_{\rm R}(s)=2D(s/2)-D(s).
\label{eq:app_shadow_richardson}
\end{equation}

\begin{table}[t]
\caption{One-sided convergence of the endpoint-regularized numerical
derivative of the area-equivalent shadow radius at \(f_{\rm sol}=0^+\).}
\label{tab:app_shadow_fit_stability}
\begin{ruledtabular}
\begin{tabular}{ccc}
\(s\) & \(D(s)\) & \(D_{\rm R}(s)\)\\
\hline
\(10^{-3}\) & \(-4.1347708198\) & \(-4.1322265034\)\\
\(10^{-4}\) & \(-4.1324811355\) & \(-4.1322267791\)\\
\(10^{-5}\) & \(-4.1322522167\) & \(-4.1322267817\)\\
\(10^{-6}\) & \(-4.1322293258\) & ---\\
\end{tabular}
\end{ruledtabular}
\end{table}
The Richardson values stabilize at \(-4.13222678\); their change between
\(s=10^{-4}\) and \(10^{-5}\) is \(2.6\times10^{-9}\), while the wider
\(10^{-3}\) comparison bounds residual window curvature by
\(2.8\times10^{-7}\).  Adaptive quadrature reports an area error below
\(5\times10^{-13}M_{\rm ADM}^2\), corresponding to less than
\(2\times10^{-8}\) in the \(s=10^{-6}\) difference quotient. Tightening
the visibility-root tolerance from \(10^{-12}\) to \(10^{-14}\) changes
the derivative by less than \(10^{-8}\).  We therefore report separately:
integration error \(<2\times10^{-8}\), endpoint error \(<10^{-8}\), and
window-truncation error \(<3\times10^{-7}\), and quote the conservative
combined value \(-4.132227\pm5\times10^{-7}\).

\subsection{Convergence test}

As a representative convergence test, consider
\begin{equation}
f_{\rm sol}=0.10,
\qquad
\widehat r_{c}=3,
\qquad
j_{\rm ADM}=0.70,
\qquad
\iota=60^{\circ}.
\label{eq:app_convergence_parameters}
\end{equation}
The outer horizon is
\begin{equation}
x_{+}
=
1.4756379397.
\label{eq:app_convergence_horizon}
\end{equation}
The visible spherical-orbit endpoints are first refined from
Eq.~\eqref{eq:app_shadow_endpoint_equation}, after which the shadow
interval is sampled with \(N_{p}\) points.

\begin{table}[t]
\caption{
Convergence of the area-equivalent shadow radius for the representative
configuration in Eq.~\eqref{eq:app_convergence_parameters}. The relative
error is measured with respect to the endpoint-regularized continuous
quadrature result, \(R_A/M_{\rm ADM}=4.65524223\).
}
\label{tab:app_shadow_convergence}
\begin{ruledtabular}
\begin{tabular}{ccc}
\(N_{p}\)
&
\(R_A/M_{\rm ADM}\)
&
relative error
\\
\hline
500
&
4.65499872
&
\(5.23\times10^{-5}\)
\\
1000
&
4.65515617
&
\(1.85\times10^{-5}\)
\\
2000
&
4.65521180
&
\(6.54\times10^{-6}\)
\\
4000
&
4.65523147
&
\(2.31\times10^{-6}\)
\\
8000
&
4.65523843
&
\(8.17\times10^{-7}\)
\\
16000
&
4.65524089
&
\(2.88\times10^{-7}\)
\\
32000
&
4.65524176
&
\(1.02\times10^{-7}\)
\\
\end{tabular}
\end{ruledtabular}
\end{table}

For the same configuration, the endpoint-refined critical curve gives
\begin{equation}
\frac{
\Delta X_{\rm sh}
}{
M_{\rm ADM}
}
=
9.14248177,
\qquad
\frac{
\Delta Y_{\rm sh}
}{
M_{\rm ADM}
}
=
9.46394036.
\label{eq:app_shadow_extents}
\end{equation}
The vertical extent is particularly stable because its maximum occurs
inside the sampled orbit interval, whereas the horizontal extrema are
controlled directly by the refined branch endpoints.

Additional consistency residuals are monitored through
\begin{equation}
\epsilon_{\rm mass}
=
\left|
\frac{
M_{\rm FDM}'(r)
-
4\pi r^{2}\rho_{\rm FDM}(r)
}{
4\pi r^{2}\rho_{\rm FDM}(r)
}
\right|,
\label{eq:app_mass_residual}
\end{equation}
\begin{equation}
\epsilon_{\rm inv}
=
\max_{\mu,\nu}
\left|
g_{\mu\alpha}g^{\alpha\nu}
-
\delta_{\mu}^{\nu}
\right|,
\label{eq:app_inverse_residual}
\end{equation}
and
\begin{equation}
\epsilon_{\rm det}
=
\left|
\frac{
\det(g_{\mu\nu})
+
\Sigma^{2}\sin^{2}\theta
}{
\Sigma^{2}\sin^{2}\theta
}
\right|.
\label{eq:app_determinant_residual}
\end{equation}
Production points are retained only when the relevant algebraic and
root residuals are below the adopted numerical tolerance.

The combination of analytical derivatives, bracketed root finding,
endpoint refinement, and explicit resolution doubling prevents the
horizon and shadow results from depending on an unverified local
optimizer or on a single radial discretization.

\begin{acknowledgments}
The authors gratefully acknowledges the Department of Theoretical Physics at Lomonosov Moscow State University. 
\end{acknowledgments}

\end{document}